\documentclass[aps,pre,showpacs,showkeys,onecolumn,10pt]{revtex4-1}
\usepackage[dvips]{graphicx}
\usepackage[dvips]{color}
\usepackage{hyperref,breakurl,amsmath,amssymb,pst-node,eepic}
\usepackage{bbold}

\newcommand{\RE}{\Re e}
\newcommand{\IM}{\Im m}

\begin{document}

\title{Quaternionic R transform and non-hermitian random matrices}

\author{Zdzislaw Burda}
\email{zdzislaw.burda@agh.edu.pl}
\affiliation{Faculty of Physics and Applied Computer Science, AGH University of Science and Technology,  al. Mickiewicza 30, PL--30059 Krak{\'o}w, Poland}
\author{Artur Swiech}
\email{swiech@thp.uni-koeln.de}
\affiliation{Institute for Theoretical Physics, \\ University of Cologne, 
Z{\"u}lpicher Stra{\ss}e 77, D-50937 K{\"o}ln, Germany}

\begin{abstract}
Using the Cayley-Dickson construction we rephrase and review the 
non-hermitian diagrammatic formalism [R. A. Janik, M. A. Nowak, G. Papp and I. Zahed, Nucl.Phys. B \textbf{501}, 603 (1997)], that generalizes the free probability calculus
to asymptotically large non-hermitian random matrices. The main object in this generalization is a quaternionic extension of the R transform which is a generating function for planar (non-crossing) cumulants. We demonstrate
that the quaternionic R transform generates all connected averages of 
all distinct powers of $X$ and its hermitian conjugate $X^\dagger$: $\langle\langle \frac{1}{N} \mbox{Tr} X^{a} X^{\dagger b} X^c \ldots \rangle\rangle$ for $N\rightarrow \infty$. We show that the R transform 
for gaussian elliptic laws is given by a simple linear quaternionic map 
$\mathcal{R}(z+wj) = x + \sigma^2 \left(\mu e^{2i\phi} z + w j\right)$
where $(z,w)$ is the Cayley-Dickson pair of complex numbers 
forming a quaternion $q=(z,w)\equiv z+ wj$. This map has five real parameters $\RE x$, $\IM x$, $\phi$, $\sigma$ and $\mu$. We use the R transform to calculate the limiting eigenvalue densities of several products of gaussian random matrices.


\end{abstract}

\maketitle

\section{Introduction}

Random matrices play an important role in physics,
engineering and mathematics \cite{m,cpv,ckn,gmw,abf,vt,agz}. While dealing with random matrices one often faces the problem of calculating eigenvalue spectra 
of sums and products of random matrices. This problem can be partially 
solved in the case of infinite invariant random matrices using methods of
free probability which is a powerful technique developed recently 
\cite{v1,v2,vdn,s}. In practice, with some caution, one can use free 
probability results also as a good approximation for finite (large) matrices.

Free probability calculus was introduced in operator algebra \cite{v1,v2}. Later 
one has realized that it can be also applied to random matrices in calculations of moments of sums and products of large invariant hermitian random matrices \cite{s}. In short, using free probability one can express the moment generating functions for a sum $A+B$ or a product $AB$ of invariant random matrices $A$ and $B$ in terms of the moment generating functions of these matrices. If a random matrix is hermitian the knowledge of 
its moments is often sufficient to reconstruct its eigenvalue density.  
Unfortunately the product $P=AB$, even for hermitian matrices, 
is generically non-hermitian. To get around this problem some authors \cite{rs} consider a related hermitian random matrix  $P'=\sqrt{A}B\sqrt{A}$, which has identical moments  as $P$: $\mathrm{Tr} P'^k = \mathrm{Tr} P^k$. Unfortunately the eigenvalue density of $P'$, except for a special class of matrices, is not the same as of $P$. So if one is interested in the eigenvalue density of $P$ one has to either apply different methods \cite{bjw,ab} or to extend free probability calculus to the realm of non-hermitian matrices. For example, one might be interested in the eigenvalue density of the product of two independent GUE matrices. Clearly, the product is a non-hermitian matrix, its eigenvalues are complex and the eigenvalue density has a non-trivial support. Actually one can find the limiting eigenvalue density of this product to be $\rho(z) = (2\pi |z|)^{-1}$ on the unit disk and zero outside the disk \cite{bjw,ab}.

Non-hermitian random matrices have been studied in many contexts,
including chaotic scattering \cite{hil,fs1,fs2,fss}, quantum chromodynamics with chemical potential \cite{jnz,v}, networks \cite{scss,dki} , lagged correlations \cite{bt,j}, matrix-valued random walk \cite{gjjn,bgntw} and many others \cite{ra,sg}. 
As a result of these intensive studies one has developed analytic techniques to calculate the eigenvalue spectra of such matrices. In the heart of these techniques lies an electrostatic analogy \cite{scss,fs2}. In this analogy eigenvalues of the random matrix play the role of positions of electric charges on two-dimensional plane and the Green's function of electric field. Given the Green's function one can calculate the eigenvalue density from the Gauss law \cite{scss,fs2,rj,jnpz1}. The Green's function can be viewed 
as an subordinated element of a two-by-two matrix being an extended 
Green's function. There were two independent approaches proposed for this extension:
the first one was directly derived \cite{rj,jnpz1,jnpz2,jnpwz} from the regularization of the electrostatic potential \cite{scss,fs2,hil} and the other one was obtained by a method of hermitization \cite{fz1,fz2}. Both the constructions are equivalent to each other and both can be formulated in terms of quaternions as one can see from the today's perspective. 

The quaternionic nature of these equations was anticipated in \cite{fz1} and worked out in \cite{jn1,jn2} where a very deep geometrical understanding of the regularization procedure was achieved. More precisely, in the calculations one usually introduces a regulator $\epsilon$. The modulus $|\epsilon|$ can be interpreted as a distance from the complex plane. This means that calculations are in fact performed in the quaternionic space in the neighborhood of the complex plane and only towards the end of the calculations, when the limit $\epsilon\rightarrow 0$ is taken, the problem is projected  back to the complex plane. The idea is similar to that used in the Sokhotski's formula (\ref{eq:one-dim_delta}) for one-dimensional delta function on the real axis except that now it is applied to two-dimensional delta function on the complex plane. In order to exploit this similarity it is convenient to recall the Cayley-Dickson construction. In this construction complex numbers are defined as pairs of real numbers and quaternions as pairs of complex numbers. In the one-dimensional Sokhotski's formula one sends the second element of the Cayley-Dickson pair (imaginary part) to zero $\epsilon\rightarrow 0$ and regains one-dimensional delta function on the real axis while for the two-dimensional case one sends the second element of the Cayley-Dickson pair forming quaternion to zero and recovers the delta function on the complex plane \cite{fz1,jn1,jn2}.

Yet before the importance of the underlying quaternionic structure of the extended resolvents was recognized \cite{jn1,jn2} one was able to derive the corresponding extended R transform (or equivalent blue function, sometimes used in physical literature) and to formulate the addition law for non-hermitian matrices \cite{rj,jnpz1,jnpz2,jnpwz,fz1}. This law was then successfully applied to a number of problems, while the multiplication law for non-hermitian matrices was formulated much later \cite{bjn}. 

Although quaternionic picture emerged already a decade ago \cite{fz1,jn1,jn2}
and has been applied 
to various problems \cite{bgntw,r,bcc,j2}, it has not been commonly adopted. 
The reason for this is probably related to the fact, that the synthetic 
picture of what is the R transform for non-hermitian matrices was still missing.
It was not clear how the R transform was related to the moments of the random matrix
or what was the classification of the R transform in the simplest case of 
gaussian laws, etc. We fill up this gap here.

We show that the quaternionic structure of the resolvent and the R transform
is capable to incorporate all mixed moments of random matrix and its hermitian 
conjugate: $\langle \frac{1}{N} \mbox{Tr} X^{a} X^{\dagger b} X^c \ldots \rangle$,
for an asymptotic class of random matrix models define by the probability measure 
of the type (\ref{muV}) for $N\rightarrow \infty$. In contrast to hermitian case,
though, these moments cannot be constructed from the moments of the eigenvalue 
density $\rho(\lambda)$ as they also depend on the left and right eigenvector 
correlations of the matrix $X$ \cite{cm,jnnpz}. So in a sense, the Green's function
and the R transform encode also information about the eigenvector correlations.

The link between free probability and random matrices was established through
planar combinatorics \cite{s}. In fact, the relation of random matrices to 
planar combinatorics was discovered long before free probability, by gaussian perturbation theory of invariant matrix models, which maps the problem of
calculating moments of random matrices to enumeration of planar Feynman 
diagrams \cite{th,bipz}. Planar Feynman diagrams can be enumerated by field theoretical methods, in particular by Dyson-Schwinger equations, which relate various moment generating functions \cite{bipz,biz}. These equations are
very powerful, as one can use them to derive all relations between generating functions known from free probability. The idea is very universal and it can be also applied to non-hermitian matrices to derive relations between the quaternionic
R transforms for independent random matrices $A$, $B$ and the corresponding
sums $A+B$ \cite{rj,jnpz1,jnpz2,jnpwz} and products $AB$ \cite{bjn}. 
We follow this approach here.

While presenting the material we first recall the hermitian case, which is well known and directly linked to free probability, and then we use it as a reference point while showing construction of the corresponding objects for non-hermitian matrices. 
We begin with the Green's function and then turn to the R transform. Once we have recalled the definition of R transform for the non-hermitian matrices \cite{jn1,jn2}, we will classify all possible forms of this transform for gaussian elliptic ensembles \cite{gi}. 
Then we extend the discussion to non-gaussian invariant measures  (\ref{muV}) and link them to planar Feynman diagrams \cite{bipz,biz}. We recall the addition law 
\cite{rj,jnpz1,jnpz2,jnpwz} and the multiplication law for non-hermitian matrices \cite{bjn} and use them to evaluate eigenvalue density for a couple of examples of products of gaussian random matrices from elliptic ensembles. We shortly summarize the paper in the last section. In Appendix \ref{AA} we recall the Cayley-Dickson construction of quaternions. In Appendix \ref{AB} we discuss quaternionic representation of the two-dimensional delta function, in Appendix \ref{AC} we recall the corresponding representation of the sum of delta functions located at the eigenvalues of a given non-hermitian matrix \cite{fs2} and finally in Appendix D we discuss 
planar Feynman diagrams \cite{th,bipz,biz} and recall how to use the diagrammatic
technique to derive the basic relations (\ref{GF_R},\ref{GRquat}) between the Green's 
function and the R transform. 

\section{Quaternionic resolvents for non-hermitian matrices}

In this section we discuss Green's functions (resolvents) for random matrix ensembles
in the limit of large matrix size. We begin with the standard case of hermitian matrices 
and then generalize it to non-hermitian matrices. The generalization
is closely related to the Cayley-Dickson construction of quaternions. We briefly recall this construction in Appendix \ref{AA}. For hermitian matrices eigenvalues are real and the corresponding resolvents are complex, while for non-hermitian matrices eigenvalues are complex and the corresponding Green's functions are quaternionic.

Our goal is to calculate eigenvalue distribution of a random matrix $H$ which is defined by a probability measure $d\mu(H)$. Assume that $H$ is hermitian $H=H^\dagger$,
so it has real eigenvalues. In the limit of infinite matrix size $N\rightarrow \infty$,  the eigenvalue density of $H$ is given by 
\begin{equation}
\rho\left(x\right) =
\lim_{N\rightarrow \infty} \left\langle \frac{1}{N} \sum_{i=1}^{N}\delta\left(x-\lambda_{i}\right) \right\rangle
\equiv \lim_{N\rightarrow \infty} \int \frac{1}{N} \sum_{i=1}^{N}\delta\left(x-\lambda_{i}\right) d \mu(H) ,
\end{equation}
where $\lambda_{i}$'s are eigenvalues of $H$. It is convenient to
define the Green's function (resolvent) which is a complex function of
a complex argument $z=x+ i\epsilon$ 
\begin{equation}
\label{eq:GF}
G\left(z\right)= \lim_{N\rightarrow \infty} 
\left\langle \frac{1}{N} \mathrm{Tr} (z\mathbb{1}-H)^{-1} \right\rangle = 
\lim_{N\rightarrow \infty}  
\left\langle \frac{1}{N} \sum_{i=1}^N \frac{1}{z-\lambda_i} \right\rangle \ .
\end{equation}
The symbol $\mathbb{1}$ stands for the identity matrix of size $N$. 
The eigenvalue density can be calculated from the resolvent (\ref{eq:GF}) 
\begin{equation}
\label{eq:rhogreens}
\rho\left(\lambda\right)=- \frac{1}{\pi} \lim_{\epsilon\rightarrow0^{+}} \IM \, G\left(\lambda+i \epsilon\right) \ ,
\end{equation}
as follows from the Sokhotski's representation of the one-dimensional Dirac delta function
\begin{equation}
\delta(x) = - \frac{1}{\pi} \lim_{\epsilon\rightarrow0^{+}} \IM \frac{1}{x + \epsilon i} =
\frac{1}{\pi} \lim_{\epsilon\rightarrow0^{+}}  \frac{\epsilon}{x^2 + \epsilon^2}
\label{eq:one-dim_delta} \ .
\end{equation} 
The limit $\epsilon\rightarrow 0^+$ projects the problem to real axis. All eigenvalues are located on this axis (\ref{eq:rhogreens}). 

One can now implement an analogous strategy for non-hermitian matrices. Eigenvalues are complex now, so one has to go beyond the complex plane. Using the Cayley-Dickson construction we first introduce a second auxiliary complex variable, $w$, to define a quaternion $q=(z,w)=z+wj$ (see Appendix \ref{AA}) and a quaternionic Green's function $\mathcal{G}(q)$. Once the Green's function is determined one can send the second part of the quaternion to zero, $w \rightarrow 0$, and project it to the complex plane where eigenvalues reside.
This is analogous to sending the imaginary part in the construction of eigenvalue density for hermitian matrices (\ref{eq:rhogreens}).

Let us discuss this in detail. We use the notation of quaternions and of quaternionic matrices as in Appendix \ref{AA}. We start with the quaternionic representation of the Dirac delta on the complex plane 
\begin{equation}
\label{eq:two-dim_delta}
\delta^{\left(2\right)}\left(z\right)= \frac{1}{\pi} \lim_{w \rightarrow 0} 
 \frac{\partial}{\partial\bar{z}} F \frac{1}{z+wj} 
= \frac{1}{\pi} \lim_{w \rightarrow 0} 
 \frac{\partial}{\partial\bar{z}} F (z,w)^{-1} 
=\frac{1}{\pi} \lim_{w \rightarrow 0}  \frac{|w|^{2}}{\left(\left|z\right|^{2}+|w|^{2}\right)^{2}}
\ .
\end{equation}
where $F$ stands for the first Cayley-Dickson part of quaternion. It is analogous to
the real part $\RE$ of complex number. This representation is an counterpart of the one-dimensional delta function representation (\ref{eq:one-dim_delta}). The delta function is obtained by calculating the inverse quaternion close to the complex plane and eventually taking the limit by sending the second part of the quaternion to zero. The difference as compared to Eq. (\ref{eq:one-dim_delta}) is that the imaginary part $\IM$ is replaced by the first part $F$ and that there is an additional derivative. In Appendix \ref{AB} we explicitly demonstrate that the expression (\ref{eq:two-dim_delta}) indeed yields a representation of the two-dimensional delta function. The representation (\ref{eq:two-dim_delta}) provides us with a guideline as to how to define the quaternionic Green's function and how to calculate the eigenvalue distribution from it.
 
The quaternionic resolvent for a non-hermitian matrix $X$ is defined in a similar way as for hermitian matrices (\ref{eq:GF}) \cite{rj,jnpz1}. The Green's function $\mathcal{G}(q)$ and its argument $q$ are now quaternions \cite{jn1,jn2}. Denote the first part of the Green's function by $G(q)$ and the second part by $\Gamma(q)$, $\mathcal{G}(q)=(G(q),\Gamma(q))$ where $q=(z,w)$.
In analogy to Eq. (\ref{eq:GF}) we have
\begin{equation}
\label{eq:QGF}
\mathcal{G}\left(q\right) \equiv (G(q),\Gamma(q)) = \lim_{N\rightarrow \infty}  
\left\langle \frac{1}{N} \mathrm{Tr_b} (z\mathbb{1}-X,w\mathbb{1})^{-1} \right\rangle \ ,
\end{equation}
where $\mathrm{Tr_b}$ is quaternionic trace defined in (\ref{CDM}).
In the Pauli matrix representation (\ref{PMR}) this equation reads
\begin{equation}
\label{eq:QMR}
\mathcal{G}(q) = 
\left( \begin{array}{rc} 
G(q) & \Gamma(q) \\
-\bar{\Gamma}(q) & \bar{G}(q) 
\end{array} \right) =
\lim_{N\rightarrow \infty} 
\left\langle \frac{1}{N} \mathrm{Tr_b} 
\left( \begin{array}{cc} 
z\mathbb{1}-X & w \mathbb{1} \\
-\bar{w} \mathbb{1} & \bar{z}\mathbb{1} - X^\dagger 
\end{array} \right)^{-1} 
\right\rangle \ .
\end{equation}
In this representation the quaternionic trace $\mathrm{Tr_b}$ is equivalent to taking trace of each of four $N\times N$ blocks separately.
The matrix on the right hand side can be inverted 
\begin{equation}
\left(\begin{array}{cc} 
z\mathbb{1}-X & w \mathbb{1} \\
- \bar{w} \mathbb{1} & \bar{z}\mathbb{1} - X^\dagger 
\end{array} \right)^{-1} =
\left(\begin{array}{cc} 
(\bar{z}\mathbb{1}-X^\dagger) H_L^{-1} & -w H_R^{-1} \\
 \bar{w} H_L^{-1} & (z\mathbb{1} - X) H_R^{-1} 
\end{array} \right)
\end{equation}
where
\begin{equation}
\label{HLHR}
\begin{split}
H_L & = (z\mathbb{1}-X)(\bar{z}\mathbb{1}-X^\dagger)+|w|^2\mathbb{1} \\
H_R & = (\bar{z}\mathbb{1}-X^\dagger)(z\mathbb{1}-X)+|w|^2\mathbb{1} \ .
\end{split}
\end{equation}
We see that the first part of the quaternionic Green's function is
\begin{equation}
G(q) = F \mathcal{G}(q) = 
\lim_{N\rightarrow \infty} \frac{1}{N} 
\left\langle \mathrm{Tr} (\bar{z}\mathbb{1}-X^\dagger) H_L^{-1} \right\rangle 
= \lim_{N\rightarrow \infty} \frac{1}{N} 
\left\langle \mathrm{Tr} \frac{\bar{z}\mathbb{1}-X^\dagger}
{(z\mathbb{1}-X)(\bar{z}\mathbb{1}-X^\dagger)+|w|^2\mathbb{1}}\right\rangle
\end{equation}
and its derivative 
\begin{equation}
\label{dGX}
\begin{split}
\frac{\partial}{\partial\bar{z}} G(q) & =
\lim_{N\rightarrow \infty} \frac{1}{N} 
\left\langle \mathrm{Tr} H_R^{-1} |w|^2 H_L^{-1} \right\rangle \\ &
=\lim_{N\rightarrow \infty} \frac{1}{N} \left\langle \mathrm{Tr} \frac{|w|^2}{\left((z\mathbb{1}-X)(\bar{z}\mathbb{1}-X^\dagger)+|w|^2\mathbb{1}\right)
\left((\bar{z}\mathbb{1}-X^\dagger)(z\mathbb{1}-X)+|w|^2\mathbb{1}\right)}
\right\rangle \ . 
\end{split}
\end{equation}
In the limit of the second part of the quaternion $q=(z,w)$ tending to zero,
$w\rightarrow 0$, the expression in the brackets takes the form of a sum 
of Dirac deltas located at the eigenvalues of the matrix $X$ 
\begin{equation}
\label{Grho}
\lim_{w\rightarrow 0}\frac{1}{\pi} \frac{\partial}{\partial\bar{z}} 
G\left((z,w)\right) = \lim_{N\rightarrow \infty} \left\langle \frac{1}{N} \sum_{i=1}^{N}\delta^{(2)}\left(z-\lambda_{i}\right) \right\rangle = \rho(z) 
\end{equation}
so the eigenvalue density is regained in this limit similarly as in the one-dimensional case (\ref{eq:rhogreens}). The mechanism of localization of the expression in the brackets (\ref{dGX}) at eigenvalues of the matrix $X$ is analogous  to that in the representation (\ref{eq:two-dim_delta}) of two-dimensional delta function. The details are given in Appendix \ref{AC} where we recall the argument given in \cite{fs2}.  

\section{Quaternionic R transform}
As in the previous section let us begin by recalling the hermitian case which is well known. It will serve as a reference point.  The $1/z$-expansion of the Green's function (\ref{eq:GF}) generates moments 
\begin{equation}
m_{n}= \lim_{N\rightarrow \infty} \left\langle \frac{1}{N} \mathrm{Tr} H^n \right\rangle
\end{equation}
of the random matrix $H$~:
\begin{equation}
\label{1z_exp}
G\left(z\right)=
\sum_{n=0}^\infty \frac{m_{n}}{z^{n+1}} \ .
\end{equation}
The R transform is defined by the following equation \cite{vdn}
\begin{equation}
\label{GF_R}
G\left(z\right)=\frac{1}{z-R\left(G\left(z\right)\right)} \ .
\end{equation}
The R transform is a generating function for planar connected moments 
$\kappa_{n}$ of the matrix $H$~:
\begin{equation}
\label{R_GF}
R\left(z\right)=\sum_{n=1}^{\infty}\kappa_{n}z^{n-1} \ .
\end{equation}
The diagrammatic derivation \cite{bipz,biz} of the relation between the Green's function and the R transform is presented in Appendix D. The connected moments $\kappa_{n}$ are also known as ''free cumulants'' 
or ''non-crossing cumulants''. They are related to the moments 
$m_{n}$ as follows:  $\kappa_{1} = m_{1}$, $\kappa_{2}=m_{2}-m^2_{1}$, $\kappa_{3}=m_{3}-3m_{2}m_{1} + 2m_{1}^3$, 
$\kappa_{4} = m_{4} - 4m_{1} m_{3} - 2m_{2}^2 + 10 m_{1}^2 m_{2} - 4m_{1}^4$, {\em etc}. These relations can be derived by expanding (\ref{GF_R}). 
Note that the first three relations are identical as the corresponding relations 
between standard cumulants and moments known from 
classical probability. The reason why the R transform is important is that 
it is additive for sums of independent random matrices as discussed in the next section.

Now consider a non-hermitian random matrix $X$.
The situation is a bit more complicated since the $1/q$-expansion of the quaternionic Green's function (\ref{eq:QGF}) generates moments of differently ordered powers of the matrix $X$ and its hermitian conjugate $X^\dagger$.  

It is convenient to rewrite the resolvent (\ref{eq:QGF}) as follows \cite{jn1,jn2}
\begin{equation}
\label{Gq}
\mathcal{G}(q) = 
\lim_{N\rightarrow \infty} \left\langle \frac{1}{N} \mathrm{Tr_b} (\mathcal{Q} - \mathcal{X})^{-1} \right\rangle 
\end{equation}
where 
\begin{equation}
\label{calX}
\mathcal{X} = \left( \begin{array}{cc} \mathcal{X}_{11} & 
\mathcal{X}_{12} \\ \mathcal{X}_{21} & \mathcal{X}_{22} \end{array}\right) =
\left( \begin{array}{cc} X & 0 \\ 0 & X^\dagger \end{array}\right) 
\end{equation}
and $\mathcal{Q} = q \otimes \mathbb{1}$ where $\mathbb{1}$ is 
the $N \times N$ identity matrix and
\begin{equation}
\label{q_rep}
q =   \left(\begin{array}{cc}  z   & w \\
                        -\bar{w} & \bar{z} \end{array} \right) 
\end{equation}
is a matrix representing the quaternion $q=(z,w)$ (\ref{2x2}).
We will index elements of the quaternionic matrix $q$ by Greek letters $q_{\alpha\beta}$
and consistently the positions of the blocks $\mathcal{X}_{\alpha\beta}$ in the matrix $\mathcal{X}$ (\ref{calX}) and in other block matrices.

The $1/q$ expansion of the Green's function can be written as
\begin{equation}
\begin{split}
\mathcal{G}(q) & = 
\lim_{N\rightarrow \infty} \left\langle \frac{1}{N} \mathrm{Tr_b} (\mathcal{Q} - \mathcal{X})^{-1} \right\rangle \\ & = \lim_{N\rightarrow \infty}
\left\langle \frac{1}{N} \mathrm{Tr_b} \mathcal{Q}^{-1} \right\rangle + \lim_{N\rightarrow \infty}
\left\langle \frac{1}{N} \mathrm{Tr_b} \mathcal{Q}^{-1} \mathcal{X}  \mathcal{Q}^{-1} \right\rangle + \lim_{N\rightarrow \infty}
\left\langle \frac{1}{N} \mathrm{Tr_b} \mathcal{Q}^{-1} \mathcal{X}  \mathcal{Q}^{-1}
\mathcal{X} \mathcal{Q}^{-1} \right\rangle + \ldots  
\end{split}
\end{equation}
or in the index notation as
\begin{equation}
\mathcal{G}(q)_{\alpha\zeta} = 
q^{-1}_{\alpha\zeta} + 
\sum_{\beta\gamma} \lim_{N\rightarrow \infty} \left\langle \frac{1}{N} \mathrm{Tr} q^{-1}_{\alpha\beta} 
\mathcal{X}_{\beta\gamma}  q^{-1}_{\gamma\zeta} \right\rangle +
\sum_{\beta\gamma\delta\epsilon} \lim_{N\rightarrow \infty}
\left\langle \frac{1}{N} \mathrm{Tr} q^{-1}_{\alpha\beta} 
\mathcal{X}_{\beta\gamma}  q^{-1}_{\gamma\delta} 
\mathcal{X}_{\delta\epsilon}  q^{-1}_{\epsilon\zeta}
\right\rangle + \ldots \ . 
\end{equation}
We can now define moments $m^{(n)}$
of order $n$ as multidimensional arrays with the following elements
\begin{equation}
m^{(n)}_{\alpha_1\alpha_2\ldots\alpha_{2n-1}\alpha_{2n}} = 
\lim_{N\rightarrow \infty}
\left\langle \frac{1}{N} \mathrm{Tr} \mathcal{X}_{\alpha_1 \alpha_2} \ldots 
\mathcal{X}_{\alpha_{2n-1} \alpha_{2n}} \right\rangle \ ,
\end{equation}
so we have 
\begin{equation}
\label{eq:QGF_expansion}
\mathcal{G}(q)_{\alpha\zeta} = 
q^{-1}_{\alpha\zeta} + 
\sum_{\beta\gamma} m^{(1)}_{\beta\gamma} 
q^{-1}_{\alpha\beta} q^{-1}_{\gamma\zeta} +
\sum_{\beta\gamma\delta\epsilon} m^{(2)}_{\beta\gamma\delta\epsilon}
q^{-1}_{\alpha\beta} q^{-1}_{\gamma\delta} 
q^{-1}_{\epsilon\zeta} + \ldots \ . 
\end{equation}
The last equation can be simplified since the off-diagonal 
blocks of the matrices $\mathcal{X}$ (\ref{calX}) are zero,
so the only non-trivial moments are those involving only diagonal blocks:
\begin{equation}
M^{(n)}_{\alpha_1\ldots \alpha_n} = 
m^{(n)}_{\alpha_1\alpha_1\ldots\alpha_{n}\alpha_{n}} = 
\lim_{N\rightarrow \infty}
\left\langle \frac{1}{N} \mathrm{Tr} \mathcal{X}_{\alpha_1 \alpha_1} \ldots 
\mathcal{X}_{\alpha_{n} \alpha_{n}} \right\rangle \ .
\end{equation}
Combining that with Eq. (\ref{eq:QGF_expansion}) we arrive at
\begin{equation}
\mathcal{G}(q)_{\alpha\zeta} = 
q^{-1}_{\alpha\zeta} + 
\sum_\beta M^{(1)}_{\beta} 
q^{-1}_{\alpha\beta} q^{-1}_{\beta \zeta} +
\sum_{\beta\gamma} M^{(2)}_{\beta\gamma}
q^{-1}_{\alpha\beta} q^{-1}_{\beta\gamma} q^{-1}_{\gamma\zeta} + \ldots \ . 
\end{equation} 
This equation is analogous to Eq. (\ref{1z_exp}) except that now the moments are given by multidimensional arrays which encode information about all
mixed moments. For example $\left(M_{5}\right)_{12112} = \lim_{N\rightarrow \infty} \left\langle \frac{1}{N} \mathrm{Tr} XX^\dagger X^2 X^\dagger \right\rangle$. The first and the second moments are 
\begin{equation}
M^{(1)}_{1} = \lim_{N\rightarrow \infty} \left\langle \frac{1}{N} \mathrm{Tr} X \right\rangle,
\quad
M^{(1)}_{2} = \lim_{N\rightarrow \infty} \left\langle \frac{1}{N} \mathrm{Tr} X^\dagger \right\rangle
\end{equation}
and 
\begin{equation}
\begin{split}
M^{(2)}_{11} & = \lim_{N\rightarrow \infty} \left\langle \frac{1}{N} \mathrm{Tr} XX \right\rangle \ , \quad 
M^{(2)}_{12}  = \lim_{N\rightarrow \infty} \left\langle \frac{1}{N} \mathrm{Tr} XX^\dagger \right\rangle
\\
M^{(2)}_{21} & = \lim_{N\rightarrow \infty} \left\langle  \frac{1}{N} \mathrm{Tr} X^\dagger X \right\rangle, \quad 
M^{(2)}_{22}  = \lim_{N\rightarrow \infty} \left\langle \frac{1}{N} \mathrm{Tr} X^\dagger X^\dagger \right\rangle ,
\end{split}
\end{equation}
respectively. Generally the array representing the $n$-th moment has $2^n$ elements, but 
not all of them are independent. For example
$M^{(2)}_{11}=\overline{M^{(2)}_{22}}$,  
$M^{(2)}_{21}=M^{(2)}_{12}$.
 
The quaternionic R transform is defined in the same way as the R transform for hermitian matrices (\ref{GF_R}) \cite{rj,jnpz1,jn1,jn2}
\begin{equation}
\label{GRquat}
\mathcal{G}\left(q\right)=\frac{1}{q-\mathcal{R}\left(\mathcal{G}
\left(q\right)\right)} \ 
\end{equation}
except that it is a quaternionic function. It generates planar cumulants
$K^{(n)}$ which are now multidimensional arrays in contrast to the hermitian case where
they are real numbers (\ref{R_GF})
\begin{equation}
\label{RK}
\mathcal{R}\left(q\right)_{\alpha\zeta}= K^{(1)}_{\alpha} \delta_{\alpha\zeta} + 
K^{(2)}_{\alpha\zeta} q_{\alpha\zeta} + 
\sum_{\beta} K^{(3)}_{\alpha\beta\zeta} q_{\alpha\beta} q_{\beta\zeta} + 
\sum_{\beta\gamma} K^{(4)}_{\alpha\beta\gamma\zeta} q_{\alpha\beta}  q_{\beta\gamma} q_{\gamma\zeta} \ldots  .
\end{equation}
The first cumulant is 
\begin{equation}
\label{K1}
K^{(1)}_{1} = \lim_{N\rightarrow \infty} 
\left\langle \frac{1}{N} \mathrm{Tr} X \right\rangle, \quad
K^{(1)}_{2} = \lim_{N\rightarrow \infty} \left\langle \frac{1}{N} \mathrm{Tr} X^\dagger \right\rangle
\end{equation}
and the second one is
\begin{equation}
\label{K2}
\begin{split}
K^{(2)}_{11} & = \lim_{N\rightarrow \infty} \left\langle \frac{1}{N} \mathrm{Tr} (X-x\mathbb{1})(X -x\mathbb{1})\right\rangle, \quad 
K^{(2)}_{12} = \lim_{N\rightarrow \infty} \left\langle  \frac{1}{N} \mathrm{Tr} 
(X-x\mathbb{1}) (X -x\mathbb{1})^\dagger \right\rangle
\\
K^{(2)}_{21} & = \lim_{N\rightarrow \infty} \left\langle  \frac{1}{N} \mathrm{Tr} (X-x\mathbb{1})^\dagger (X-x\mathbb{1}) \right\rangle, \quad 
K^{(2)}_{22} = \lim_{N\rightarrow \infty} \left\langle  \frac{1}{N} \mathrm{Tr} (X-x\mathbb{1})^\dagger (X-x\mathbb{1})^\dagger \right\rangle \ .
\end{split}
\end{equation}
We used here a shorthand notation $x=K^{(1)}_{1}$ for brevity.

\section{Gaussian elliptic ensembles}

In this section we consider the class of gaussian elliptic random matrices
which are defined by the condition that the third and higher cumulants 
vanish: $K^{(3)} = K^{(4)} = \ldots = 0$. The R transform (\ref{RK}) is a linear 
function in this case
\begin{equation}
\mathcal{R}\left(q\right)_{\alpha\zeta}= K^{(1)}_{\alpha} \delta_{\alpha\zeta} + 
K^{(2)}_{\alpha\zeta} q_{\alpha\zeta} \ . 
\end{equation}
The last equation can be rewritten as a matrix equation
\begin{equation}
\mathcal{R}(q) = K^{(1)} + K^{(2)} \circ q \ ,
\end{equation}
where the symbol $\circ$ denotes the Hadamard product and 
$K^{(1)}$, $K^{(2)}$ are two-by-two matrices representing
the first two cumulants 
\begin{equation}
\label{K1_K2}
K^{(1)} = \left( \begin{array}{cc} x & 0 \\ 0 & \bar{x} \end{array} \right) \ ,
\quad
K^{(2)} = \sigma^2 \left( \begin{array}{cc} \mu e^{2i\phi}  & 1 \\ 1 & 
\mu e^{-2i\phi} \end{array} \right) \ . 
\end{equation}
The first cumulant is parametrized by two real parameters: $\RE x$ and
$\IM x$ while the second one by three: $\sigma$, $\mu$ and $\phi$.
The eigenvalue density of the corresponding 
random matrix is uniform on an ellipse on the complex plane.
The parameter $x$ is the position of the ellipse center, 
$\sigma^2=\sigma_1^2+\sigma_2^2$ is the sum of squares of lengths of semi-axes, 
$\mu$ is an eccentricity parameter defined
as $\mu = (\sigma_1^2-\sigma_2^2)/(\sigma_1^2+\sigma_2^2)$, and 
$\phi$ is the angle between the first semi-axis and the real positive semi-axis.                  
We define a family of standardized elliptic laws by setting $x=0$ and $\sigma=1$ and $\phi=0$. All elliptic laws can be obtained from the standardized one by shifting it by $x$, rescaling by a factor $\sigma$ and rotating by $\phi$. The standardized family is 
parametrized by one parameter $\mu$. The R transform is
\begin{equation}
\mathcal{R} \left(\left( \begin{array}{cc} z  & w \\ -\bar{w} & 
\bar{z} \end{array} \right)\right) 
 = \left( \begin{array}{cc} \mu  & 1 \\ 1 & 
\mu \end{array} \right) \circ \left( \begin{array}{cc} z  & w \\ -\bar{w} & 
\bar{z} \end{array} \right) = \left( \begin{array}{cc} \mu z  & w \\ -\bar{w} & 
\mu \bar{z} \end{array} \right) \ .
\end{equation} 
Using the notation with quaternionic units (\ref{CD0}) we
can write this equation as
\begin{equation}
\label{stand}
\mathcal{R}(z+wj) = \mu z + w j\ .
\end{equation} 
We see that the R transform is given by a very simple linear map which
multiplies the first Cayley-Dickson part (\ref{CD}) of the quaternion 
by $\mu$: $F(\mathcal{R}(q)) = \mu F(q)$ and leaves
the second part intact  $S(\mathcal{R}(q)) = S(q)$. For hermitian matrices
$X^\dagger=X$, $\mu=1$ and $\mathcal{R}(q)=q$. For anti-hermitian matrices
$X^\dagger=-X$, $\mu=-1$ and $\mathcal{R}(z+wj)=-z+wj$. For Ginibre matrices \cite{g}
$\mu=0$ and $\mathcal{R}(z+wj)=wj$. The R transform for a general elliptic
law reads
\begin{equation}
\label{nonstand}
\mathcal{R}(z+wj) = x + \sigma^2 \left(\mu e^{2i\phi} z + w j\right)\ .
\end{equation} 
One should note that the rotation acts only on the
first part of the quaternion, while the rescaling on both.

One can easily derive the probability measure for elliptic 
gaussian random matrices. Elliptic random matrices can be constructed from gaussian
hermitian matrices. Let $H_1$ and $H_2$ be two independent
$N\times N$ gaussian hermitian matrices defined by the following product 
probability measure
\begin{equation}
\label{P12}
d\mu(H_1,H_2) \propto DH_1 DH_2 
e^{-\frac{N}{2} \mathrm{Tr} \left( H_1^2 + H_2^2\right) }\ ,
\end{equation}
where $DH_1$ and $DH_2$ are flat measures for hermitian matrices
each having $N^2$ real degrees of freedom. We skipped the normalization 
constant in the last equation.
As the measure factorizes, we have
\begin{equation}
\left\langle \frac{1}{N} \mathrm{Tr} H_a \right\rangle =  0 \ , \quad
\left\langle \frac{1}{N} \mathrm{Tr} H_a H_b \right\rangle = \delta_{ab} \ .
\end{equation}
for $a=1,2$, $b=1,2$. Let us define $X$ as a linear
combination of these matrices
\begin{equation}
X = x\mathbb{1} + e^{i\phi} \left(\sigma_1 H_1 + i \sigma_2 H_2\right) \ .
\end{equation}
This combination is generically non-hermitian and has $2N^2$ degrees of freedom. 
Its hermitian conjugate is
\begin{equation}
X^\dagger = \bar{x}\mathbb{1} + e^{-i\phi} \left(\sigma_1 H_1 - 
i \sigma_2 H_2\right) \ .
\end{equation}
It is easy to calculate the one-point and two-point correlation 
functions (\ref{K1}) and (\ref{K2}) for $X$ and $X^\dagger$. One
gets the R transform (\ref{K1_K2}) or equivalently (\ref{nonstand}).
In order to obtain the measure for $X$ one can invert the two 
last equations for $H_1$ and $H_2$ and insert the result to (\ref{P12}).
Without loss of generality we set $x=0$. We have
\begin{equation}
H_1 = \frac{e^{-i\phi}X + e^{+i\phi}X^\dagger}{2\sigma_1} \ , \quad
H_2 = \frac{e^{-i\phi}X - e^{+i\phi}X^\dagger}{2 i\sigma_2}
\end{equation}
and
\begin{equation}
\label{muX}
d\mu_0(X) \propto DX 
\exp\left(-\frac{N}{\sigma^2(1-\mu^2)} \mathrm{Tr}\left( 
X X^{\dagger} -
\frac{\mu}{2}\left(e^{-2i\phi} X^2 + e^{+2i\phi} 
X^{\dagger2}\right)\right)\right) 
\end{equation}
where $DX$ is a flat measure for non-hermitian matrices which has $2N^2$ real
degrees of freedom. For $x\ne 0$ the matrix $X$ should be 
replaced by $X-x\mathbb{1}$. It is the most general form of the gaussian
elliptic measure. For any $\mu \in (-1,1)$ the distribution is uniform on 
an ellipse. The limit $\mu\rightarrow \pm 1$ should be taken
carefully. In this limit the exponent in (\ref{muX}) changes to a delta function
$\delta(e^{-i\phi} X \mp e^{+i\phi} X^\dagger)$ which is responsible for
a reduction of the number of  degrees of freedom from $2N^2$ to $N^2$. 
The width of the ellipse shrinks to zero and the resulting 
eigenvalue density has a one dimensional support being an interval on the complex plane. 
The eigenvalue density is not uniform on this interval but it is given by the 
Wigner semicircle law. It is worth noting, that those are only two limiting cases, while density experiences an interesting crossover regime when $\mu\sim\pm1\mp\frac{1}{N}$ \cite{fks}.

\section{Non-gaussian ensembles}

In this section we go beyond the gaussian regime. Consider 
random matrices defined by the following invariant probability measure 
\begin{equation}
\label{muV}
d\mu(X) \propto DX e^{-N \mathrm{Tr} V(X,X^\dagger)}
\end{equation}
where $V(X,X^\dagger)$ is a two-matrix polynomial such that
$\mathrm{Tr} V(X,X^\dagger)$ is real and bounded from below. 
This means that there exist a real number $r$ 
such that $\mathrm{Tr} V(X,X^\dagger) > r$ for any matrix $X$. Such measure is invariant w.r.t. transformation $X\rightarrow UXU^{\dagger}$, but such invariance is not sufficient to express the measure only in terms of eigenvalues like in hermitian case. The left and right eigenvectors are not simply complex conjugates of each other, but their correlations carry an non-trivial information. The polynomial $V(X,X^\dagger)$ is constructed as a sum of terms
of powers of $X$ and $X^\dagger$ 
\begin{equation}
\label{chain}
g X^{k_1} X^{\dagger j_1} \ldots X^{k_n} X^{\dagger j_n}
\end{equation}
with complex coefficients $g$ called coupling constants.
Different terms may have different coupling constants. Terms which can be obtained from
each other by a cyclic permutation of matrices have identical trace and 
are thus equivalent from the point of view of the measure (\ref{muV}). 
It is therefore convenient to divide all terms 
into equivalence classes of products of $X$ and $X^\dagger$ which 
can be obtained from each other by a cyclic permutation. 
For example the following products $X^2 X^{3\dagger} X^2$, $X^4 X^{3\dagger}$,
$X^\dagger X^4 X^{2\dagger}$ belong to the same equivalence class and
have the same trace $\mathrm{Tr} X^2 X^{3\dagger} X^2 = \mathrm{Tr} X^4 X^{3\dagger} =
\mathrm{Tr} X^\dagger X^4 X^{2\dagger}$. We call these 
equivalence classes periodic chains. 
We also define dual chains. A chain is said to be dual to a given chain 
if it is obtained by swapping positions of $X$'s and $X^\dagger$'s in the chain. 
The requirement that the 
trace of the potential is a real number means that the potential $V(X,X^\dagger)$ must 
contain pairs of dual terms with complex conjugate coupling constants.
For example the terms $g X^2 X^\dagger$ and $\bar{g} X^{\dagger 2} X$ are mutually dual with proper coupling constants. 
For self-dual chains, as for instance $g X^2 X^{2\dagger}$, we have $g=\bar{g}$, therefore those coupling constants must be real.
For example the most general form of the third order terms in the potential is
$g_1 X^3 + \bar{g}_1 X^{\dagger 3} + g_2 X^2 X^\dagger + \bar{g}_2 X^{\dagger 2} X$
with two complex coupling constants denoted here by $g_1$ and $g_2$. The fourth
order potential has two dual pairs $g_3 X^4 + \bar{g}_3 X^{\dagger 4}$ 
and  $g_4 X^3 X^\dagger + \bar{g}_4 X^{\dagger 3} X$, as well as two 
distinct self-dual terms $g_5 X^2 X^{\dagger 2}$ and 
$g_6 X X^\dagger X X^\dagger$ with real coupling constants $g_5$ and $g_6$.

\begin{figure}
\centering
\includegraphics[scale=1]{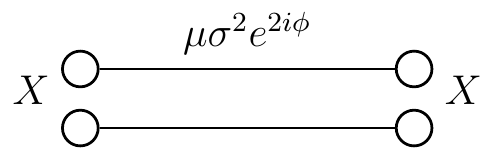}
\ ,\ 
\includegraphics[scale=1]{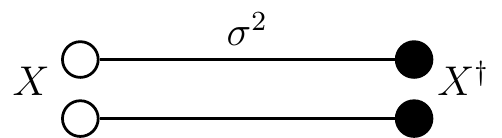}
\ ,\ 
\includegraphics[scale=1]{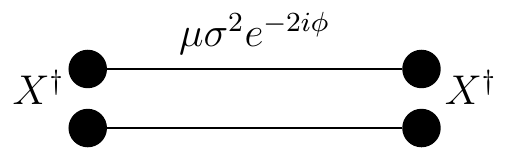}
\caption{Three possible propagators coming from gaussian part $d\mu_{0}\left(X\right)$ of the probability measure.}
\label{fig:Propagators}
\end{figure}

\begin{figure}
\centering
\includegraphics[scale=1]{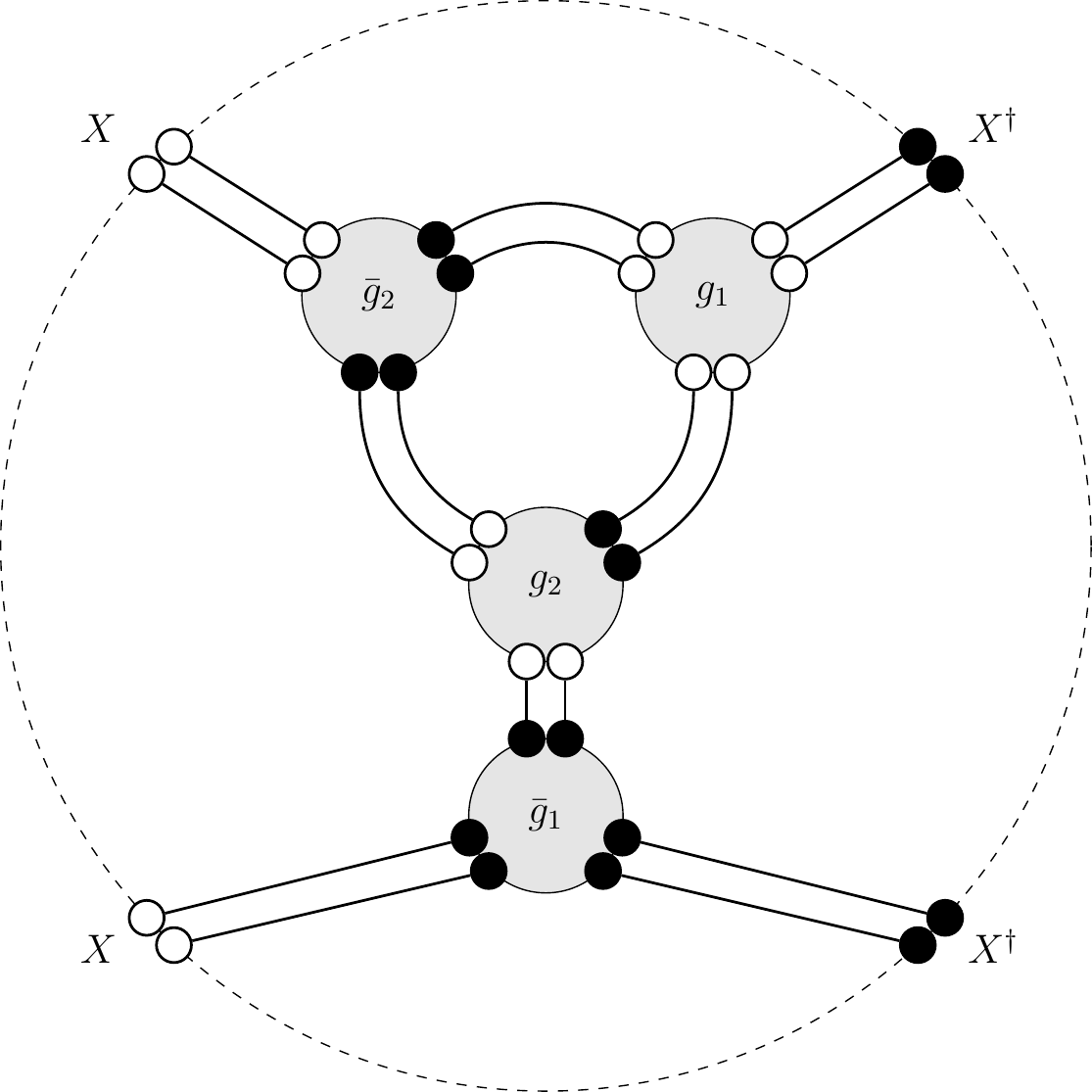}
\caption{Example of a diagram consisting of all four kinds of third order vertices (light gray circles), contributing to cumulant of order 4: $\left\langle\left\langle\mathrm{Tr}X^{2}X^{\dagger2}\right\rangle\right\rangle$. The big dashed circle emphasises symmetry with respect to cyclic permutations under trace.}
\label{fig:Diagram}
\end{figure}

For convenience we assume, without loss of generality, that the 
measure is centered, that is $\langle \frac{1}{N} \mathrm{Tr} X \rangle = 0$,
so we can skip linear terms in the potential. Next step is to 
split the measure into the gaussian part (\ref{muX})
and to treat the residual part as a perturbation of the gaussian measure \cite{bipz}
\begin{equation}
d \mu(X) = d \mu_0(X) e^{-N \mathrm{Tr} V_R(X,X^\dagger)} = 
d \mu_0(X) \left( 1 - N \mathrm{Tr} V_R(X,X^\dagger) 
+ \frac{1}{2} N^2 \mathrm{Tr}^2 V_R(X,X^\dagger) + \ldots \right)
\end{equation}
where $d\mu_0(X)$ is the elliptic gaussian measure (\ref{muX}) and 
$V_R$ is the remaining part of the potential which contains the third 
or higher order terms. In this way, the average of an observable 
$O(X)$ is calculated as an average over the gaussian measure
$d\mu_0(X)$ but of a new observable modified by terms coming from
the perturbative expansion of the residual part 
\begin{equation}
\langle O(X) \rangle = 
\left\langle O(X) \left( 1 - N \mathrm{Tr} V_R(X,X^\dagger) 
+ \frac{1}{2} N^2 \mathrm{Tr}^2 V_R(X,X^\dagger) + \ldots \right)\right\rangle_0 \ .
\end{equation}
Now the calculation is reduced to gaussian integration. Thanks to the Wick's theorem it can be mapped onto the problem of Feynman diagram enumeration which
simplifies in the limit $N\rightarrow \infty$ to enumeration of
planar diagrams \cite{th,bipz}. We will not give the details here, the interested reader 
can find them in \cite{bjn}. We only give the Feynman rules for this model.
Diagrams are constructed by drawing lines between vertices. The lines, called
propagators, are deduced from the gaussian part $d\mu_0(X)$ (\ref{muX}).
There are three different cases shown in Fig. \ref{fig:Propagators} which correspond to 
two-point functions $\langle XX\rangle_0$, $\langle XX^\dagger\rangle_0$
$\langle X^\dagger X^\dagger\rangle_0$. The vertices 
come from the expansion of the residual part. They correspond
to different periodic chains. As an example, we show in Fig. \ref{fig:Diagram} a diagram consisting two third order vertices contributing to $\left\langle XXX^{\dagger}X^{\dagger}\right\rangle$. For each such diagram, there will be a dual one obtained by replacing matrices, $X\leftrightarrow X^{\dagger}$, and vertices, $g_{i}\leftrightarrow \bar{g}_{i}$, to their dual ones. The sum of all contributions from connected diagrams with $n$ external 
lines is equal to planar cumulant of order $n$. Sometimes it is called a non-crossing cumulant or free cumulant of order $n$. One can symbolically denote
cumulants as connected averages $\langle\langle \frac{1}{N} \mathrm{Tr} X^a X^{\dagger b} X^c \ldots \rangle\rangle$. A cumulant  of order $n$ is an array which contains connected averages for all distinct periodic chains of length $n=a+b+c+\ldots$. The R transform is a generating function of such cumulants (\ref{RK}). The main relation (\ref{GF_R}) between the Green function and R transform can be derived from the Dyson-Schwinger equations
for planar diagrams. The details can be found in \cite{bjn}.

The R transform has an important property. 
The R transform of a sum $A+B$ of two independent random matrices, each defined
by a probability measure of the type (\ref{muV}), is additive \cite{rj,jnpz1}
\begin{equation}
\label{addL}
\mathcal{R}_{A+B}(q) = \mathcal{R}_A(q) + \mathcal{R}_B(q) \ .
\end{equation}
This property can be easily understood in terms of Feynman diagrams. The measure $d\mu(A,B)=d\mu(A)d\mu(B)$ factorizes for independent matrices and also
its gaussian part does $d\mu_0(A,B)=d\mu_0(A)d\mu_0(B)$.
As a consequence, the gaussian part does not contain mixed $AB$-terms 
and therefore the mixed propagators vanish $\langle AB \rangle_0=\langle AB^\dagger \rangle_0 = \ldots =0$. This means that there are no lines in the Feynman diagrams which would directly connect 
vertices of type $A$ and $B$. In effect, any connected diagram
contains either vertices solely of type $A$ or vertices solely of type $B$.
In the former case they come from $\mathcal{R}_A(q)$ and in the latter one from $\mathcal{R}_B(q)$. Thus the generating function for connected diagrams $\mathcal{R}_{A+B}(q)$ splits into separate contributions from diagrams 
of type $A$ and of type $B$: $\mathcal{R}_A(q) + \mathcal{R}_B(q)$. 
This is the standard diagrammatic interpretation of the additivity of cumulant generating functions.

One can work out consequences of the absence of mixed propagators 
also for the product $AB$ of independent matrices. We do not show this 
calculation here and refer the interested reader to \cite{bjn}. The multiplication
law can be written in the form of three equations.
\begin{equation}
\label{eq:multlawgeneral}
\begin{aligned}
\mathcal{R}_{AB}\left(\mathcal{G}_{AB}\right)&=\left[\mathcal{R}_{A}\left(\mathcal{G}_{B}\right)\right]^{L}
\left[\mathcal{R}_{B}\left(\mathcal{G}_{A}\right)\right]^{R}, \\
\left[\mathcal{G}_{A}\right]^{R}&=\mathcal{G}_{AB}\left[\mathcal{R}_{A}\left(\mathcal{G}_{B}\right)\right]^{L}, \\
\left[\mathcal{G}_{B}\right]^{L}&=\left[\mathcal{R}_{B}\left(\mathcal{G}_{A}\right)\right]^{R}\mathcal{G}_{AB},
\end{aligned}
\end{equation}
Since now the R transforms are quaternionic (or matrix-valued) the order of multiplication matters. The quaternions $\mathcal{G}_{A}=\mathcal{G}_{A}(\hat{z})$,
$\mathcal{G}_{B}=\mathcal{G}_{B}(\hat{z})$, $\mathcal{G}_{AB}=\mathcal{G}_{AB}(\hat{z})$ in the last equation correspond to the Green's functions for $A$, $B$ and $AB$ projected to the complex plane, that is calculated for quaternion $\hat{z}=(z,0)$ having the second part equal zero. Quaternions tagged in the last equation by $L$ or $R$ correspond to left or right rotated copies of quaternions, given by the following transformations:
\begin{align}
\left[\mathcal{Q} \right]^{L}&=U \mathcal{Q} U^{\dagger}, \nonumber \\
\left[\mathcal{Q} \right]^{R}&=U^{\dagger} \mathcal{Q} U.
\end{align}
where $U=\mathrm{diag}\left(e^{i\phi/4},e^{-i\phi/4}\right)$ is a unitary matrix constructed from the phase of the complex number $z=r e^{i\phi}$. 

For hermitian matrices the quaternionic set of Eqs. (\ref{eq:multlawgeneral}) 
simplifies to a set of equations for R transforms $R_A$, $R_B$ and $R_{AB}$ being complex functions
\begin{equation}
\label{eq:Rform}
\begin{aligned}
R_{AB}\left(z\right)&=R_{A}\left(w\right)R_{B}\left(v\right),\\
v&=zR_{A}\left(w\right),\\
w&=zR_{B}\left(v\right) .
\end{aligned}
\end{equation}
Moreover, if additionally the first cumulants of $A$ and $B$ are non-zero 
the last set of equations can be concisely written in this case in terms of the S transform \cite{v2}
\begin{equation}
S_{AB}(z) = S_A(z) S_B(z)
\label{eq:Sform}
\end{equation}
which is related to the R transform as follows \cite{bjn}
\begin{equation}
S_A(z) = \frac{1}{R_A(zS_A(z))}  \quad {\rm or} \quad R_A(z) = \frac{1}{S_A(zR_A(z))} \ . 
\label{eq:SR}
\end{equation} 
Unfortunately, there is no corresponding S transform representation 
of the multiplication law for non-hermitian matrices.

We finish this section by giving the transformation 
law for the R transform under multiplication of the matrix by a complex number:
$A \rightarrow A'=\alpha A$. Let us denote by $\hat{\alpha}$ 
a quaternion whose first Cayley-Dickson part is equal $\alpha$ 
and the second part is zero: $\hat{\alpha}=(\alpha,0)$.
One can see \cite{jn1,jn2} directly from the definition of the quaternionic Green's function (\ref{eq:QGF}) that
\begin{equation}
\mathcal{G}_{\alpha A} (q) = \mathcal{G}_A (\hat{\alpha}^{-1} q) \hat{\alpha}^{-1} \ .
\end{equation} 
If follows from Eq. (\ref{GF_R}) that the R transform obeys the following transformation law under multiplication of the matrix by a complex number
\begin{equation}
\label{alphaT}
{\mathcal R}_{\alpha A}(q) = \hat{\alpha} {\mathcal R}_A(q\hat{\alpha}) \ .
\end{equation}
Additionally if we include constant shifts in the transformation law 
$A \rightarrow A' = x \mathbb{1} + \alpha A$ we obtain
\begin{equation}
{\mathcal R}_{A'}(q) = \hat{x} + 
\hat{\alpha} {\mathcal R}_A(q\hat{\alpha}) \ .
\end{equation}
In particular when we apply this transformation to the standardized elliptic gaussian
law (\ref{stand}) we obtain the general form (\ref{nonstand}).

As an example of the application of this transformation law let us derive
the R transform for $X=\frac{1}{\sqrt{2}}\left(A + iB\right)$ where $A$ and $B$ are
standardized independent gaussian hermitian matrices: $\mathcal{R}_A(q) = \mathcal{R}_B(q)=q$. Using the addition law (\ref{addL}) 
and the transformation (\ref{alphaT}) for rescaling we easily find in agreement with \cite{jn1,jn2}
\begin{equation}
\mathcal{R}_X((z,w)) = \frac{1}{2} (z,w) + \frac{1}{2} (i,0)(z,w)(i,0) = (0,w) \ ,
\end{equation}
where the argument of the R transform is written as a Cayley-Dickson pair $q=(z,w)$.
In the last step of the calculation we applied Eq. (\ref{qprod}). As expected, we reconstructed the R transform for the Ginibre matrices. For pedagogical reasons we rewrite the last equation in the matrix representation (\ref{2x2}) which is better known
\begin{equation}
\mathcal{R}_X\left( \left( \begin{array}{cc}  z   & w \\
-\bar{w} & \bar{z}\end{array} \right) \right) 
= \frac{1}{2} 
\left(\begin{array}{cc} z & w \\
-\bar{w} & \bar{z} \end{array} \right) 
+ 
\frac{1}{2} 
\left(\begin{array}{cc} i & 0 \\
0 & -i \end{array} \right)
\left(\begin{array}{cc} z & w \\
-\bar{w} & \bar{z} \end{array} \right) 
\left(\begin{array}{cc} i & 0 \\
0 & -i \end{array} \right)
= 
\left(\begin{array}{cc} 0 & w \\
-\bar{w} & 0 \end{array} \right) . 
\end{equation}

\section{Applications of quaternionic R transform}

In this section, as an illustration, we give several examples of how to calculate eigenvalue densities of sums and products of gaussian elliptic random matrices using the
quaternionic R transform. 

\subsection{Sums of gaussian random matrices}

\begin{figure}
\centering
\includegraphics[scale=1]{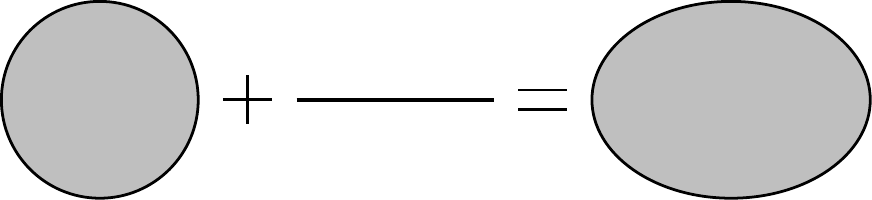}\ ,
\caption{Schematic representation of addition law in case of sum of Ginibre and GUE matrices, which results in Elliptic matrix.}
\label{fig:Ginibre+GUE=Elliptic}
\end{figure}

It is easy to see that a sum $C=A+B$ of gaussian elliptic matrices $A$ and $B$ is again an elliptic gaussian random matrix. Denote the R transforms (\ref{nonstand}) for $A$ and $B$ as  
$\mathcal{R}_A(z+wj) = x_A + \sigma^2_A \left(\mu_A e^{2i\phi_A} z + w j\right)$ and
$\mathcal{R}_B(z+wj) = x_B + \sigma^2_B \left(\mu_B e^{2i\phi_B} z + w j\right)$. As follows from the addition law (\ref{addL}) the R transform for $C$ has exactly the same form $\mathcal{R}_C(z+wj) = x_C + \sigma^2_C \left(\mu_C e^{2i\phi_C} z + w j\right)$
with $x_C = x_A+x_B$, $\sigma^2_C = \sigma^2_A + \sigma^2_B$, 
$\mu_C e^{i\phi_C} = (\mu_A \sigma^2_A e^{i\phi_A} + \mu_B \sigma^2_B e^{i\phi_B})/(\sigma_A^2+\sigma_B^2)$. In other words, the elliptic gaussian laws are stable under addition. In particular if we add a Ginibre matrix $X$ ($x_A=0, \mu_A=0, \sigma_A=1, \phi_A=0$) to a hermitian GUE matrix $H$ \cite{m} ($x_A=0,\mu_B=1,\sigma_B=1,\phi_B=0$) we obtain a matrix $C=X+H$ with the R transform $\mathcal{R}_C(z+wj)= z + 2 w j$. The corresponding eigenvalue density is uniform on an ellipse located at the origin of the complex plane with the longer semi-axis of length $\sqrt{2}$ on the real axis and the shorter one of length $1$ on the imaginary axis. This is schematically shown in Fig. \ref{fig:Ginibre+GUE=Elliptic}. 

\subsection{Products of gaussian random matrices} 
 
The eigenvalue density of a product of gaussian elliptic random matrices can be calculated from the general multiplication law Eqs. (\ref{eq:multlawgeneral}) and the Eq. (\ref{GRquat})
\begin{equation}
\label{GABcp}
\mathcal{G}_{AB}\left(\hat{z}\right)=\frac{1}{\hat{z}-\mathcal{R}_{AB}\left(\mathcal{G}_{AB}\left(\hat{z}\right)\right)} \ 
\end{equation}
which relates the Green's function and the R transform for the product $AB$. From these equations we obtain the Green's function $\mathcal{G}_{AB}\left(\hat{z}\right)$ on the complex plane that is for quaternions having the second part equal zero $\hat{z}=(z,0)$. Then we extract the first part of the quaternionic Green's function
\begin{equation}
G_{AB}(z) = F \mathcal{G}_{AB}\left(\hat{z}\right)
\end{equation}
which is a complex function $G_{AB}(z)$ and eventually we use Eq. (\ref{Grho}) 
to calculate the eigenvalue density 
\begin{equation}
\rho_{AB}(z) = \frac{1}{\pi} \frac{\partial G_{AB}(z)}{\partial \bar{z}}.
\end{equation}
Let us give a couple of examples. For convenience we introduce a shorthand notation. We denote the standardized elliptic matrix with the eccentricity parameter $\mu$ by $E(\mu)$ \cite{gi}. The matrix $E(\mu)$ has the following quaternionic R transform $\mathcal{R}_{E(\mu)}(z+wj) = \mu + wj$, or in the matrix representation 
\begin{equation}
\mathcal{R}_{E(\mu)}\left(\left(
\begin{array}{rr}
z & w \\
-\bar{w} & \bar{z} \end{array}\right)\right) = 
\left(
\begin{array}{rr}
\mu z & w \\
-\bar{w} & \mu \bar{z} \end{array}\right).
\end{equation}
We reserve a special notation for the standardized Ginibre matrices $X=E(0)$ and the standardized hermitian GUE matrix by $H=E(1)$. Let us note that the Ginibre ensemble is invariant under rotation $X \rightarrow X' = e^{i\phi} X$. Indeed, using the transformation law (\ref{alphaT}) we see that
\begin{equation}
\label{inv_phase}
\mathcal{R}_{X'}\left(\left(
\begin{array}{rr} z & w \\-\bar{w} & \bar{z} \end{array}\right)\right)=
\left(\begin{array}{cc} e^{i\phi} & 0 \\ 0  & e^{-i\phi} \end{array}\right)
\left(\begin{array}{rr} 0 & w \\-\bar{w} & 0 \end{array}\right)
\left(\begin{array}{cc} e^{i\phi} & 0 \\ 0  & e^{-i\phi} \end{array}\right) =
\left(\begin{array}{rr} 0 & w \\-\bar{w} & 0 \end{array}\right) = 
\mathcal{R}_X\left(\left(
\begin{array}{rr} z & w \\-\bar{w} & z \end{array}\right)\right)
\end{equation}
the R transform stays intact unter the multiplication of the Ginibre matrix by the phase factor $e^{i\phi}$. Now we are ready to consider examples.

We use the Pauli matrix representation for practical calculations, because complex algebra is familiar to all readers and is well implemented in many software packages, which is not the case for quaternions.

The first example is the product of $A=a\mathbb{1}+bX_{1}$ and $B=c\mathbb{1}+dX_2$
\begin{equation}
AB=\left(a\mathbb{1}+bX_1\right)\left(c\mathbb{1}+dX_2\right),
\end{equation}
where $a,b,c,d\in\mathbb{C}$, $b,d \ne 0$ and $X_1,X_2$ are independent standardized Ginibre matrices. The question we ask here is: what is the eigenvalue spectrum of the product? For such a product one can easily see, using the invariance of the Ginibre ensemble under the multiplication $X \rightarrow X' = e^{i\phi} X$ (\ref{inv_phase}), that the following redefinition of parameters:
\begin{equation}
s=\left|\frac{a}{b}\right|, \quad t=\left|\frac{c}{d}\right|, \quad
u=|bd| e^{i\mathrm{Arg}(ac)},
\end{equation}
reduces the calculations to the problem of finding the eigenvalue density of a matrix
\begin{equation}
u\left(s\mathbb{1}+X_1\right)\left(t\mathbb{1}+X_2\right),
\end{equation}
where $s,t\in\mathbb{R}_{+}$. The parameter $u$ scales the eigenvalue density with $\left|u\right|$ and rotates it around the center of the complex plane by $\mathrm{Arg}\left(u\right)$ angle. This means that it is sufficient
to consider products of the type
\begin{equation}
AB =\left(s\mathbb{1}+X_1\right)\left(t\mathbb{1}+X_2\right) 
\end{equation}
without loss of generality. Using the addition law (\ref{addL}) for matrices $A=s\mathbb{1}+X$ and $B=t\mathbb{1}+X$ we have 
\begin{equation}
\label{eq:RA}
\mathcal{R}_{A}\left(\left(
\begin{array}{cc}
w_{B} & v_{B} \\
-\bar{v}_{B} & \bar{w}_{B} \\
\end{array}
\right)\right)=\left(
\begin{array}{cc}
s & v_{B} \\
-\bar{v}_{B} & \bar{s} \\
\end{array}
\right),
\end{equation}
and 
\begin{equation}
\label{eq:RB}
\mathcal{R}_{B}\left(\left(
\begin{array}{cc}
w_{A} & v_{A} \\
-\bar{v}_{A} & \bar{w}_{A} \\
\end{array}
\right)\right)=\left(
\begin{array}{cc}
t & v_{A} \\
-\bar{v}_{A} & \bar{t} \\
\end{array}
\right),
\end{equation}
respectively. Since $s$ and $t$ are real we may omit their complex conjugation. Inserting Eqs. (\ref{eq:RA},\ref{eq:RB}) to the multiplication law formulas (\ref{eq:multlawgeneral}) and writing $z$ in the polar form $z=re^{i\phi}$ we get the R transform for the product:
\begin{align}
\mathcal{R}_{AB}\left(\mathcal{G}_{AB}\right)=\left(
\begin{array}{cc}
st - \bar{v}_{A}v_{B}e^{i\phi} & tv_{A}e^{-i\phi/2}+sv_{B}e^{i\phi/2} \\
-t\bar{v}_{A}e^{i\phi/2}-s\bar{v}_{B}e^{-i\phi/2} & st-v_{A}\bar{v}_{B}e^{-i\phi}
\end{array}
\right).
\end{align}
Now using Eq. (\ref{GABcp}) we obtain the quaternionic Green's function on the complex plane. Its first Cayley-Dickson part reads
\begin{equation}
\label{eq:s+X_t+X_greens}
G_{AB}(z)=-\frac{st-re^{-i\phi}-v_{A}\bar{v}_{B}}{s^{2}t^{2}+r^{2}+t^{2}\left|v_{A}\right|^{2}+s^{2}\left|v_{B}\right|^{2}+r(v_{A}\bar{v}_{B}+\bar{v}_{A}v_{B})+\left|v_{A}\right|^{2}\left|v_{B}\right|^{2}-2str\cos\phi}.
\end{equation}
Now we have to find $v_{A}$ and $v_{B}$ as a function of complex argument $z$ (or equivalently $r$ and $\phi$). To this end we utilize the second and third quaternionic equations in (\ref{eq:multlawgeneral}) that are equivalent to the following four independent complex equations
\begin{align}
t\left|v_{A}\right|^{2}+e^{i\phi}\left(w_{A}r+\bar{v}_{A}v_{B}\left(s+w_{A}\right)\right)&=t(1+sw_{A}),\label{eq:s+X_t+X_first}\\
s\left|v_{B}\right|^{2}+e^{i\phi}\left(w_{B}r+v_{A}\bar{v}_{B}\left(t+w_{B}\right)\right)&=s(1+tw_{B}),\label{eq:s+X_t+X_second}\\
e^{i\phi}\left(-s\bar{w}_{A}v_{B}+v_{B}\left|v_{A}\right|^{2}+v_{A}r-v_{B}\right)&=tv_{A} \left(s+\bar{w}_{A}\right),\label{eq:s+X_t+X_third}\\
e^{-i\phi}\left(-tw_{B}v_{A}+v_{A}\left|v_{B}\right|^{2}+v_{B}r-v_{A}\right)&=sv_{B}\left(t+w_{B}\right).\label{eq:s+X_t+X_fourth}
\end{align}
The last two equations have a trivial solution: $v_{A}=v_{B}=0$. It corresponds to the holomorphic part of Green's function, $G\left(z\right)=(z-st)^{-1}$, valid outside the eigenvalue domain.

By some algebraic manipulations in Eq. (\ref{eq:s+X_t+X_third}) we may express $v_{B}$ as
\begin{equation}
v_{B}=v_{A}\frac{\alpha}{\beta-\left|v_{A}\right|^{2}},
\end{equation}
which means that combinations $v_{A}\bar{v}_{B}$ and $\bar{v}_{A}v_{B}$ do not depend on phases of $v_{A}$ and $v_{B}$. Indeed an inspection of the expression for the Green's function (\ref{eq:s+X_t+X_greens}) shows, that it depends on $v_A$ and $v_B$ only
through the moduli $\left|v_{A}\right|$,$\left|v_{B}\right|$. Their phases are also irrelevant in Eqs. (\ref{eq:s+X_t+X_third},\ref{eq:s+X_t+X_fourth}). In fact, this is true for any product of elliptic ensembles. For the rest of the discussion those phases will be set to $0$, so $v_{A}=\bar{v}_{A}$, $v_{B}=\bar{v}_{B}$ and $v_{A},v_{B}\in\mathbb{R_{+}}$. Solving Eqs. (\ref{eq:s+X_t+X_first},\ref{eq:s+X_t+X_second},\ref{eq:s+X_t+X_third},\ref{eq:s+X_t+X_fourth}) for $v_{A}$ and $v_{B}$ yields:
\begin{align}
v_{A}(-1+v_{A}^{2} + s^{2}) (v_{B}^{2} + t^{2}) + (-1 + 2 v_{A}^{2}) v_{B} r + v_{A} r^{2} - 2 stv_{A}r \cos\phi = 0,\label{eq:s+X_t+X_finalfirst}\\
v_{B}(-1+v_{B}^{2} + t^{2}) (v_{A}^{2} + s^{2}) + (-1 + 2 v_{B}^{2}) v_{A} r + v_{B} r^{2} - 2 stv_{B}r \cos\phi = 0.\label{eq:s+X_t+X_finalsecond}
\end{align}
Note that those equations are invariant under the simultaneous change of sings $v_A \rightarrow -v_A$ and $v_B \rightarrow -v_B$ which is a remnant of the 
symmetry of $v$'s mentioned above. In effect if  $v_{A}$ and $v_{B}$ is a solution to these equations,  then also $v_{A}'=-v_{A}$ and $v_{B}'=-v_{B}$ is a solution that gives the same Green's function, leading to the same eigenvalue density.

The eigenvalue density is represented as a function with a compact support on the complex plane. Let us focus on the outline (contour) of this support. We may calculate it by matching the holomorphic ($v_{A}=v_{B}=0$) and non-holomorphic solutions. To do so, we solve quadratic Eq. (\ref{eq:s+X_t+X_finalfirst}) for $v_{B}$, and pick the leading term of the solution, that vanishes in the  limit $v_{A}\rightarrow0$. Then by inserting this solution into Eq.(\ref{eq:s+X_t+X_finalfirst}), and performing the limit $v_{A}\rightarrow0$ we are left with the following equation
\begin{equation}
\begin{aligned}
\label{eq:s+X_t+X_contour}
&(s^{2}t^{2}-s^{4}t^{2}-s^{2}t^{4}+s^{4}t^{4})+(2s^{3}t\cos\phi +2st^{3}\cos\phi-4s^{3}t^{3}\cos\phi)r+\\+&(-1-s^{2}-t^{2}+2s^{2}t^{2}+4s^{2}t^{2}\cos^{2}\phi)r^{2}-4st\cos\phi\ r^{3}+r^{4}=0,
\end{aligned}
\end{equation}
which may be treated as a fourth order polynomial equation for $r$ and solved for any value of $\phi$ giving the contour of the eigenvalue distribution.

\begin{figure}
\includegraphics[width=0.45\textwidth]{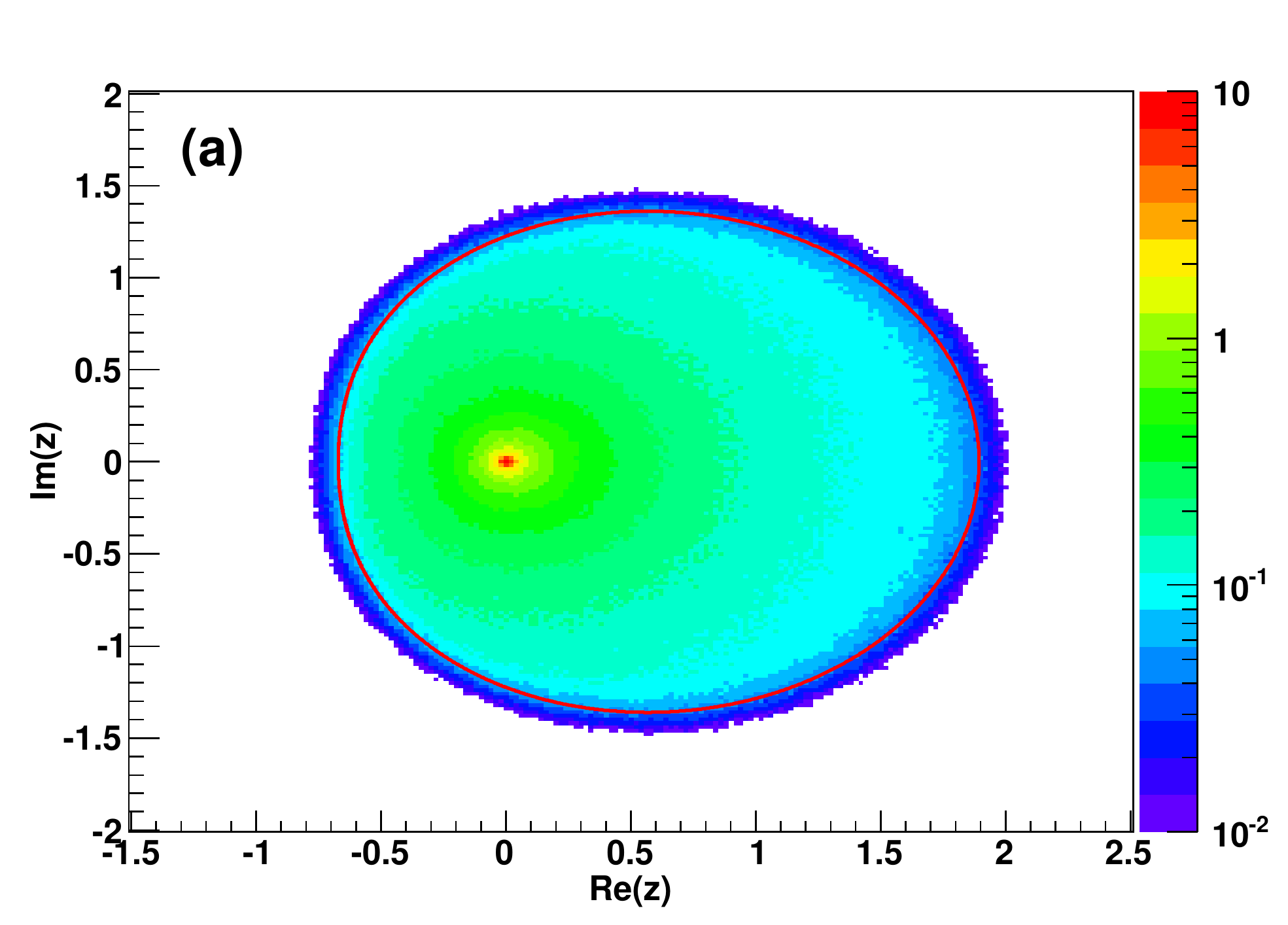}
\includegraphics[width=0.45\textwidth]{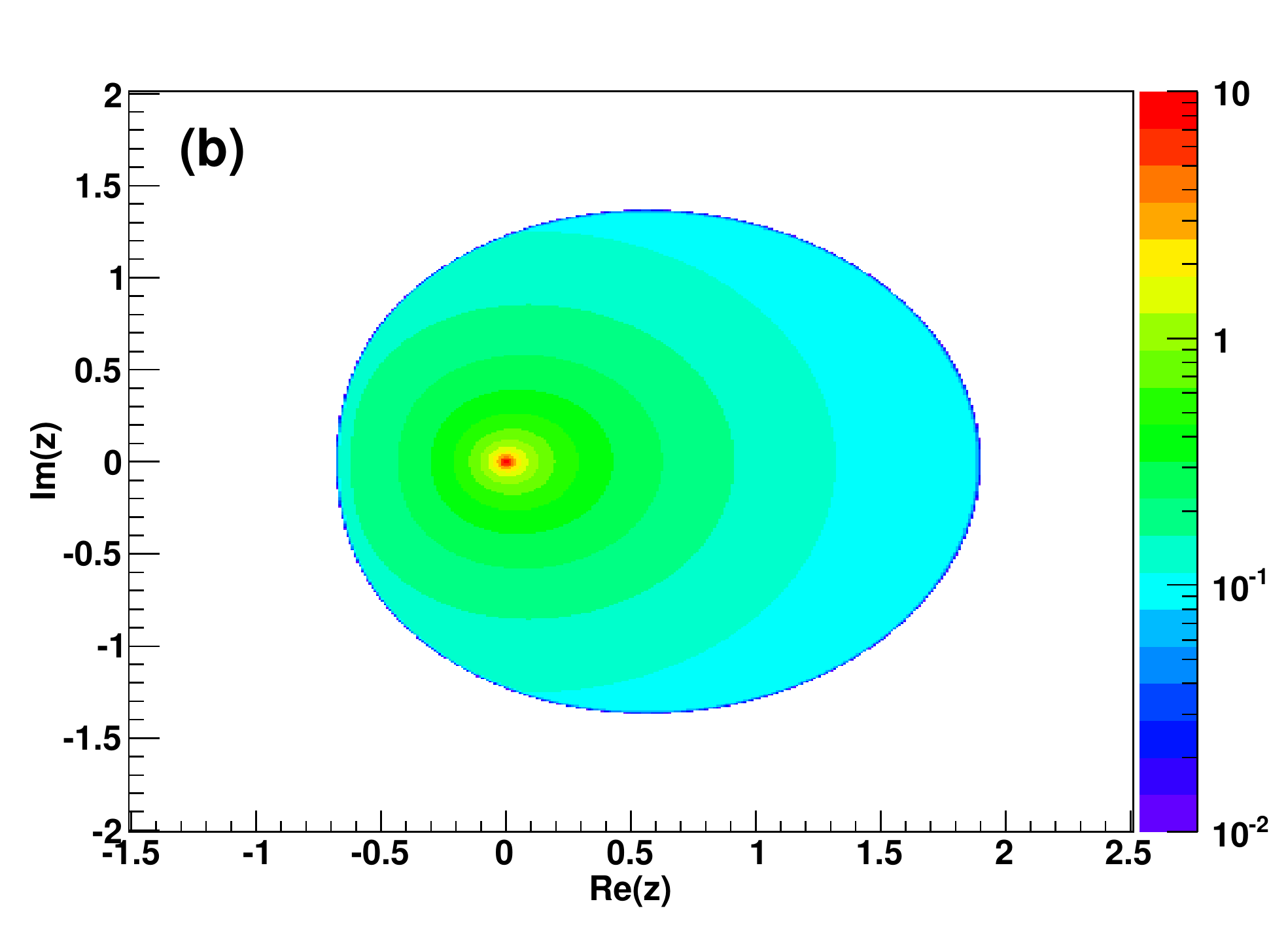}
\includegraphics[width=0.45\textwidth]{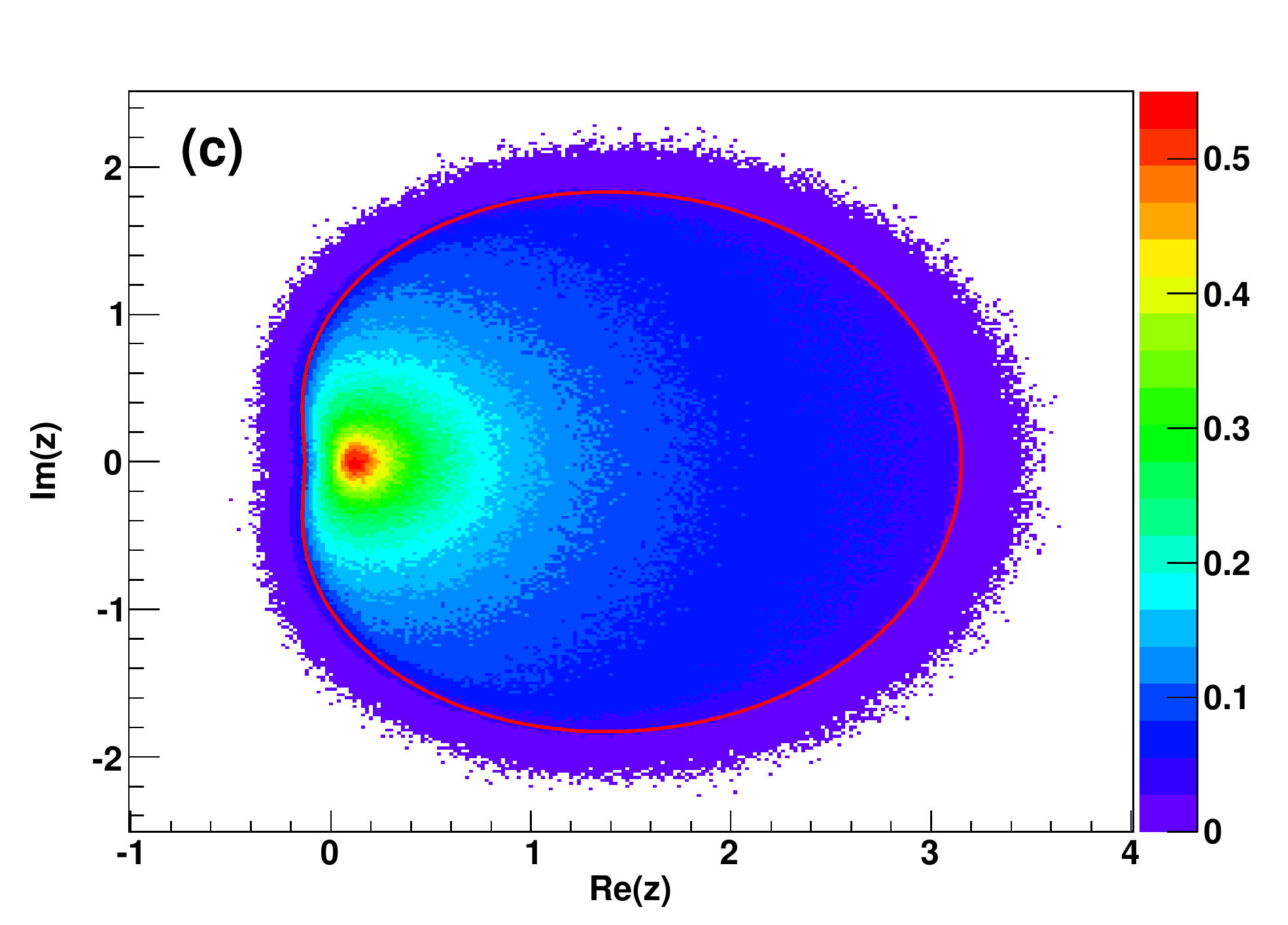}
\includegraphics[width=0.45\textwidth]{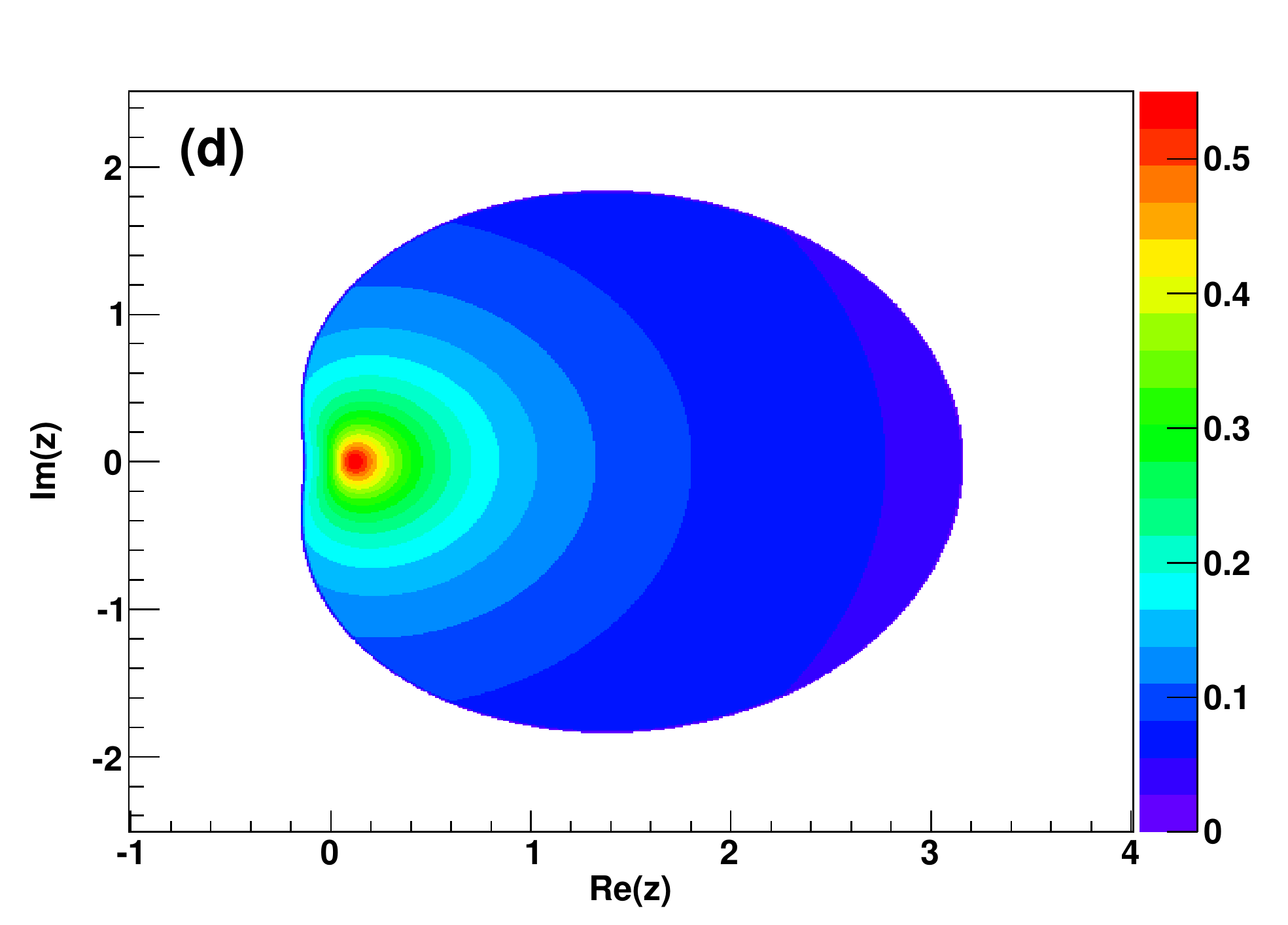}
\includegraphics[width=0.45\textwidth]{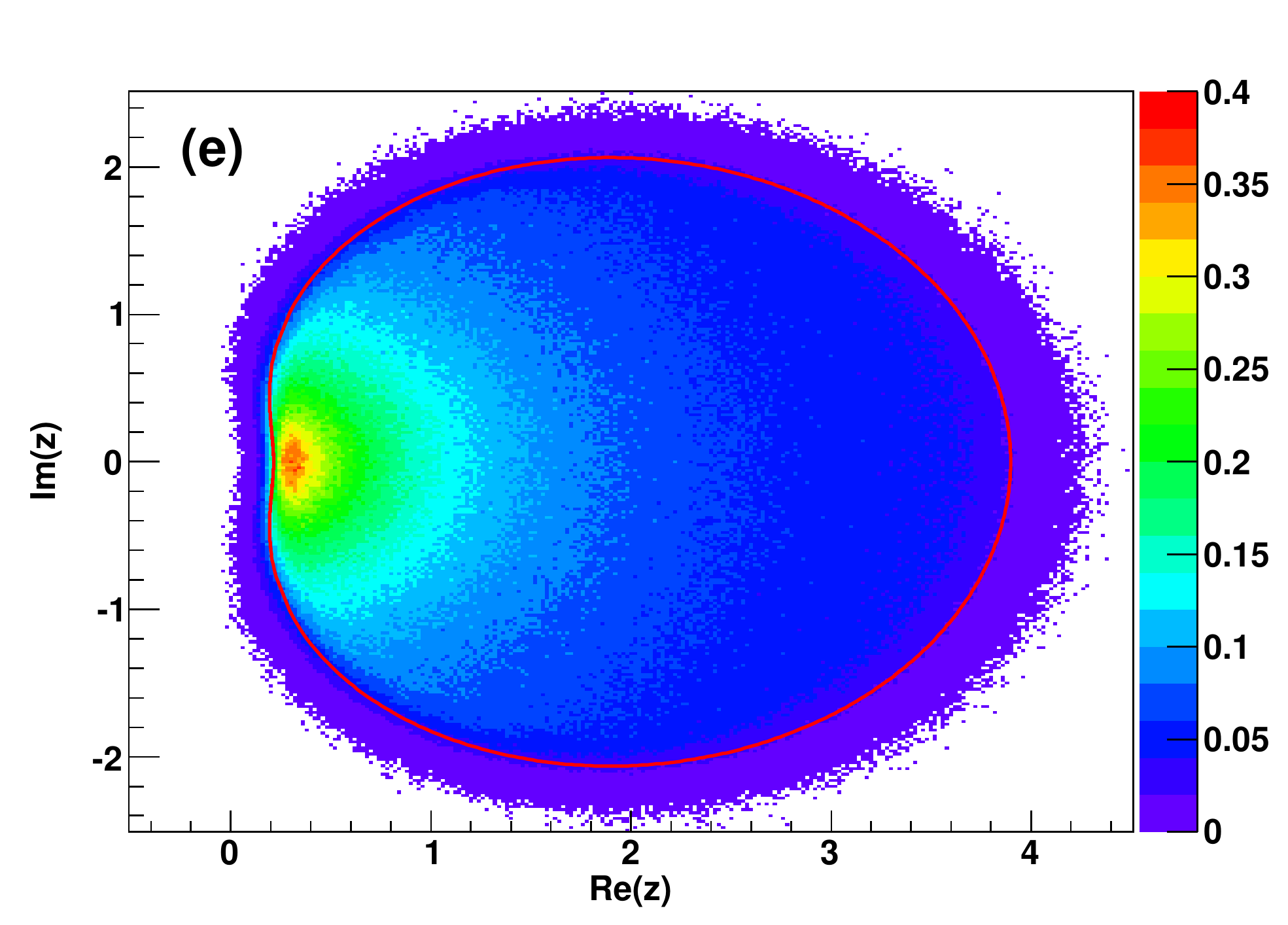}
\includegraphics[width=0.45\textwidth]{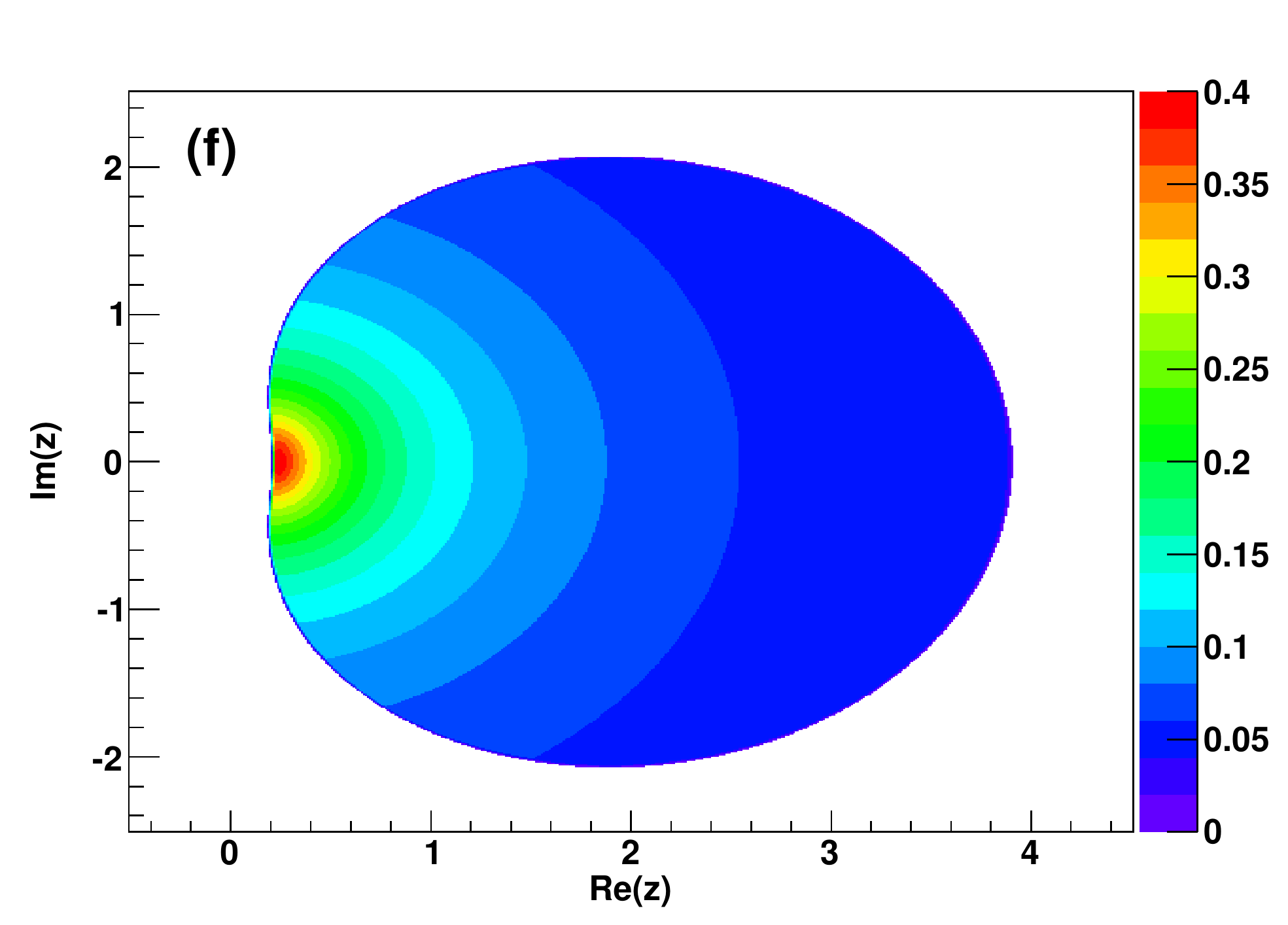}
\caption{(Color online) Numerical eigenvalue distributions for matrices $(s\mathbb{1}+X)(t\mathbb{1}+X)$ $100\times 100$ for (a) $s=0.50$,$t=0.75$ (c) $s=0.9$,$t=1.2$ and (e) $s=1.2$,$t=1.3$ compared with theoretically evaluated ones in (b),(d) and (f) respectively. The edge of the distribution is shown on the plots (a),(c) and (e) (red curve). Each histogram is made from $10^{7}$ eigenvalues.}
\label{fig:s+X_t+X}
\end{figure}

Now the remaining part is to numerically solve real polynomial Eqs. (\ref{eq:s+X_t+X_finalfirst},\ref{eq:s+X_t+X_finalsecond}), e.g. on a lattice inside the previously calculated domain, plug the results into Eq. (\ref{eq:s+X_t+X_greens}) and calculate eigenvalue density according to Eq. (\ref{Grho}). The results for several different values of $s$ and $t$ are shown in Fig. \ref{fig:s+X_t+X}, together with comparison to numerical simulations. The agreement is very good. Due to the finite size
effects the edges of the support are smeared and instead of a sharp edge on the contour line there is a crossover region of width that decreases with $N$ where the eigenvalue density falls off to zero.

The equation (\ref{eq:s+X_t+X_contour}) is general in the sense
that it holds for any $t$ and $s$. The special symmetric case $t=s$ 
can be solved in a simpler way.  The symmetry $A\longleftrightarrow B$
which holds in this case significantly reduces the complexity of the problem 
because one may at the beginning of the calculation set $w_{A}=w_{B}$ and $v_{A}=v_{B}$
and reduce the number of unknown variables. 
Two special cases: $t=s=0$ and $t=s=1$ of this symmetric problem have been discussed previously \cite{bjn} and can be used as a cross-check of our general formula  (\ref{eq:s+X_t+X_contour}). The case $t=s=0$ corresponds to the product of
two standardized Ginibre matrices with the contour given by the equation \cite{bjw}
\begin{equation}
r^{2}-1=0 \ .
\end{equation}
The case $t=s=1$ corresponds to the product of two standardized Ginibre matrices shfited by unit matrix and has the contour given by the equation 
\begin{equation}
r^{2}-4\cos\phi\ r+4\cos^{2}\phi\ -1=(r-2\cos\phi\ -1)(r-2\cos\phi\ +1)=0.
\end{equation}
The first one gives a circle and the second one 
a Pascal lima\c{c}on, in agreement with \cite{gjjn,bjn}.

\begin{figure}
\includegraphics[width=0.45\textwidth]{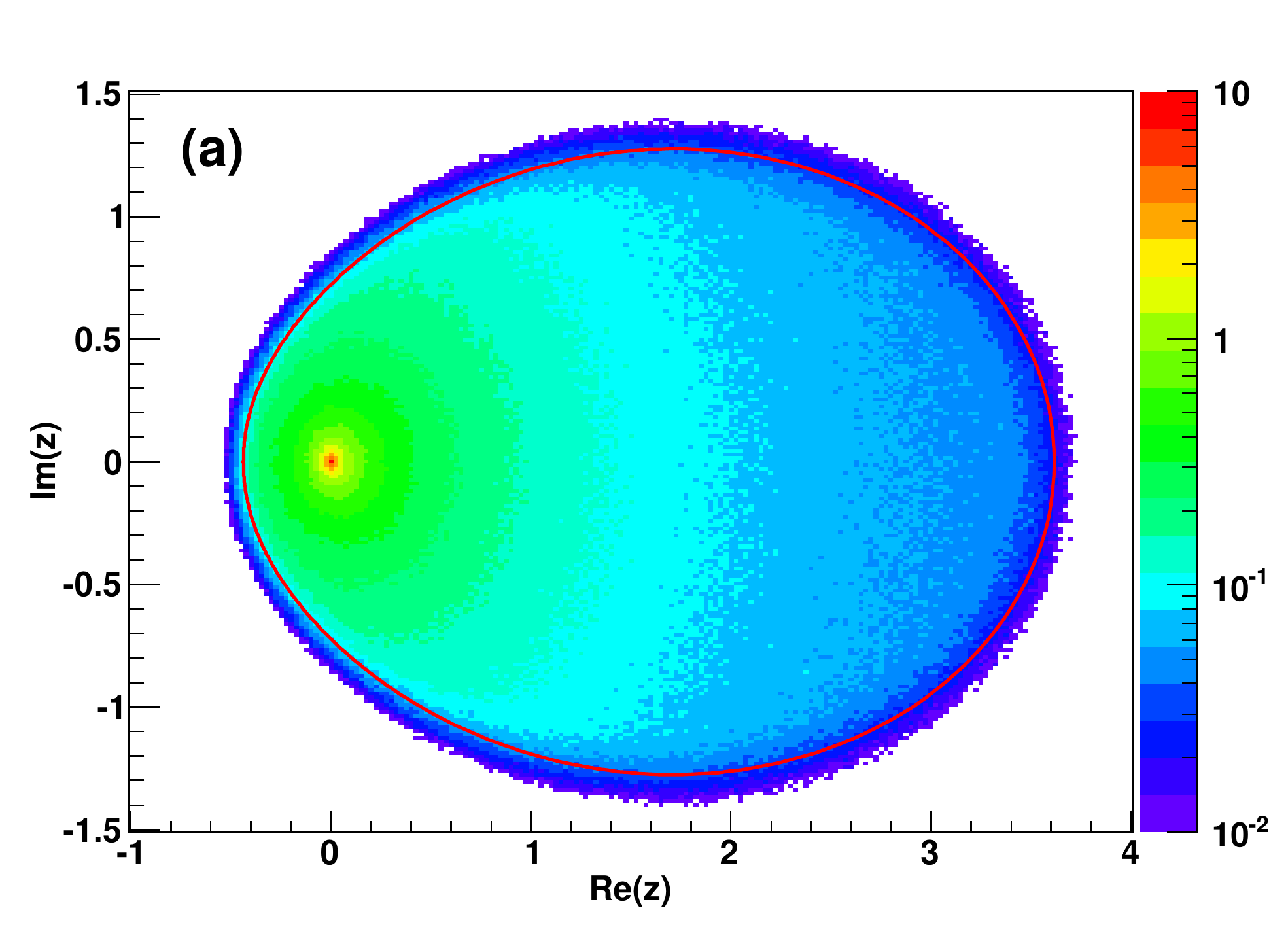}
\includegraphics[width=0.45\textwidth]{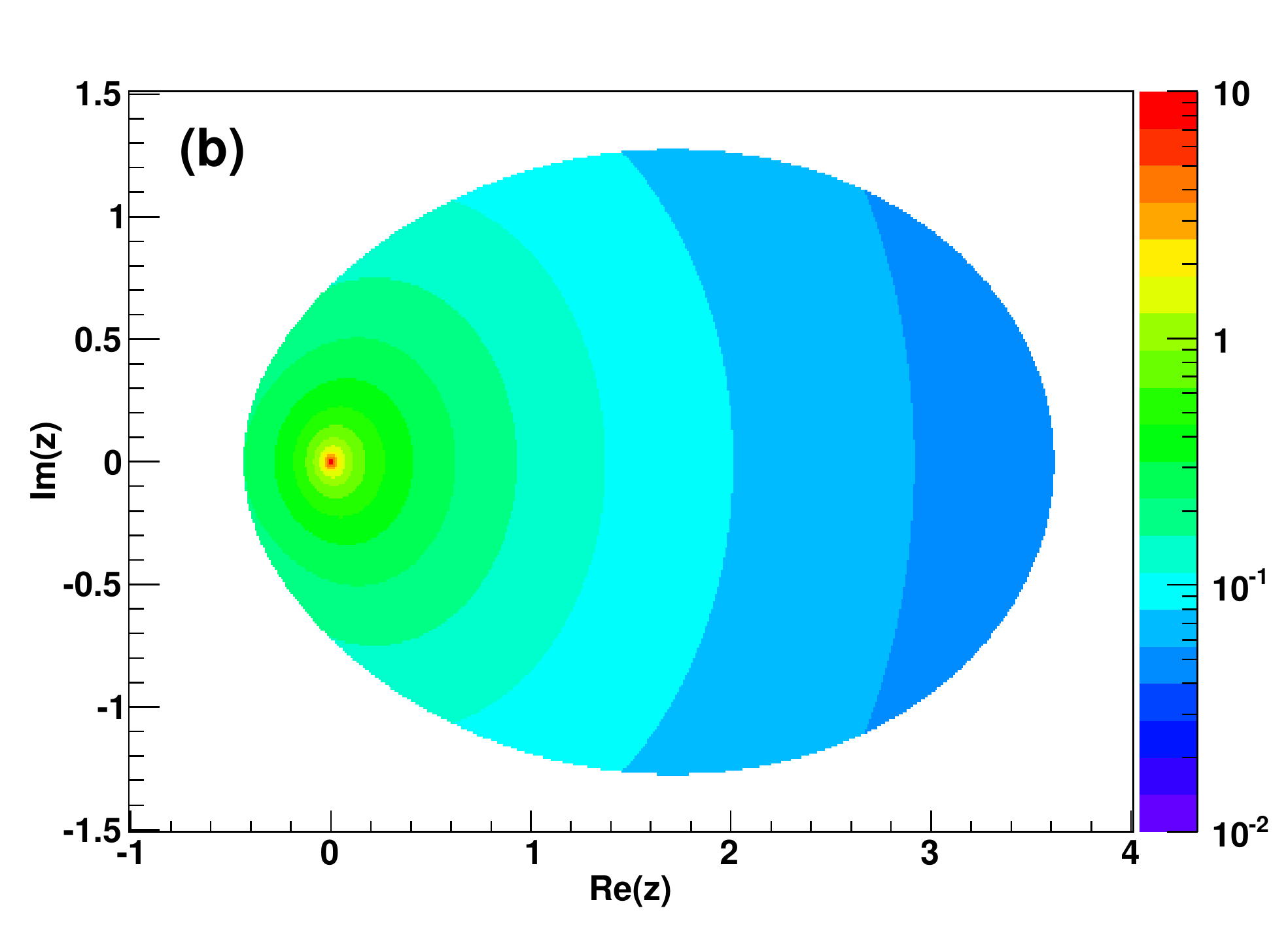}
\includegraphics[width=0.45\textwidth]{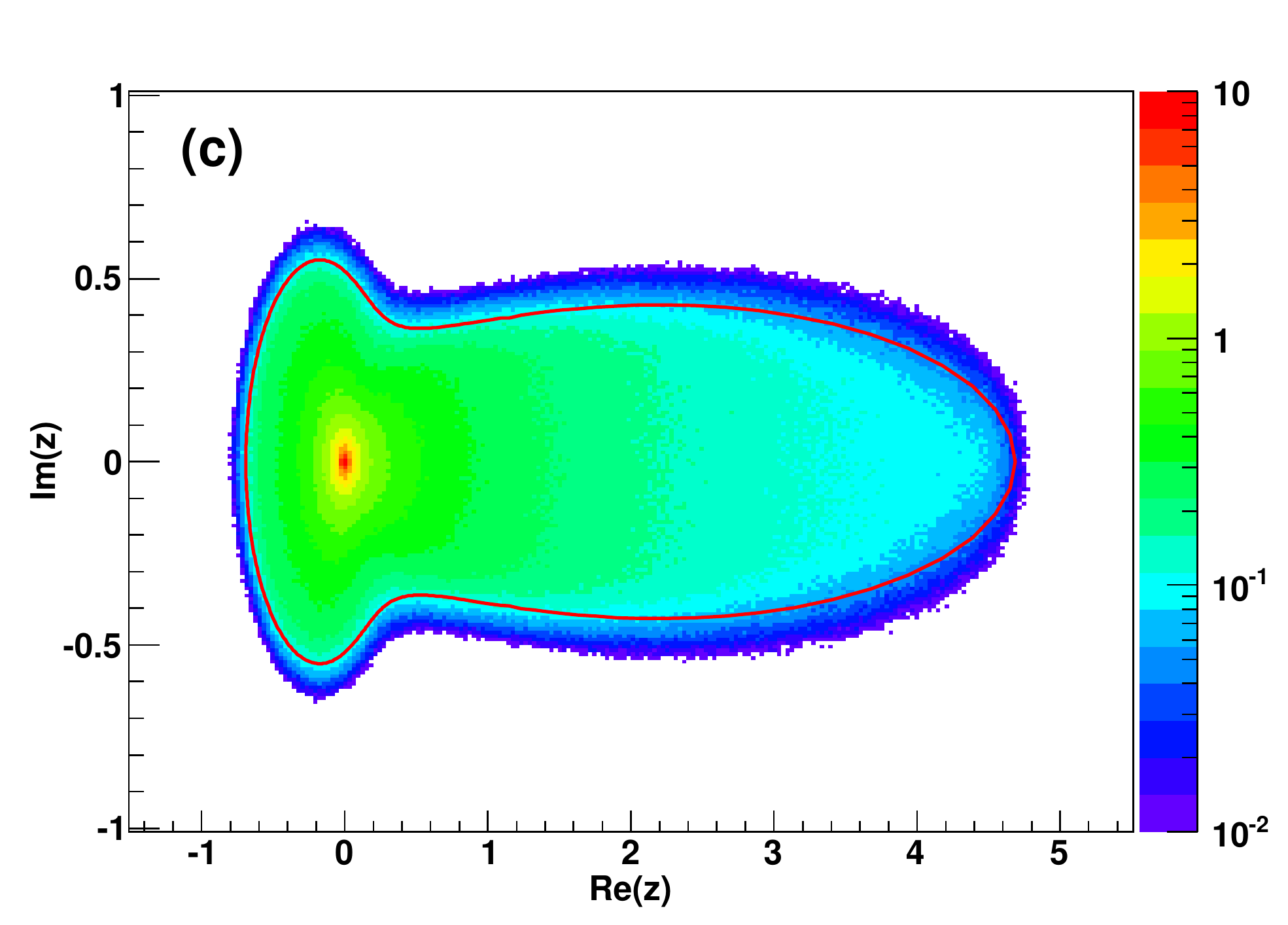}
\includegraphics[width=0.45\textwidth]{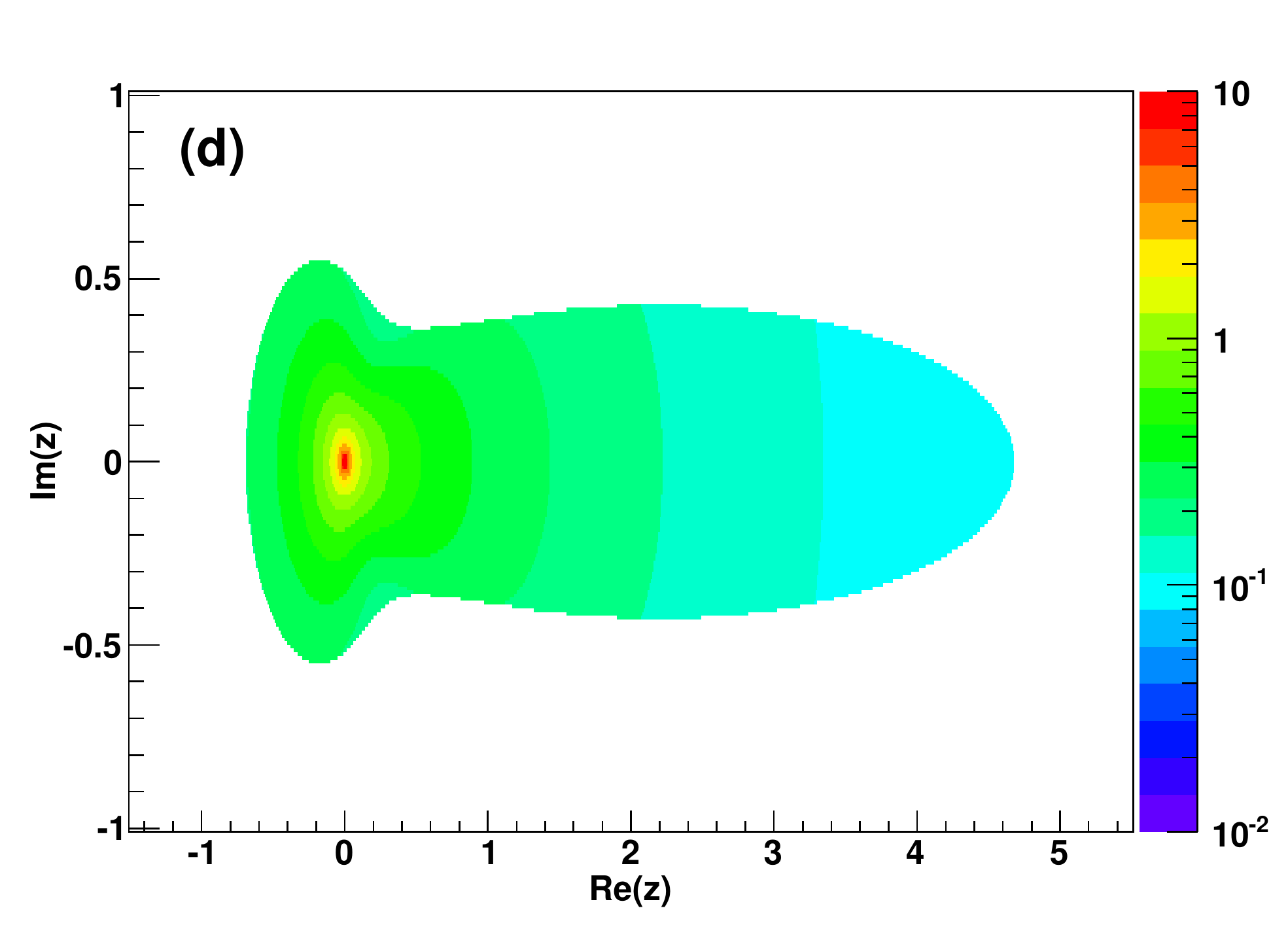}
\includegraphics[width=0.45\textwidth]{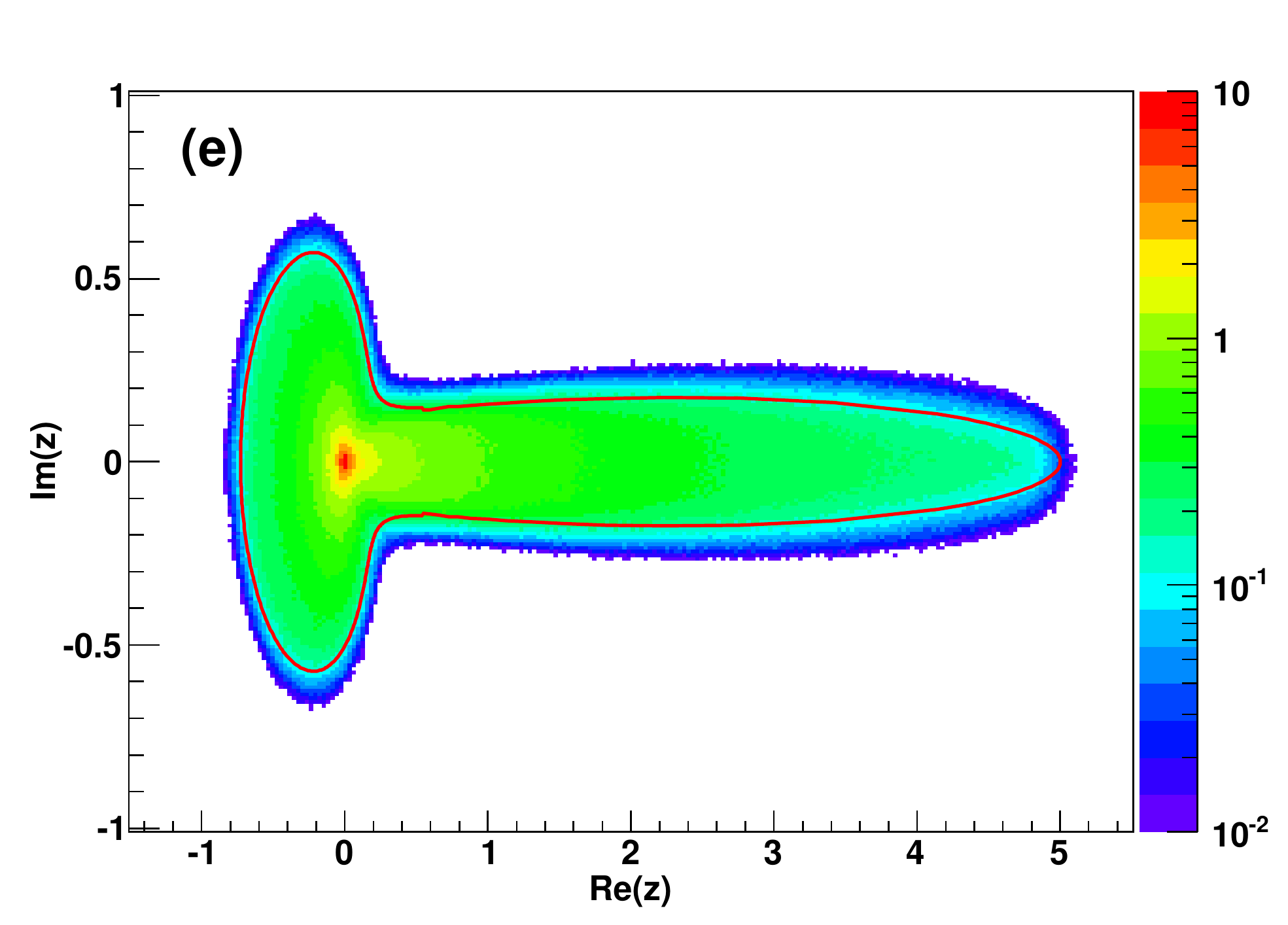}
\includegraphics[width=0.45\textwidth]{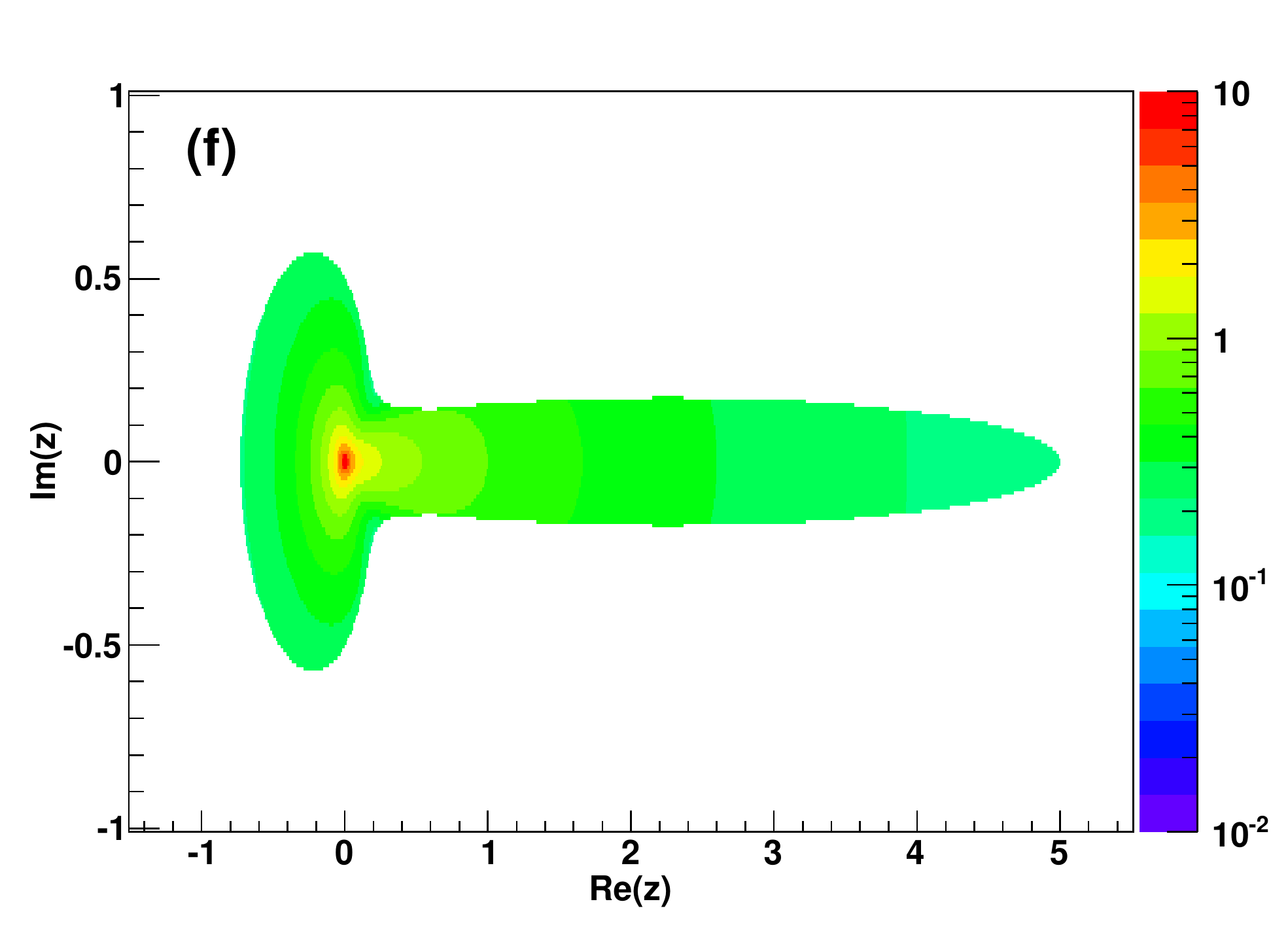}
\caption{(Color online) Numerical eigenvalue distribution for matrices $(\mathbb{1}+E_{1})(\mathbb{1}+E_{2})$ $100\times 100$ for (a)$\mu=1/3$ (c) $\mu=4/5$ and (e) $\mu=23/25$ compared with theoretically evaluated ones in (b), (d) and (f) respectively. The edge of the distribution is shown on the plots (a), (c) and (e) (red curve). Each histogram is made from $10^{7}$ eigenvalues.}
\label{fig:1+E_1+E}
\end{figure}

The second example we consider is a product 
\begin{equation}
AB=(\mathbb{1}+E_{1}(\mu))(\mathbb{1}+E_{2}(\mu)),
\end{equation}
of two independent identically distributed shifted elliptic matrices $A=\mathbb{1}+E_{1}(\mu)$ and $B=\mathbb{1}+E_{2}(\mu)$. We follow exactly the same procedure as in the first example except that now we exploit the invariance of the problem
under the exchange $E_{1}(\mu)\longleftrightarrow E_{2}(\mu)$ to make a symmetric Ansatz
$w_{A}=w_{B}=w$ and $v_{A}=v_{B}=v$ at the beginning of the calculations. This simplifies the problem remarkably. The R transform for a shifted elliptic matrix is given by (\ref{nonstand})
\begin{equation}
\mathcal{R}\left(\left(
\begin{array}{cc}
w & v \\
-\bar{v} & \bar{w}
\end{array}
\right)\right)=\left(
\begin{array}{cc}
1+w\mu & v \\
-\bar{v} & 1+\bar{w}\mu
\end{array}
\right), \label{eq:Relliptic}
\end{equation}
and for the product:
\begin{equation}
\mathcal{R}_{AB}\left(\mathcal{G}_{AB}\right)=\left(
\begin{array}{cc}
\left(1+w\mu\right)^{2}-\left|v\right|^{2}e^{i\phi} & v\left(e^{-i\phi/2}\left(1+w\mu\right)+e^{i\phi/2}\left(1+\bar{w}\mu\right)\right) \\
-\bar{v}\left(e^{-i\phi/2}(1+w\mu)+e^{i\phi/2}(1+\bar{w}\mu)\right) & \left(1+\bar{w}\mu\right)^{2}-\left|v\right|^{2}e^{-i\phi}
\end{array}
\right).
\end{equation}
The complex Green's function is therefore equal
\begin{equation}
G_{AB}\left(z\right)=\frac{e^{i\phi}\left(1+\bar{w}\mu\right)^{2}-r-\left|v\right|^{2}}
{
\left(r-e^{i\phi}-\bar{w}^{2}e^{i\phi}\mu^{2}-2e^{i\phi}\bar{w}\tau\right)\left(\left(1+w\mu\right)^{2}-re^{i\phi}\right)
-\left|v\right|^{2}e^{i\phi}\left(
2\left(1+r+\mu\left(w+\bar{w}+\mu\left|w\right|^{2}\right)\right)+\left|v\right|^{2}\right)}.
\end{equation}
The four complex equations for $w$ and $v$ reduce to two
\begin{align}
e^{i\phi}\left(wr+\left(1+w+\bar{w}\mu\right)\left|v\right|^{2}\right)=\left(1+w\mu\right)\left(1+w+w^{2}\mu-\left|v\right|^{2}\right),\label{eq:1+E_1+E_first}\\
ve^{i\phi}\left(-1+r+\left|v\right|^{2}-\bar{w}\left(1+\bar{w}\mu\right)\right)=v\left(1+w\mu\right)\left(1+w\mu +\bar{w}\right).\label{eq:1+E_1+E_second}
\end{align}
$v=0$ is again a trivial solution. A non-trivial solution is obtained by the procedure of matching holomorphic to non-holomorphic solutions by linear approximation in $v$ in the very same way as in the previous example. We find equations describing the boundary of the eigenvalue distribution
\begin{align}
re^{i\phi}w=\left(1+w\mu\right)\left(1+w+w^{2}\mu\right),\label{eq:1+E_1+E_first_boundary}\\
e^{i\phi}\left(-1+r-\bar{w}\left(1+\bar{w}\mu\right)\right)=\left(1+w\mu\right)\left(1+w\mu +\bar{w}\right),\label{eq:1+E_1+E_second_boundary}
\end{align}
which may be solved for any value of $\phi\in[0,2\pi)$ giving the desired contour. 
Then one may numerically solve the polynomial Eqs. (\ref{eq:1+E_1+E_first},\ref{eq:1+E_1+E_second}) inside the eigenvalue domain. The results for different values of $\mu$, are presented in Fig. \ref{fig:1+E_1+E} supported by numerical simulation of random matrices. Theoretical predictions match the simulation up to finite size effects.

\begin{figure}
\includegraphics[width=0.45\textwidth]{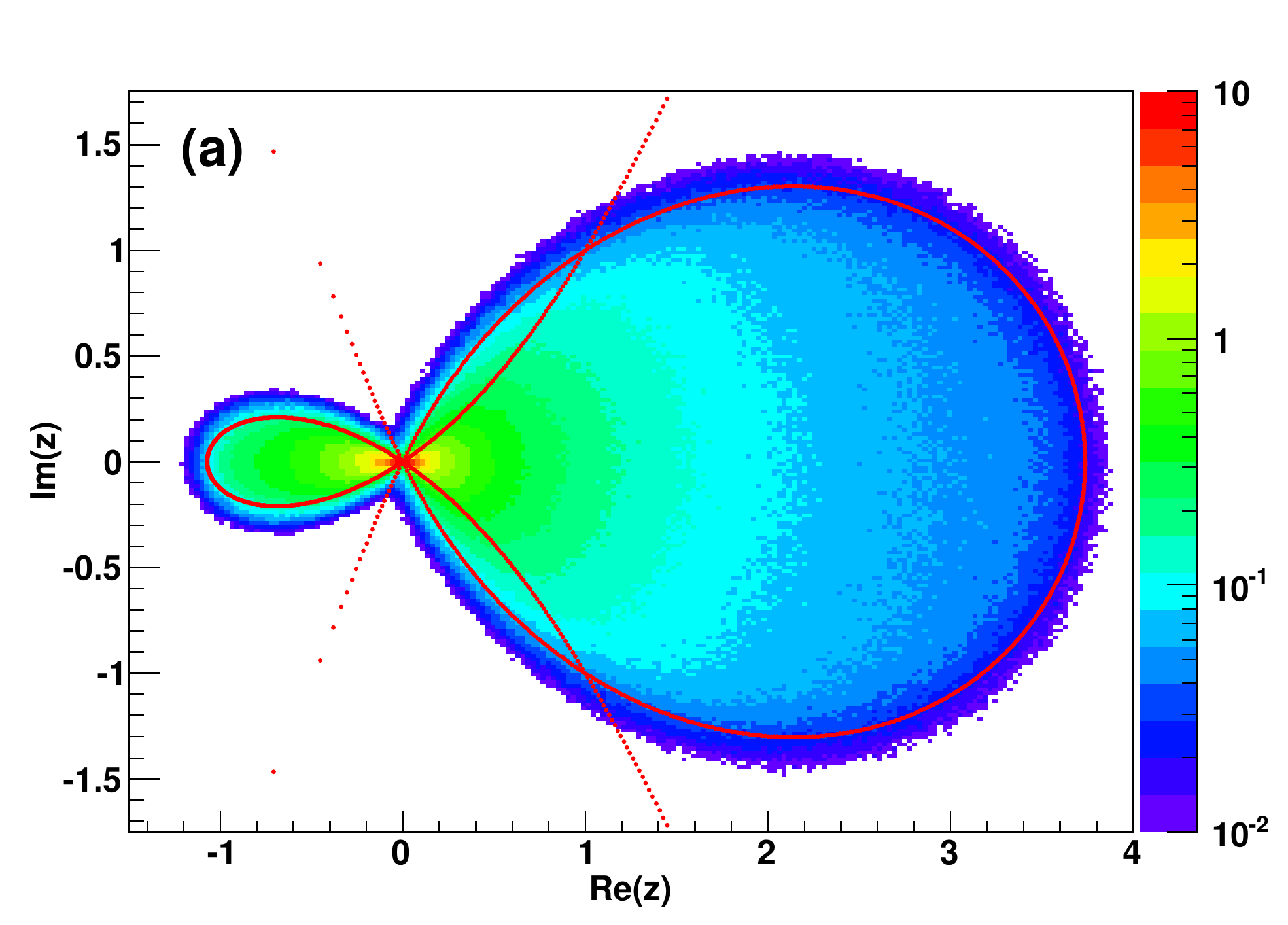}
\includegraphics[width=0.45\textwidth]{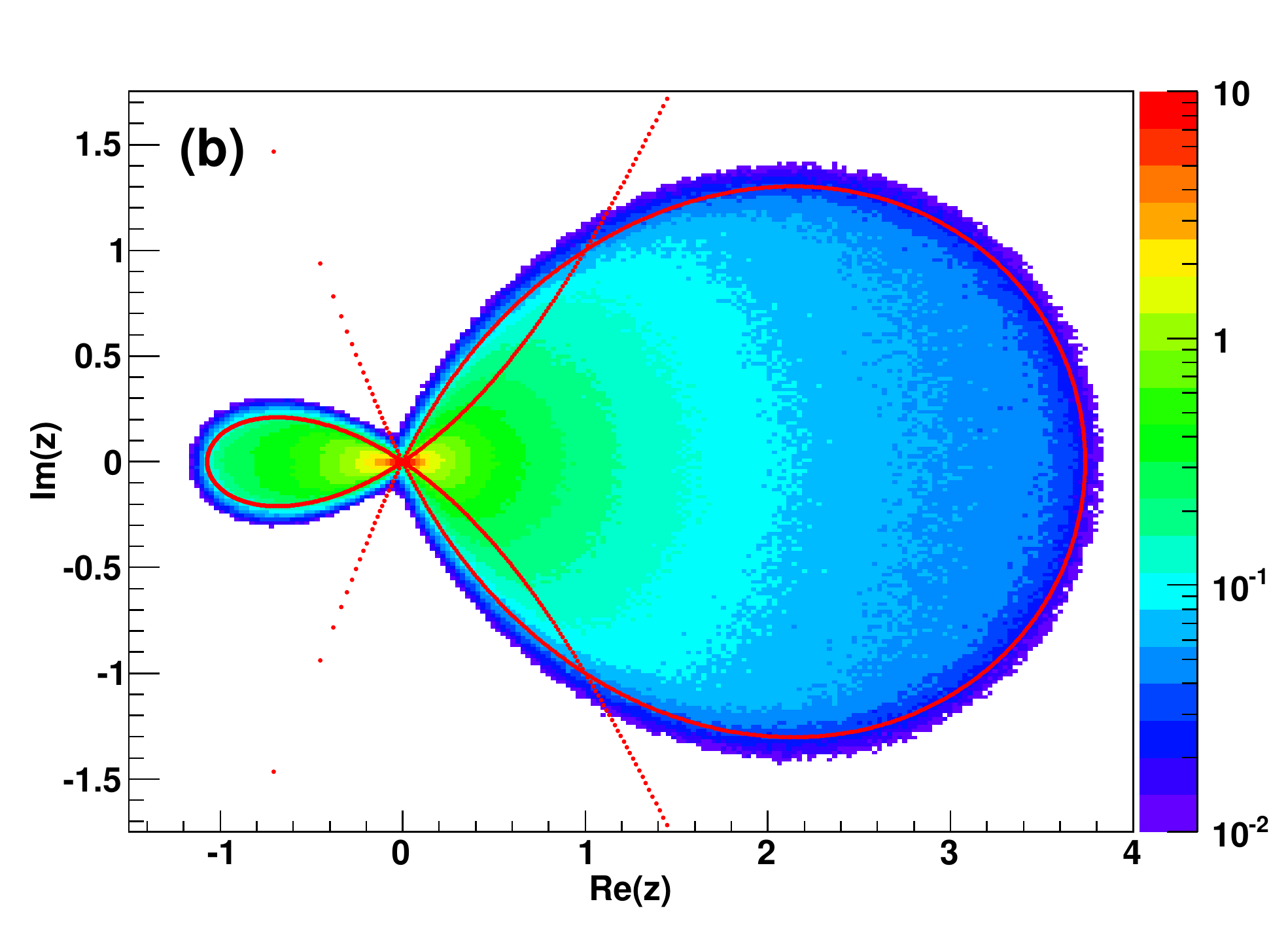}
\includegraphics[width=0.45\textwidth]{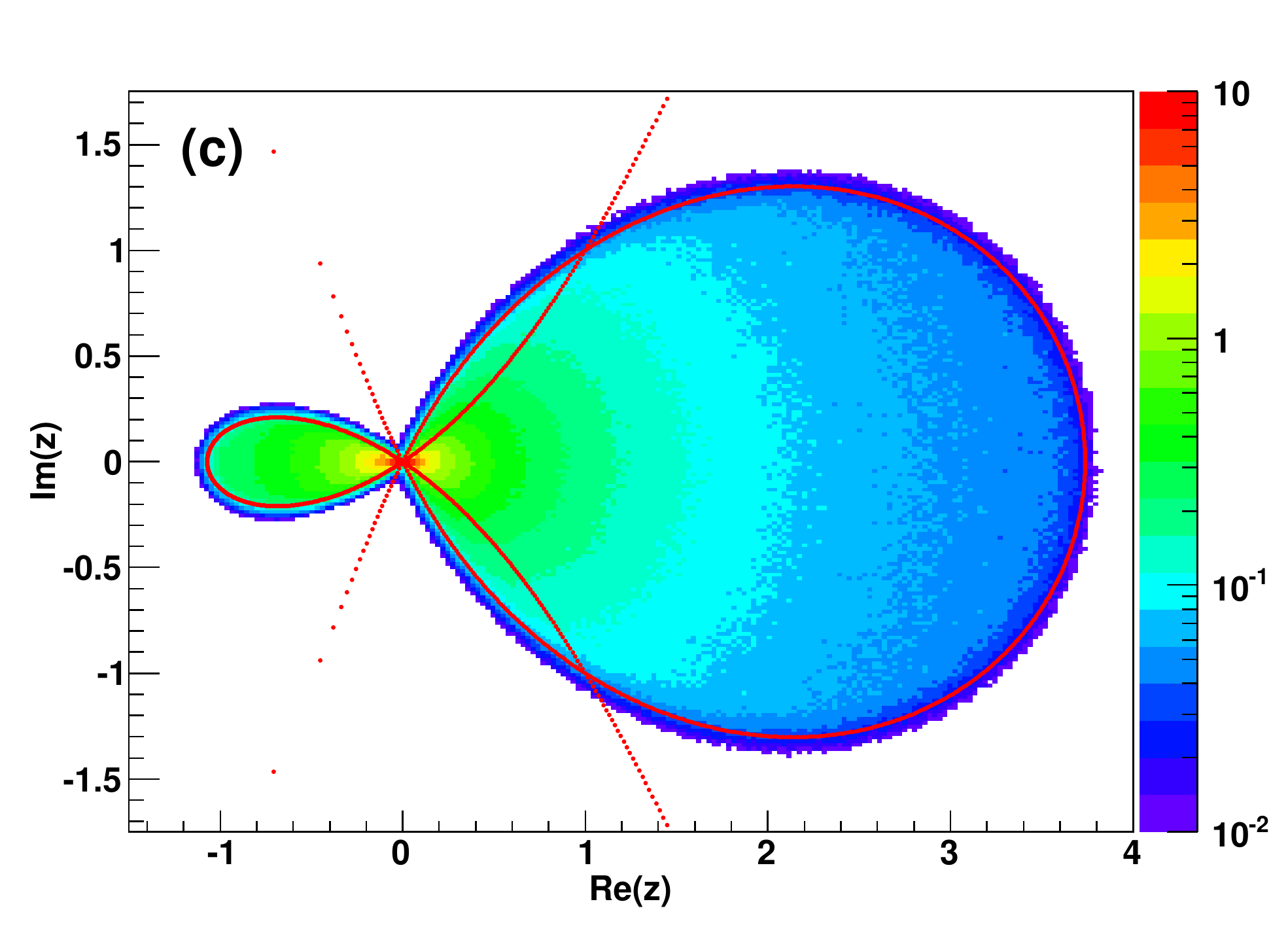}
\caption{(Color online) Numerical eigenvalue distribution for matrices $(\mathbb{1}+H)(\mathbb{1}+X)$ of size (a) $50\times 50$ (b) $100\times 100$ and (c) $200\times 200$ compared with theoretical prediction for the edge of the distribution (red points). Each histogram is made from $10^{7}$ eigenvalues. The eigenvalue domain is determined by closed regions given by analytic solutions of Eqs. (\ref{eq:1+H_1+X_r}) and (\ref{eq:1+H_1+X_wB}).}
\label{fig:1+H_1+X}
\end{figure}

Finally the last example to be analyzed here is the product of the form:
\begin{equation}
AB=(\mathbb{1}+H)(\mathbb{1}+X),
\end{equation}
of a shifted standardized hermitian GUE matrix $A=\mathbb{1}+H$ and a shifted standardized Ginibre matrix $B=\mathbb{1}+X$. The R transforms are special cases of the elliptic one (\ref{eq:Relliptic}), for $\mu=1$ and $\mu=0$ respectively.
The calculation procedure is analogous to the previous ones. The R transform for the product is:
\begin{equation}
\mathcal{R}_{AB}\left(\mathcal{G}_{AB}\right)=\left(
\begin{array}{cc}
1+w_{B}-\bar{v}_{A}v_{B}e^{i\phi} & e^{-i\phi/2}\left(v_{A}+w_{B}v_{A}+v_{B}e^{i\phi}\right)\\
-e^{i\phi/2}\left(\bar{v}_{A}+\bar{w}_{B}\bar{v}_{A}+\bar{v}_{B}e^{-i\phi}\right) & 1+\bar{w}_{B}-v_{A}\bar{v}_{B}e^{-i\phi}
\end{array}
\right).
\end{equation}
The Green's function reads
\begin{align}
&G_{AB}\left(z\right)=\\
&=\frac{1-re^{-i\phi}+\bar{w}_{B}-e^{-i\phi}v_{A}\bar{v}_{B}}{2\mathrm{Re}\left(re^{i\phi}\left(1+\bar{w}_{B}\right)\right)- \left(r^{2}+\left(2\mathrm{Re}\left(w_{B}\right)+\left|w_{B}\right|^{2}\right)\left(1+\left|v_{A}\right|^{2}\right)+2r\mathrm{Re}\left(v_{A}\bar{v}_{B}\right)+\left(1+\left|v_{A}\right|^{2}\right)\left(1+\left|v_{B}\right|^{2}\right)\right)}.
\end{align}
The solution $v_{A}=v_{B}=0$ again gives a holomorphic Green's function. Then, in similar way to the previous two cases it is sufficient to consider linear terms to obtain eigenvalue domain. One is left with three independent equations relating $w_{A}$, $w_{B}$, $r$ and $\phi$
\begin{align}
r&=\frac{\left(1+w_{B}+w_{B}^{2}\right)e^{-i\phi}}{w_{B}},\label{eq:1+H_1+X_r}\\
w_{A}&=w_{B}\left(1+w_{B}\right),\\
0&=-1+\left|w_{B}\right|^{2}+w_{B}e^{2i\phi}\left(1+w_{B}+\bar{w}_{B}\right). \label{eq:1+H_1+X_wB}
\end{align}
It is pointless to give explicit solutions for $w$'s, as they are given by long and cumbersome formulas. One can nevertheless implement them to a numerical code to evaluate the position $r$ of the density border for any $\phi\in[0,2\pi)$ through Eq. (\ref{eq:1+H_1+X_r}). It turns out that the solution has two branches, each forming a closed loop around a different portion of the eigenvalue density. The loops touch at a single point at the origin $z=0$. The result and the comparison to numerical simulations for finite matrices is presented in Fig. \ref{fig:1+H_1+X}. The finite N results based on the MC simulation converge to the theoretically evaluated domain while the size of matrix is being increased.

\section{Conclusions}

One can systematically define the Green's function and the R transform for non-hermitian random matrices given by invariant measures of the form (\ref{muV}) and calculate
the corresponding eigenvalue densities in the limit $N\rightarrow \infty$. The Green's function and the R transform have an analogous structure to those for hermitian matrices known from the standard free probability calculus, but they are quaternionic. We demonstrated how to derive these functions using the Cayley-Dickson construction. 

The addition and multiplication laws for invariant non-hermitian matrices can be deduced from Feynman perturbation theory for matrix models with the invariant measures. These laws provide us with a practical method to calculate the limiting eigenvalue densities for sums and products of non-hermitian invariant random matrices in the limit $N\rightarrow \infty$.

If one combines the method discussed in this paper with the trick of linearization 
\cite{gjjn,hst} one can also apply it to determine eigenvalue densities of more complicated matrix polynomials. So far this method could only be applied to 
polynomials which were hermitian by construction 
as for instance $AB+BA +A+B$ or $ABC+BCA+CAB$ for  $A$, $B$ and $C$ being hermitian. 
This required a special adaptation of the linearization method \cite{a,bms}. 
Now, this constraint can be removed  and one can treat non-hermitian polynomials 
as well, as for instance $AB+A+B$ or $ABC + BCA$.

As mentioned before, the non-hermitian products of random matrices appear in multiple contexts including iterative stochastic evolution of linear systems \cite{n,f,abk}, capacity of complicated MIMO networks \cite{mu,awk} or analysis of multi-dimensional data with asymmetric correlations, like e.g. time series of stock prices \cite{j,bt,blmp,lr} and others. Many of such problems were tackled with methods of free probability in the hermitian case. Now, by the use of the quaternionic R-transform, they can be extended towards the non-hermitian regime which is often computationally harder, but physically better motivated.

\section*{Acknowledgments}
The work was supported by the Grant DEC-2011/02/A/ST1/00119 of the National Centre of Science and German Research Foundation via Collaborative Research Center/Transregio 12. The authors thank R. Janik, A. Jarosz and M.A. Nowak for discussions.

\appendix 
\section{} 
\label{AA}

In this Appendix we recall the Cayley-Dickson construction 
of quaternions and discuss an extension of this construction to matrices.
 
Quaternions $q$ are linear combinations
\begin{equation}
q = a + b i + c j + d k
\end{equation}
of quaternionic units $1,i,j,k$ with real coefficients $a,b,c,d$. 
The quaternionic units form the Hamilton basis which is defined by the 
the following multiplication rules
\begin{equation}
i^2 = j^2 = k^2 = i j k = -1 \ .
\label{eq:HB}
\end{equation}
Quaternions can be represented as ordered pairs $(z,w)$ of complex numbers $z = a + b i$, $w= c + d i$
\begin{equation}
\label{CD0}
q = a + b i + (c + d i) j =z + w j \equiv (z,w)\ .
\end{equation} 
The conjugate of $q$ is 
$q^* = a - b i - c j - d k = \bar{z} - w j = (\bar{z},-w)$.
Quaternion addition is defined by
\begin{equation}
(z,w)+(v,y)=(z+v,w+y) \ ,
\end{equation} 
and multiplication by
\begin{equation}
\label{qprod}
(z,w) (v,y) = (zv - w\bar{y},zy+w\bar{v}) \ .
\end{equation}
It is useful to define auxiliary functions $F$ and $S$ which allow one to
extract the first and the second element of the Cayley-Dickson pair $q=(z,w)$
\begin{equation}
\label{CD}
z=F(q) \ , \ w = S(q) \ . 
\end{equation} 
These functions play an analogous role for quaternions as the
real and imaginary parts for complex numbers. 
We refer to them as to "first part" and "second part" of quaternion. 
Quaternions have a matrix representation in terms of Pauli matrices $\sigma_{i}$. A quaternion $q = a+bi+cj+dk= (z,w)$ 
can be mapped to two-by-two complex matrix (we use the same letter to denote this matrix) by
\begin{equation}
\label{2x2}
q =  a \sigma_0 + b (i\sigma_3) + c (i\sigma_2) + d (i\sigma_1) =
\left(\begin{array}{cc}  a + i b   & c + i d \\
                        -c + i d   & a - i b \end{array} \right) = 
\left(\begin{array}{cc}  z   & w \\
                        -\bar{w} & \bar{z} \end{array} \right) \ .
\end{equation}
The first row of the quaternion matrix representation (\ref{2x2}) can be identified with the Cayley-Dickson pair $(z,w)$. It is easy to check that the matrices $\sigma_0, i\sigma_3, i\sigma_2, i \sigma_1$ form the Hamilton basis (\ref{eq:HB})
\begin{equation}
(i\sigma_3)^2 = (i\sigma_2)^2 = (i\sigma_1)^2 = 
(i\sigma_3)(i\sigma_2)(i\sigma_1) = -\sigma_0 \ .
\end{equation} 
Quaternion addition corresponds in this representation 
to matrix addition while quaternion multiplication to matrix multiplication:
\begin{equation}
\left(\begin{array}{cc}  z   & w \\
                        -\bar{w} & \bar{z} \end{array} \right)
\left(\begin{array}{cc}  v   & y \\
                        -\bar{y} & \bar{v} \end{array} \right) =
\left(\begin{array}{cc}  zv - w\bar{y} & zy+w\bar{v} \\
                        -\bar{z}\bar{y}-\bar{w}v & \bar{z}\bar{v} - \bar{w}y
                        \end{array} \right)      
\end{equation}
as can be seen by comparing the upper rows of matrices in the last equation
with the Cayley-Dickson pairs in (\ref{qprod}). 
The conjugate quaternion corresponds to the hermitian conjugate 
$q^* = q^\dagger$. The quaternion norm squared is $||q||^2 = q q^\dagger = \det q$ 
and the inverse quaternion  $q^{-1} = q^{\dagger}/\det q$. 

We now superimpose the quaternionic structure on $N\times N$ matrices. 
Given four hermitian $N\times N$ matrices $A,B,C,D$ we construct two
non-hermitian matrices $X = A+ i B$ and $Y=C + iD$ and a $2N \times 2N$ 
quaternionic matrix 
\begin{equation}
\label{PMR}
\mathcal{Q} =  A \otimes \sigma_0 +  B \otimes i\sigma_3 + C \otimes 
i\sigma_2 + D \otimes i\sigma_1 =
\left(\begin{array}{cc}  A + i B   & C + i D \\
                        -C + i D &   A - i B \end{array} \right) =
\left(\begin{array}{cc}  X  &  Y \\
                        -Y^\dagger &   X^\dagger  \end{array} \right) \equiv (X,Y) \ .
\end{equation}
The upper row of blocks specify the whole matrix, so we can use the notation $\mathcal{Q} = (X,Y)$ is as in the Cayley-Dickson construction. As before, we define the functions $F$ and $S$ to extract the first and the second block of the pair
\begin{equation}
X = F({\cal Q}) \ , \ Y = S({\cal Q}) \ .
\end{equation}
Additionally we denote a trace in space of quaternionic matrices, which acts as block-trace operation in Pauli matrix representation, by
$\mathrm{Tr_{b}} \mathcal{Q} = (\mathrm{Tr} X, \mathrm{Tr} Y)$.
It projects quaternionic matrices to quaternions
\begin{equation}
\label{CDM}
\mathrm{Tr_{b}} \mathcal{Q} =
\left(\begin{array}{cc}  \mathrm{Tr} X  & \mathrm{Tr} Y \\
                        - \mathrm{Tr} Y^\dagger &   \mathrm{Tr} X^\dagger  \end{array} \right)
\ .
\end{equation}
Equipped with this matrix extension of quaternions one can naturally define pertinent objects to handle non-hermitian matrices including the quaternionic resolvent and
the R transform. 

\section{} 
\label{AB}
We show here that Eq. (\ref{eq:two-dim_delta}) can be treated as 
a representation of the two-dimensional delta function. We have 
\begin{equation}
\frac{1}{\pi} \frac{\partial}{\partial \bar{z}} F \frac{1}{z+wj} = 
\frac{1}{\pi} \frac{\partial}{\partial \bar{z}} F \frac{\bar{z} - w j}{|z|^2 + |w|^2} = \frac{1}{\pi} \frac{\partial}{\partial \bar{z}}  \frac{\bar{z}}{|z|^2 + |w|^2}
=\frac{1}{\pi}\frac{|w|^{2}}{\left(\left|z\right|^{2}+|w|^{2}\right)^{2}} \ .
\end{equation}
The resulting function is a circularly symmetric bell-shaped function located at the origin of the $z$-complex plane. The width of the peak at half maximum is 
of order $|w|$ and the height of order $1/|w|^2$. For any non-zero value of $w$ the integral of this function over the whole $z$-complex plane is equal one 
\begin{equation}
\int \frac{1}{\pi}\frac{|w|^{2}}{\left(\left|z\right|^{2}+|w|^{2}\right)^{2}} d^2 z =1 \ .
\end{equation}
and it is independent of $w$. In the limit $w\rightarrow 0$ 
the volume under the peak stays constant while the width
of the peak tends to zero. The whole function and its integral 
gets concentrated in a single point so it is the delta function.

We note that also the second part of the inverse quaternion can be used to
define the two-dimensional Dirac delta
\begin{equation}
\frac{1}{\pi} \frac{\partial}{\partial \bar{w}} S \frac{1}{z+wj} = 
- \frac{1}{\pi} \frac{\partial}{\partial \bar{w}}  \frac{w}{|z|^2 + |w|^2} =
\frac{w}{\bar{w}} \frac{1}{\pi}
\frac{|w|^{2}}{\left(\left|z\right|^{2}+|w|^{2}\right)^{2}} \ .
\end{equation}
We see that when we fix the argument $\phi$ of $w = e^{i\phi} |w|$
and take the limit $|w| \rightarrow 0^+$ we obtain the delta function
multiplied by a phase factor $e^{2i\phi} \delta^{(2)}(z)$. 
For $\phi=0$ this expression reduces to the delta function. 
This representation is closer in spirit to the standard representation of the one-dimensional delta (\ref{eq:one-dim_delta}) since  
the second part $S$ for quaternions is like the imaginary part $\IM$ 
for 
complex numbers. This representation is however more difficult to use in practice since one has to extract the dependence on $w$ and
calculate the derivative before one takes the limit.

\section{}
\label{AC}
One can show \cite{fs2} that the expression in the brackets in Eq. (\ref{dGX}) 
behaves in the limit $w \rightarrow 0$ as a sum of delta functions located
at eigenvalues of $X$ 
\begin{equation}
\label{HwH}
\lim_{N\rightarrow \infty} \frac{1}{N} \mathrm{Tr} H_R^{-1} |w|^2 H_L^{-1} 
\longrightarrow  \frac{1}{N} \sum_{i=1}^{N}\delta^{(2)}\left(z-\lambda_{i}\right) \ .
\end{equation}
We repeat this reasoning here. The two matrices $H_L$ and $H_R$ (\ref{HLHR}) are hermitian, therefore they have real eigenvalues. We assume that they are non-degenerate. Denote the corresponding eigenvectors by $|L_\alpha\rangle$
and $|R_\alpha\rangle$ respectively
\begin{equation}
\begin{split}
H_L |L_\alpha\rangle & = \Lambda_\alpha |L_\alpha\rangle \\
H_R |R_\alpha\rangle & = \Lambda_\alpha |R_\alpha \rangle \ .
\end{split}
\end{equation} 
The two matrices have the same eigenvalues. Indeed, multiplying the first equation on both sides by  $(\bar{z} \mathbb{1} - X^\dagger)$ and using the identity $(\bar{z} \mathbb{1} - X^\dagger) H_L = H_R (\bar{z} \mathbb{1} - X^\dagger)$ we get 
\begin{equation}
H_R (\bar{z} \mathbb{1} - X^\dagger) |L_\alpha\rangle = \Lambda_\alpha 
(\bar{z} \mathbb{1} - X^\dagger) |L_\alpha\rangle \ ,
\end{equation}
which means that $|R_\alpha\rangle = \eta (\bar{z} \mathbb{1} - X^\dagger) |L_\alpha\rangle$ is an eigenvector of $H_R$ to the same eigenvalue $\Lambda_\alpha$. 
The coefficient $\eta$ is to fix the norm of the vector $\langle R_\alpha | R_\alpha \rangle=1$. Using the spectral decomposition of $H_L$ and $H_R$
one can cast the expression on the left hand side of Eq. (\ref{HwH}) into the form
\begin{equation}
\label{w2}
\frac{1}{N} \mathrm{Tr} H_R^{-1} |w|^2 H_L^{-1} = \frac{1}{N} \sum_{\alpha=1}^N 
\sum_{\beta=1}^N
\frac{|w|^2 \left|\langle R_\alpha|L_\beta\rangle\right|^2}{\Lambda_\alpha\Lambda_\beta} \ .
\end{equation}
The  eigenvalues $\Lambda_\alpha$ and eigenvectors $|R_\alpha\rangle$ and $|L_\alpha\rangle$ depend on $X$ but also on $w$ and $z$. In our notation
this dependence is implicit. We will display it if needed.  
When $z$ is not equal to an eigenvalue of $X$ the eigenvalues 
$\Lambda_\alpha(z,w=0)$ of $H_L$ and $H_R$ (\ref{HLHR}) 
are strictly positive. As a consequence the expression on the right hand side 
of Eq. (\ref{w2}) behaves as $|w|^2$ for small $w$ and vanishes for $w=0$. 
The situation changes when $z$ is equal to an eigenvalue of 
$X$ because in this case the smallest eigenvalue $\Lambda_0(z=\lambda_j,w)=|w|^2$ 
of $H_L$  and $H_R$ tends to zero for $w\rightarrow 0$ and then 
the expression (\ref{w2}) diverges as $1/|w|^2$ 
\begin{equation}
\frac{1}{N} \mathrm{Tr} H_R^{-1} |w|^2 H_L^{-1} \propto 
\frac{\left|\langle R_{0}|L_{0}\rangle\right|^2}{|w|^2} \ .
\end{equation}
The divergence comes from the term $\alpha=\beta=0$ of the sum (\ref{w2}).
To summarize, in the limit $w\rightarrow 0$ 
the expression on the left hand side of Eq. (\ref{HwH}) vanishes for
all values of $z$ except those equal to eigenvalues of $X$. 
A more careful analysis of the behavior of the expression (\ref{w2}) 
in the vicinity of eigenvalues of $X$ shows that 
it behaves as delta functions located at those eigenvalues. In order to see this consider $z$ close to an eigenvalue $\lambda_j$ of $X$: such that the distance $|z - \lambda_j|$
is much smaller than the eigenvalue separation. One can apply the perturbation theory to determine the lowest eigenvalue of $H_L$ and $H_R$ (\ref{HLHR}). Up to the second order one finds
\begin{equation}
\Lambda_{0} \approx 
|w|^2  + \left|\langle R_0|L_0\rangle\right|^2 |z-\lambda_j|^2 \ .
\end{equation}
Inserting $\Lambda_0$ to the diverging term ($\alpha=\beta=0$) of the sum (\ref{w2}) 
and neglecting remaining terms, which are of order $|w|^0$ or $|w|^2$,
one finds for $w \rightarrow 0$
\begin{equation}
\frac{1}{N} \mathrm{Tr} H_R^{-1} |w|^2 H_L^{-1} \approx 
\frac{1}{N} \frac{|w|^2 \left|\langle R_0|L_0\rangle\right|^2}
{\left(|w|^2  + \left|\langle R_0|L_0\rangle\right|^2 |z-\lambda_j|^2\right)^2}
\longrightarrow \frac{1}{N} \delta^{(2)}(z-\lambda_j)  \ .
\end{equation}
The same holds for $z$ in the vicinity of any eigenvalue $\lambda_i$ of $X$
so one eventually arrives at Eq. (\ref{HwH}). 

\section{} 
\label{AD}
In this appendix we recall diagrammatic derivation of Eqs. (\ref{GF_R}) and (\ref{GRquat}). We begin with hermitian matrices and towards the end
we describe how to generalize the derivation to non-hermitian ones. 

\begin{figure}
\centering
\includegraphics[scale=1]{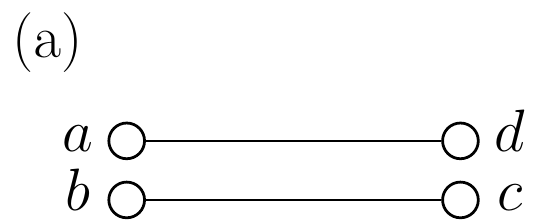} \quad
\includegraphics[scale=1]{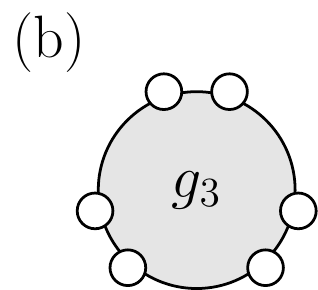} \quad
\includegraphics[scale=1]{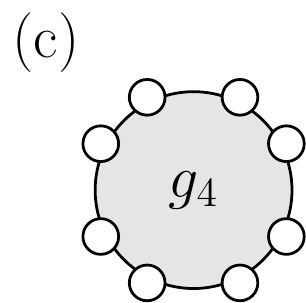}
\caption{\label{fig:AD:Propagator}Pictorial representation of the (a) propagator
(\ref{prop}) and (b,c) the first two vertices generated by the
perturbative expansion (\ref{pexp}). In this appendix we use convention
that external vertices are represented by open circles. 
In contrast, internal vertices (see below) are represented by filled circles. 
Internal vertices correspond to indices which are summed over.}
\end{figure}

\begin{figure}
\centering
\includegraphics[scale=1]{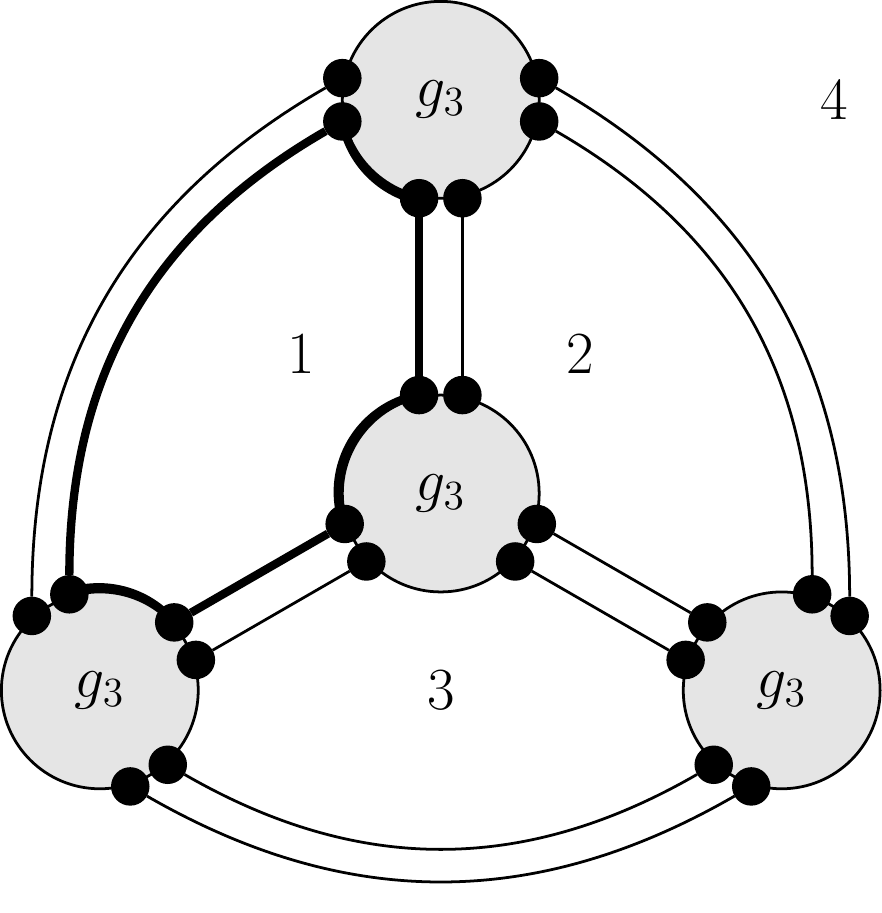}
\caption{\label{AD_tetra} Example of a planar Feynman diagram obtained
from $V=4$ cubic vertices  
and $E=6$ propagators. It has $F=4$ closed lines, like e.g. the thick one 
surrounding the region $1$. With each such closed
line one can associate a face of polyhedron constructed from the edges
of the diagram. If the diagram is drawn on a sphere, the region $4$ can 
be viewed as a compact region or the fourth face of tetrahedron.}
\end{figure}

\begin{figure}
\centering
\includegraphics[scale=0.75]{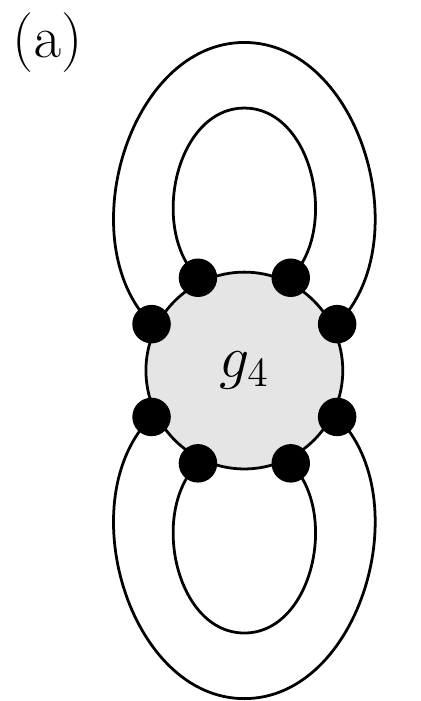}\ \ \ ,\ \ \ 
\includegraphics[scale=0.75]{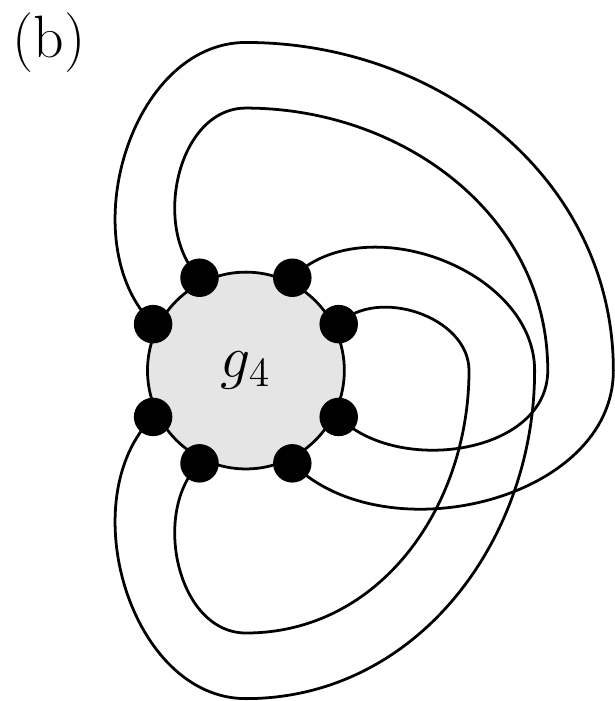}
\caption{\label{fig:AD:planar_nonplanar}Examples of (a) a planar diagram
and (b) a non-planar one.}
\end{figure}

Let us recall main ideas \cite{bipz,biz} and introduce graphical notation. 
We consider hermitian random matrices defined by the probability measure
\begin{equation}
\label{hm}
d \mu(H) \propto DH \exp -N \mathrm{Tr} V(H) = 
DH \exp - N \mathrm{Tr} \left( \frac{1}{2} H^2 + \frac{g_3}{3} H^3 + 
\frac{g_4}{4} H^4 + \ldots \right) \ ,
\end{equation}
and apply Gaussian perturbation theory to calculate statistical averages
$\langle \ldots \rangle$ with respect to this measure. One does it by spliting
the measure into the Gaussian part and the residual part
$V(H) = \frac{1}{2} H^2 + V_R(H)$, where $V_R(H) = \frac{g_3}{3} H^3 + 
\frac{g_4}{4} H^4 + \ldots$, and treating the latter as a perturbation
\begin{equation}
\label{pexp}
d \mu(H) = d \mu_0(H) 
\left(1 + N \mathrm{Tr} V_R(H) + \frac{1}{2} \left(N \mathrm{Tr} V_R(H)\right)^2 
+ \ldots\right)\ ,
\end{equation}
where $d\mu_0(H) \propto \exp - N \mathrm{Tr} H^2/2$ with a normalization
ensuring that $\int d \mu_0(H) = 1$.   
Now the problem reduces to calculating averages of powers of $H$ with respect to the
Gaussian measure $\langle \ldots \rangle_0 = \int d\mu_0(H) \ldots$. 
As follows from the Wick's theorem, averages with respect to the Gaussian measure
of higher order products of $H$ can be replaced by products of the averages of second order 
products. For example for the fourth order the theorem tells us that  
\begin{equation}
\langle H_{ab} H_{cd} H_{ef} H_{gh} \rangle_0 = 
\langle H_{ab} H_{cd} \rangle_0  \langle H_{ef} H_{gh} \rangle_0 + 
\langle H_{ab} H_{ef} \rangle_0  \langle H_{cd} H_{gh} \rangle_0 +
\langle H_{ab} H_{gh} \rangle_0  \langle H_{ef} H_{cd} \rangle_0 \ .
\end{equation}
This observation underlies the idea of Feynman diagrams which are just
graphical representation of products of factors $\langle H_{ab} H_{cd}\rangle_0$.
These factors are called propagators and are graphically represented as double 
lines between pairs of indices $ab$ and $cd$ as shown in Fig. \ref{fig:AD:Propagator}a. 
It is easy to find, by doing the Guassian integral, that the numerical value 
of the propagator for $d\mu_0(H) \propto \exp - N \mathrm{Tr} H^2/2$ is
\begin{equation}
\label{prop}
\langle H_{ab} H_{cd}\rangle_0 = \frac{1}{N} \delta_{ad} \delta_{bc} \ .
\end{equation}
This means that the only non-zero contributions of propagators are $1/N$
for $a$ equal to $d$ and $c$ to $b$. Feynman diagrams are obtained by  
connecting vertices (see Fig. \ref{fig:AD:Propagator}b,c) generated by perturbative
expansion of the residual part $V_R(H)$ (\ref{pexp}) with 
lines representing propagators (\ref{prop}). 
An example of a diagram is shown in Fig. \ref{AD_tetra}.
A contribution of the diagram to the perturbative expansion is given
by a product of contributions from propagators and vertices. 
Propagators contribute (\ref{prop}) and vertices: 
$N g_3/3$ and $N g_4/4$ etc. The factor $N$ comes from the prefactor 
before trace (\ref{pexp}). Let us concentrate on counting 
powers of $N$ for a given diagram. Each propagator contributes a factor $1/N$ (\ref{prop}), 
each vertex contributes a factor $N$ and each closed line, like the thick one drawn in Fig. \ref{AD_tetra}, contributes a factor $N$.
This is because the summation over internal indices of delta 
functions (\ref{prop}) gives for a closed line a factor $\sum_a \delta_{aa} = N$. 
Such a closed line can be viewed as a face of polyhedron made of 
edges (double lines) of the given diagram. The total power 
is thus $N^{V+F-E}$ where $V$ is the number of vertices, $F$ is the 
number of faces of the diagram viewed as polyhedron and $E$ is the number 
of edges (represented us double lines). The combination $V+F-E$ is the Euler characteristic of the polyhedron or equivalently of a two-dimensional
surface on which the diagram can be drawn without edge-crossing. 
In the example shown in Fig. \ref{AD_tetra} we have $V=4$, $F=4$ and $E=6$ 
and $V+F-E=2$. This is equal to the Euler characteristic of sphere. 
Generally, $V+F-E=2-2h$, where $h$ is the genus of 
two-dimensional surface on which diagram can be drawn without crossings. $h=0$ for sphere, $h=1$ for torus, $h=2$ for double torus, etc. 
Thus one can see that the leading contribution comes from planar diagrams, 
which can be drawn on sphere (or equivalently plane) without edge-crossing. 
Those which can be drawn on torus without edge-crossing 
are suppressed with $1/N^2$ factor, on double torus with $1/N^4$ etc. The diagrams
which can be drawn on torus without edge-crossing can also be drawn
on plane but then they have at least one edge-crossing. 
In Fig. \ref{fig:AD:planar_nonplanar} we show examples of lowest 
order diagrams which are planar ($h=0$) and non-planar ($h=1$). 
One can generalize this classification to diagrams with external lines
by considering two-dimensional surfaces with boundaries. 

The main conclusion is that in the limit $N\rightarrow \infty$ one can 
neglect contributions from non-planar diagrams which are suppressed 
with $1/N^{2h}$ powers as compared to the leading contribution coming
from planar diagrams. In effect, in this limit one can restrict 
enumeration only to planar diagrams. This observation was first made in \cite{th} and
the technique of planar diagram enumeration was developed in \cite{bipz,biz}. 

\begin{figure}
\centering
\includegraphics[scale=0.75]{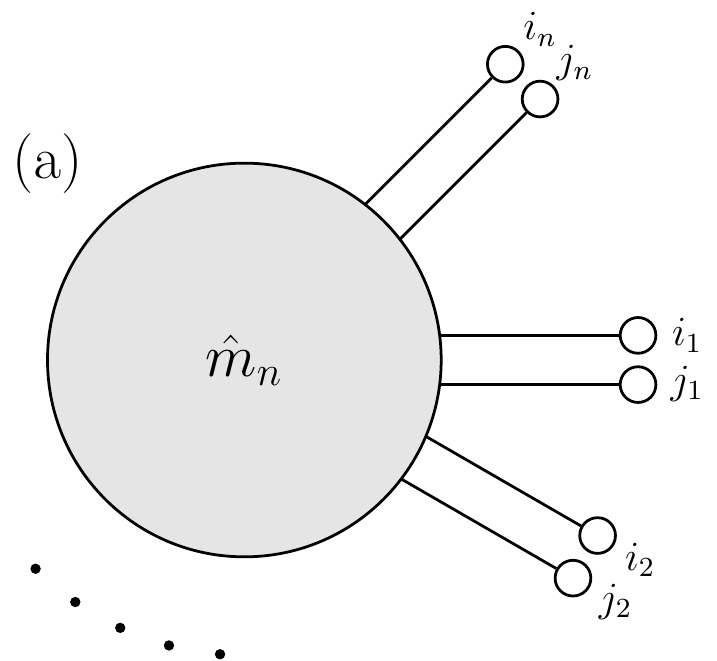}\ \ \ , \ \ \ 
\includegraphics[scale=0.75]{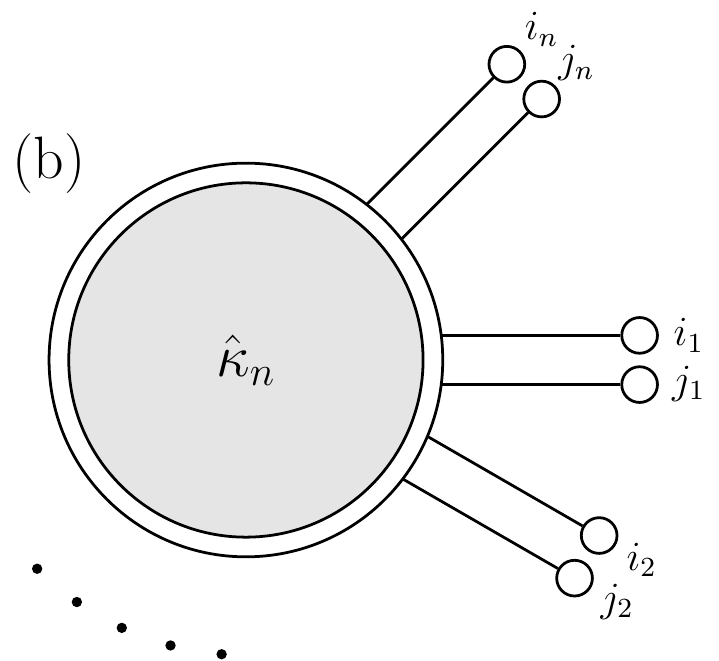}
\caption{\label{AD_f1} Pictorial representation of (a) $n$-point correlation function $\widehat{m}_{n}$ and (b) $n$-point connected correlation function $\widehat{\kappa}_{n}$.}
\end{figure}

\begin{figure}
\centering
\includegraphics[scale=0.7]{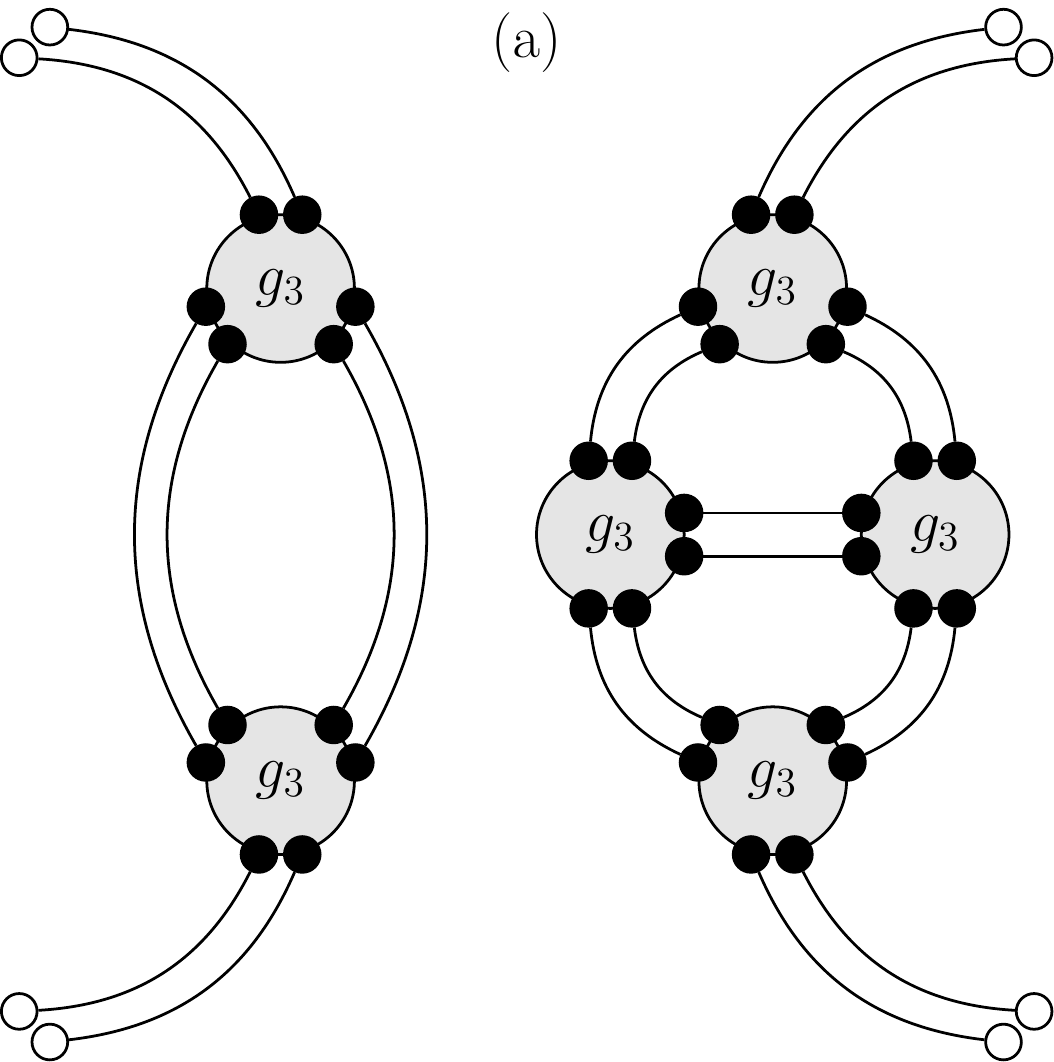}\ \ \ , \ \ \ 
\includegraphics[scale=0.7]{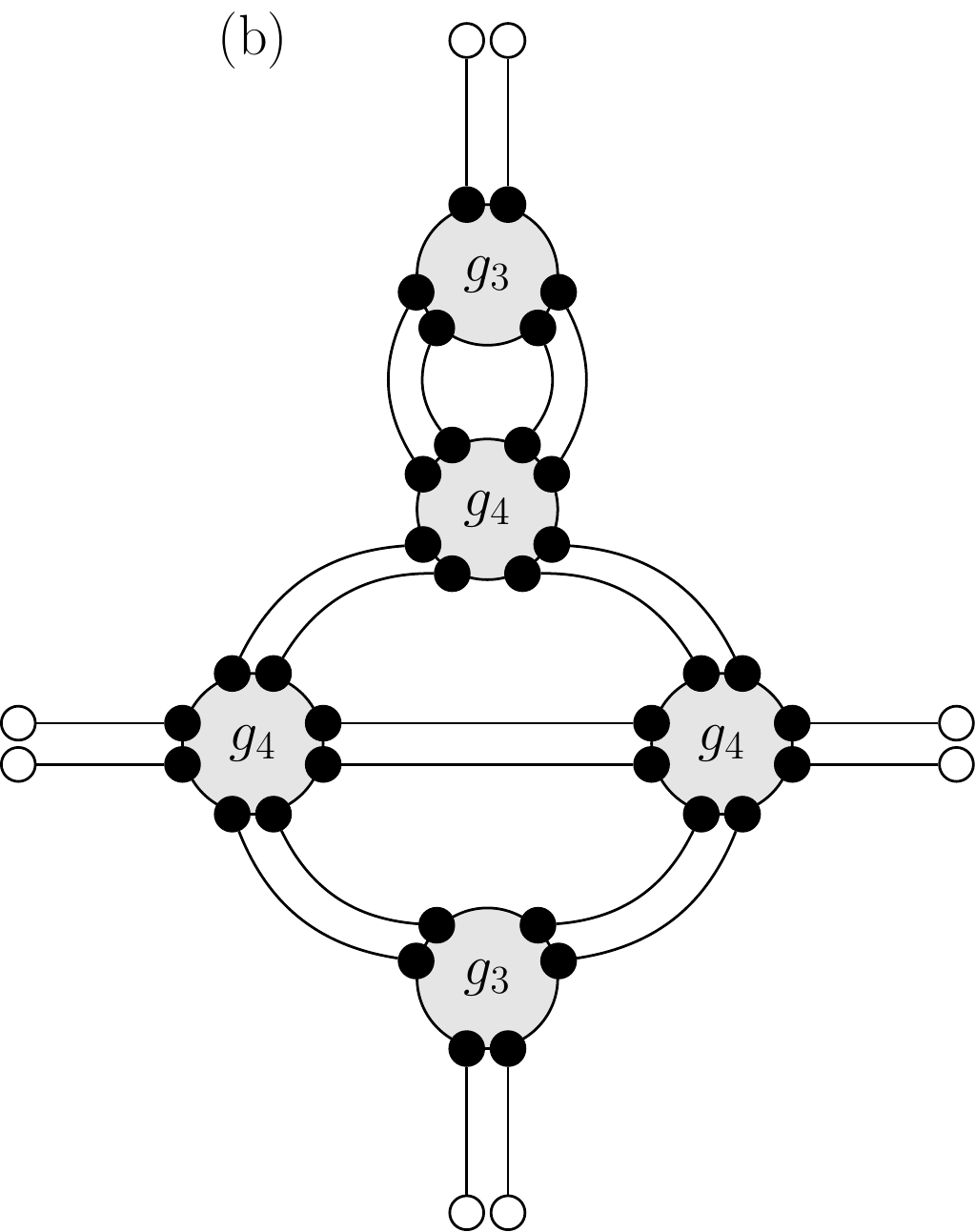}
\caption{\label{AD_f2} Examples of diagrams contributing to $4$-point correlation 
function. The diagram (b) contributes to $4$-point connected correlation function.}
\end{figure}

\begin{figure}
\centering
\includegraphics[scale=0.5]{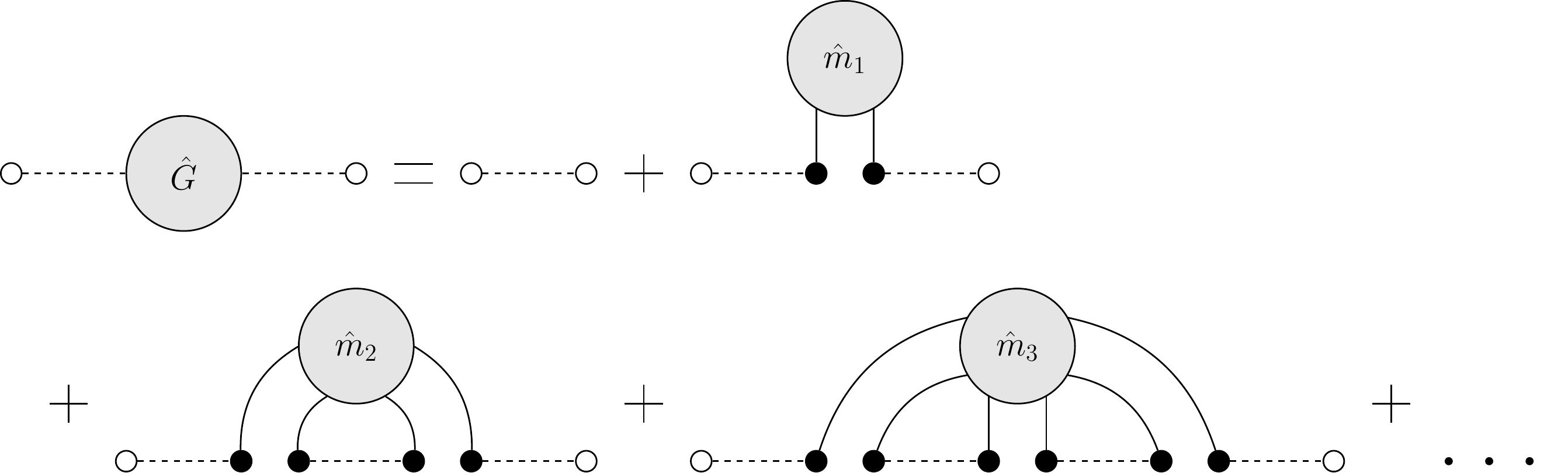}
\caption{\label{AD_f5} Diagrammatic expansion of the resolvent, Eq. (\ref{G_AD}). 
Sum over internal indices is implicit.}
\end{figure}

\begin{figure}
\centering
\includegraphics[scale=0.5]{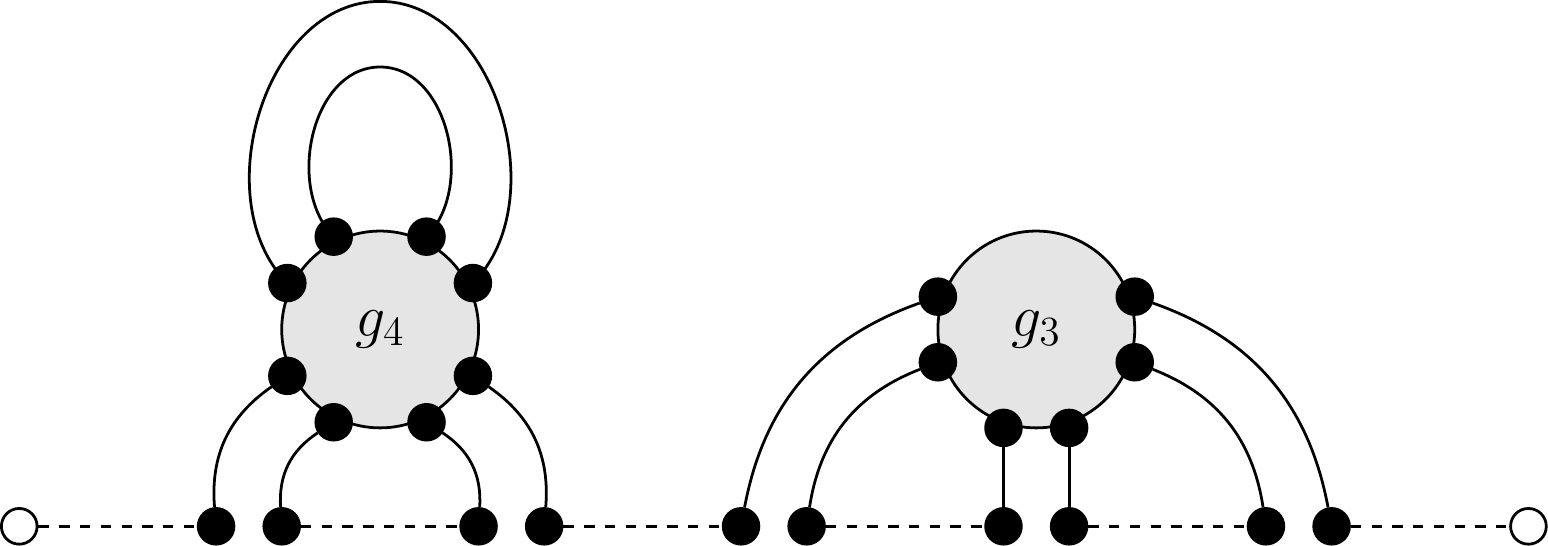}
\caption{\label{AD_f6} Example of a diagram generated by resolvent. 
Its contribution is proportional to $g_3$, $g_4$ and $z^{-6}$.}
\end{figure}

\begin{figure}
\centering
\includegraphics[scale=0.5]{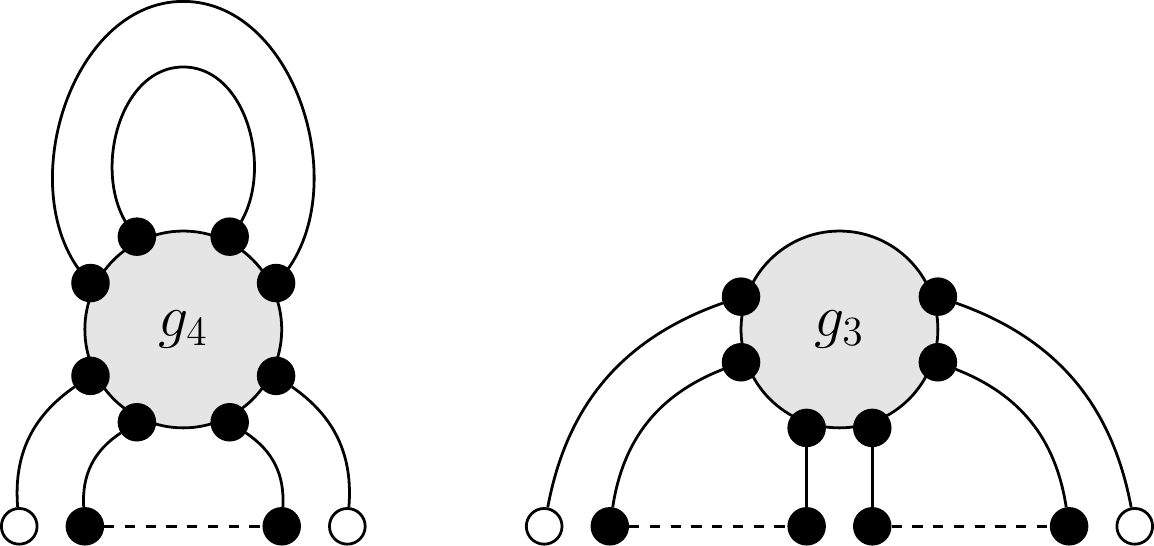}
\caption{\label{AD_f7} One-line-irreducible diagrams that are 
components of the diagram presented in Fig. \ref{AD_f6}}
\end{figure}

So let us recall the idea of graphical enumeration of planar diagrams.
Denote the $n$-point correlation function 
\begin{equation}
\left(\widehat{m}_n\right)_{i_1j_1, \ldots,i_nj_n}=\left\langle H_{i_1 j_1} H_{i_2 j_2} \ldots H_{i_n j_n}\right\rangle
\end{equation}
by a blob with $n$-external double legs as in Fig. \ref{AD_f1}a. 
In perturbation theory the $n$-point correlation function is calculated as a sum over all planar Feynman diagrams with $n$ external legs. An example of a diagram generated by the $4$-point correlation function is shown in Fig. \ref{AD_f2}a.
One also defines an $n$-point connected correlation function which generates only connected diagrams with $n$ external legs. We denote it by
\begin{equation} 
(\widehat{\kappa}_n)_{i_1j_1\ldots,i_nj_n}= \langle \langle H_{i_1 j_1} H_{i_2 j_2} \ldots H_{i_n j_n}\rangle \rangle
\end{equation}
and in the graphical representation by a blob surrounded by a double contour (Fig. \ref{AD_f1}b).
An example of a diagram generated by the $4$-point connected correlation function is shown in Fig. \ref{AD_f2}b.
Below we show how to use these functions in calculations of the Green's function (\ref{eq:GF}). It is useful to consider a generalized resolvent which is an $N\times N$ matrix of the form
\begin{equation}
\label{MG}
\widehat{G} = \left\langle\left(\widehat{Z}-H\right)^{-1} \right\rangle \ ,
\end{equation}
where the matrix $\widehat{Z}$ is an arbitrary invertible constant hermitian matrix. Clearly, by taking the normalized trace
\begin{equation}
\label{ntrG}
G = \lim_{N\rightarrow \infty} \frac{1}{N} \mathrm{Tr} \; \widehat{G}
\end{equation}
and setting $\widehat{Z} = z \mathbb{1}$ one obtains the resolvent $G(z)$ (\ref{eq:GF}). We will set $\widehat{Z} = z \mathbb{1}$ only at the end of the calculations, while they will be done for an arbitrary $\widehat{Z}$ in order to keep track of the underlying algebraic structure. Writing the resolvent (\ref{MG}) as a geometrical series one has
\begin{equation}
\label{G_AD}
\widehat{G} =  
\left\langle (\widehat{Z}-H)^{-1} \right\rangle =
  \widehat{Z}^{-1} + \left\langle \widehat{Z}^{-1} H \widehat{Z}^{-1} \right\rangle +
\left\langle \widehat{Z}^{-1} H \widehat{Z}^{-1} H \widehat{Z}^{-1} \right\rangle + 
\ldots \ .
\end{equation} 
The factors depending on $\widehat{Z}$ can be written outside the brackets since $\widehat{Z}$ is a constant matrix. What remains in the brackets are $n$-point correlation functions. For example, the contribution of the third term in the last equation to the matrix element $\widehat{G}_{af}$ reads
\begin{equation}
\label{abcdef}
\left\langle \left(\widehat{Z}^{-1} H \widehat{Z}^{-1} H \widehat{Z}^{-1}\right)_{af} \right\rangle =
\sum_{b,c,d,e} \left\langle \widehat{Z}^{-1}_{ab} H_{bc} \widehat{Z}^{-1}_{cd} H_{de} \widehat{Z}^{-1}_{ef} \right\rangle = \sum_{b,c,d,e} \widehat{Z}^{-1}_{ab} \widehat{Z}^{-1}_{cd} \widehat{Z}^{-1}_{ef} 
\left\langle H_{bc} H_{de} \right\rangle \ .
\end{equation} 
Graphical representation of equation (\ref{G_AD}) is shown in Fig. \ref{AD_f5}.
The dashed lines correspond to elements of $\widehat{Z}^{-1}$ and indices are associated with the vertices at the endpoints of the dashed line. The internal vertices, like $b,c,d$ and $e$ in Eq. (\ref{abcdef}) are summed over, while the external ones $a,f$ correspond to indices of the matrix element $\widehat{G}_{af}$. An example of a Feynman diagram generated by the resolvent $\widehat{G}$ is shown in Fig. \ref{AD_f6}. 
This diagram is built out of two diagrams shown in Fig. \ref{AD_f7}
sandwiched by dashed lines. Diagrams shown in Fig. \ref{AD_f7} belong to a class of one-line-irreducible diagrams which are defined by the condition that they cannot be disconnected by removing a single dashed line. We denote a generating function of such diagrams by $\widehat{\Sigma}$ which is an $N\times N$ matrix. Analogously, the indices of the $\widehat{\Sigma}$ matrix correspond to the two external vertices and we define $\Sigma\left(z\right)$ as a normalized trace of $\widehat{\Sigma}$:
\begin{equation}
\label{ntr}
\Sigma = \lim_{N\rightarrow \infty} \frac{1}{N} \mathrm{Tr} \; \widehat{\Sigma}.
\end{equation}
The reason why it is convenient to introduce the generating function $\widehat{\Sigma}$ is, that one can write down a set of diagrammatic equations relating $\widehat{\Sigma}$ and $\widehat{G}$ and derive from them a closed-form expression for the resolvent. These equations are often named after Dyson and Schwinger who invented a general class of such equations in field theory to sum contributions of different types of Feynman diagrams. 

\begin{figure}
\centering
\includegraphics[scale=0.5]{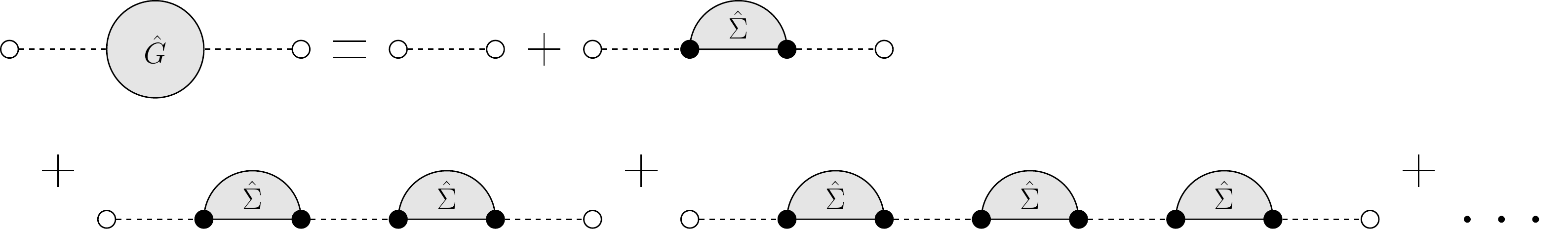}
\caption{\label{AD_f8} Pictorial representation of the first Dyson-Schwinger equation (\ref{DS1}). Each diagram may be constructed by aligning some one-line-irreducible diagrams and connecting them alternately with dashed lines.}
\end{figure}

\begin{figure}
\centering
\includegraphics[scale=0.5]{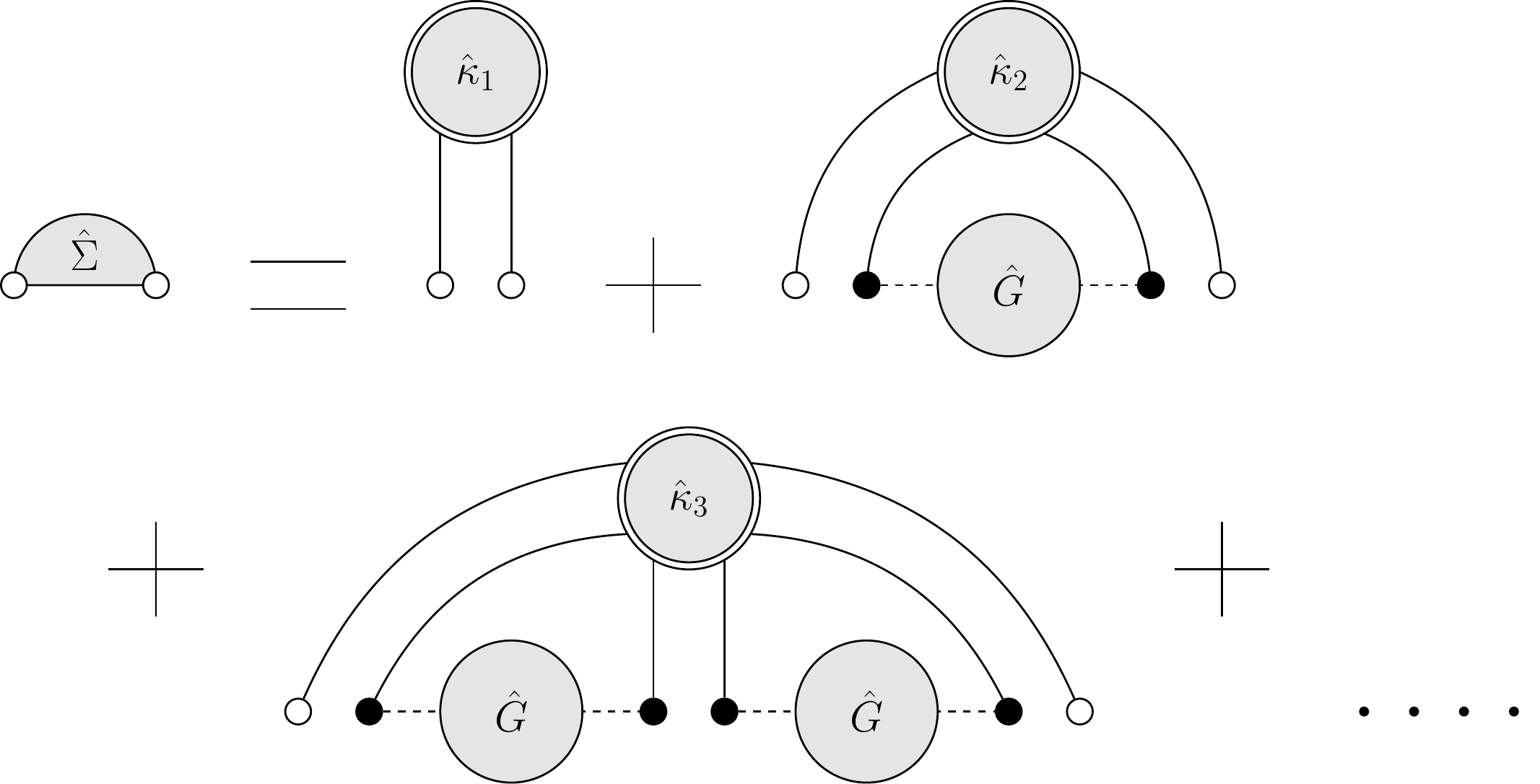}
\caption{\label{AD_f9} Pictorial representation of the second Dyson-Schwinger equation (\ref{DS2}). Each one-line-irreducible diagram may be constructed by aligning some diagrams next to each other and straddling them with a connected diagram.}
\end{figure}

In our case we will write down two Dyson-Schwinger equations to sum up all planar diagrams contributing to the resolvent $\widehat{G}$. The first Dyson-Schwinger equation exploits the fact that any diagram in $\widehat{G}$ can be constructed as an alternating chain of one-line-irreducible diagrams and dashed lines (compare Figs. \ref{AD_f6} and \ref{AD_f7}). This leads to the graphical relation shown in Fig. \ref{AD_f8} 
which is equivalent to the equation
\begin{equation}
\label{DS1}
\widehat{G} = \widehat{Z}^{-1} + \widehat{Z}^{-1} \widehat{\Sigma} \widehat{Z}^{-1} + \widehat{Z}^{-1} \widehat{\Sigma} \widehat{Z}^{-1} \widehat{\Sigma} \widehat{Z}^{-1} + \ldots = \left(\widehat{Z} - \widehat{\Sigma}\right)^{-1} \ .
\end{equation}
This is the first Dyson-Schwinger equation. There is another independent relation between 
$\widehat{G}$ and $\widehat{\Sigma}$ which exploits the fact that any one-line-irreducible diagram can be constructed by straddling $\widehat{G}$-diagrams by legs of a connected diagram as shown in Fig. \ref{AD_f9}. 
This graphical equation is equivalent to
\begin{equation}
\label{DS2}
\widehat{\Sigma}_{ab} = (\widehat{\kappa}_1)_{ab} + 
\sum_{cd} (\widehat{\kappa}_2)_{ac,db} \widehat{G}_{cd} + 
\sum_{cefd} (\widehat{\kappa}_3)_{ac,ef,db} \widehat{G}_{ce}
\widehat{G}_{fd}  + \ldots \ .
\end{equation}
This is the second Dyson-Schwinger equation. The two Dyson-Schwinger equations simplify for $\widehat{Z} = z \mathbb{1}$ ($\widehat{Z}_{ab} = z \delta_{ab}$) since in this case 
$\widehat{G} = G(z) \mathbb{1}$ and $\widehat{\Sigma} = \Sigma(z) \mathbb{1}$, where $G(z)$ and $\Sigma(z)$ are complex scalar functions. In effect, equations (\ref{DS1}) and (\ref{DS2}) reduce after taking the normalized trace to scalar equations
\begin{equation}
\label{ds1}
G(z) = \left(z - \Sigma(z)\right)^{-1}
\end{equation}
and 
\begin{equation}
\label{ds2}
\Sigma(z) = \kappa_1 + \kappa_2 G(z) + \kappa_3 G^2(z) + \ldots 
\end{equation}
where
\begin{equation}
\kappa_n = \lim_{N\rightarrow \infty} \frac{1}{N} 
\langle \langle \mathrm{Tr} H^n \rangle\rangle \ .
\end{equation}
These coefficients are connected planar (or free) cumulants as they are defined as sums over all connected planar diagrams with $n$ external legs. The generating function for planar cumulants is called R transform in free probability and it is defined as
\begin{equation}
R(z) = \kappa_1 + \kappa_2 z + \kappa_3 z^2 + \ldots = \sum_{n=1}^\infty \kappa_n z^{n-1} \ .
\end{equation}
Using the R transform one can concisely write the second Dyson-Schwinger equation (\ref{ds2}) 
\begin{equation}
\label{ds2prim}
\Sigma(z) = R(G(z)). 
\end{equation}
Inserting this into the first Dyson-Schwinger equation (\ref{ds1}) one 
eventually obtains the formula
\begin{equation}
\label{ds1prim}
G(z) = \frac{1}{z-R(G(z))}, 
\end{equation}
which was earlier mentioned in the main text (\ref{GF_R}).

Generalization of these equations to non-hermitian matrices is straightforward. The Dyson-Schwinger equations (\ref{DS1}) and (\ref{DS2}) have the same form as for hermitian matrices but a slighly more complex index structure, since the $N\times N$ hermitian matrices are considered as quaternionic matrices (or equivalently $2N\times 2N$ matrices with a particular block structure) (\ref{Gq}) and (\ref{calX}). The additional index structure takes care of parts of quaternion (positions of blocks in the $2N \times 2N$ matrices). After taking the block-trace (\ref{CDM})one gets a matrix equation for $2\times 2$ matrices (quaternions) (\ref{GRquat}) instead of scalar equation (\ref{ds1prim}).


\begin{thebibliography}{99}
\bibitem{m} M. L. Mehta, \textit{Random matrices} (Elsevier, Amsterdam, 2004). 
\bibitem{cpv} A. Crisanti, G. Paladin, A. Vulpiani, \textit{Products of random matrices
, Random matrices and their applications}, (Springer-Verlag, Berlin Heidelberg, 1993).
\bibitem{ckn} J. E. Cohen, H. Kesten and C.M. Newman (eds),
\textit{Random matrices and their applications}, Contemporary Mathematics
\textbf{50}, (Providence, RI: American Mathematical Society, 1986).
\bibitem{gmw} T. Guhr, A. M\"{u}ller-Groeling, H. A. Weidenm\"{u}ller, Phys. Rep. \textbf{299}, 189 (1998).
\bibitem{abf} G. Akemann, J. Baik, P. Di Francesco (Ed.), \textit{The Oxford Handbook of Random Matrix Theory}, (Oxford University Press, Oxford, 2011).
\bibitem{vt} A. M. Tulino and S. Verd\'u, \textit{Random Matrix Theory and Wireless Communications}, Foundations and Trends in Communications and Information Theory, \textbf{1} 1 (2004).
\bibitem{agz} G. W. Anderson, A. Guionnet, O. Zeitouni, \textit{An Introduction to Random Matrices}, (Cambridge University Press, Cambridge, 2009).
\bibitem{v1} D. V. Voiculescu, J. Funct. Anal. \textbf{66}, 323 (1986).
\bibitem{v2} D. V. Voiculescu, J. Operator Theory, \textbf{18}, 223 (1987).
\bibitem{vdn} D. V. Voiculescu, K. J. Dykema, A. Nica, \textit{Free random variables}, CRM Monograph Series, Vol. 1 (American Mathematical Society, Providence, RI, 1992).
\bibitem{s} R. Speicher, Math. Ann. \textbf{298}, 611 (1994).
\bibitem{rs} N. R. Rao and R. Speicher, Elect. Comm. in Probab. \textbf{12}, 248 (2007).
\bibitem{bjw} Z. Burda, R.A. Janik and B. Waclaw, Phys. Rev. E \textbf{81} 041132 (2010).
\bibitem{ab}  G. Akemann and Z. Burda, J. Phys. A: Math. Theor. \textbf{45}, 465201 (2012).
\bibitem{hil} F. Haake, F. Izrailev, N. Lehmann et al., Z.Phys. \textbf{B88} 359 (1992).
\bibitem{fs1} Y. V. Fyodorov, H.-J.Sommers, JETP Letters \textbf{63}, 1026 (1996).
\bibitem{fs2} Y. V. Fyodorov, H.-J.Sommers, J. Math. Phys. \textbf{38}, 1918 (1997). 
\bibitem{fss} Y. V. Fyodorov, D. V. Savin, H.-J. Sommers, Phys. Rev. \textbf{E55},
R4857 (1997). 
\bibitem{jnz} R. Janik, M. Nowak, I. Zahed,  Phys.Lett. \textbf{B392} 155 (1997).
\bibitem{v} J. J. M. Verbaarschot, Nucl.Phys. \textbf{A642} 305 (1998). 
\bibitem{bt} C. Biely, S. Thurner, Quant. Finance \textbf{8}, 705 (2008).
\bibitem{j} A. Jarosz, {\em Hermitian and non-Hermitian covariance estimators for multivariate Gaussian and non-Gaussian assets from random matrix theory}, 
arXiv:1010.5220.
\bibitem{scss} H.-J. Sommers, A. Crisanti, H. Sompolinsky, and Y. Stein, Phys. Rev. Lett. \textbf{60}, 1895 (1988).
\bibitem{dki} S. Drozdz, J. Kwapien, A. A. Ioannides, Acta Phys. Pol. {\bf B42} 987 (2011).
\bibitem{gjjn} E. Gudowska-Nowak, R. A. Janik, J. Jurkiewicz and M. A. Nowak, Nucl Phys. \textbf{B670} 479 (2003). 
\bibitem{bgntw} Z. Burda, J. Grela, M. A. Nowak, W. Tarnowski, P. Warchol, 
Phys.Rev.Lett. \textbf{113} 104102 (2014).
\bibitem{ra} K. Rajan, L.F. Abbott Phys Rev Lett. \textbf{97}, 188104 (2006).
\bibitem{sg} S.E. Skipetrov, A. Goetschy, J. Phys. A: Math. Theor. \textbf{44}, 065102 (2011).
\bibitem{rj} R. A. Janik, Ph.D. Thesis, Jagiellonian University, Cracow 1996, 
(unpublished).
\bibitem{jnpz1} R. A. Janik, M. A. Nowak, G. Papp and I. Zahed, Nucl.Phys. B \textbf{501}, 603 (1997).
\bibitem{jnpz2}
R. A. Janik, M. A. Nowak, G. Papp and I. Zahed, Acta Phys. Polon. \textbf{B28} 2949 (1997). 
\bibitem{jnpwz} 
R. A. Janik, M. A. Nowak, G. Papp, J. Wambach and I. Zahed, Phys. Rev. 
\textbf{E55} 4100 (1997).
\bibitem{fz1} J. Feinberg and A. Zee, Nucl. Phys. \textbf{B501} 643 (1997).
\bibitem{fz2} J. Feinberg and A. Zee, Nucl. Phys. \textbf{B504} 579 (1997). 
\bibitem{jn1} A. Jarosz and M. A. Nowak, \textit{A Novel Approach to Non-Hermitian Random Matrix Models}, arXiv:math-ph/0402057.
\bibitem{jn2} A. Jarosz and M.A. Nowak, J. Phys. {\bf A39} 10107 (2006).
\bibitem{bjn} Z. Burda, R. A. Janik, M. A. Nowak, Phys. Rev. E \textbf{84}, 061125 (2011).
\bibitem{r} T. Rodgers, J. Math. Phys. \textbf{51} 093304 (2010).
\bibitem{bcc} Ch. Bordenave, P. Caputo, D. Chafai, Commun. Pure Appl. Math. \textbf{67} 621 (2014).
\bibitem{j2} A. Jarosz, \textit{Summing free unitary random matrices}, arXiv:1010.5220.    
\bibitem{cm} J. T. Chalker and B. Mehlig, Phys. Rev. Lett. \textbf{81} 3367 (1998). 
\bibitem{jnnpz} R.A. Janik, W. Norenberg, M.A. Nowak, G. Papp and I. Zahed, 
Phys. Rev. \textbf{E60} 2699 (1999). 
\bibitem{th}  G. 't Hooft, Nucl. Phys. B \textit{72}, 461 (1974).
\bibitem{bipz} E. Brezin, C. Itzykson, G. Parisi and J.-B. Zuber, Commun. Math. Phys.
\textbf{59}, 35 (1978).
\bibitem{biz} D. Bessis, C. Itzykson and J.-B. Zuber, Adv. Appl. Math. \textbf{1}, 109 (1980).
\bibitem{gi} V. L. Girko, \textit{Spectral theory of random matrices}, in Russian (Nauka, Moscow, 1988).
\bibitem{g} J. Ginibre, J. Math. Phys. \textbf{6} 440–449 (1965).
\bibitem{fks} Y. V. Fyodorov, B. A. Khoruzhenko, H.-J. Sommers, 
Phys.Lett. \text{A226} 46 (1997).  
\bibitem{hst} U. Haagerup, H. Schultz, S. Thorbj\o{}rnsen, Adv. Math. \textbf{204} 1 
(2006).
\bibitem{a} G. W. Anderson, Ann. Probab. \textbf{41} 2103 (2013)
\bibitem{bms} S. Belinschi, T. Mai and R. Speicher, \textit{Analytic subordination theory of operator-valued free additive convolution and the solution of a general random matrix problem}, arXiv:1303.3196.
\bibitem{n} C. M. Newman, Commun. Math. Phys. \textbf{103} 121–126 (1986).
\bibitem{f} P. J. Forrester, {\em Asymptotics of finite system Lyapunov exponents for some random matrix ensembles}, arXiv:1501.05702.   
\bibitem{abk} G. Akemann, Z. Burda, M. Kieburg, J. Phys. A: Math. Theor. \textbf{47} 395202 (2014).    
\bibitem{mu} R. R. Mueller, IEEE Trans. Inf. Theor. \textbf{48} 2086 (2002).
\bibitem{awk} G. Akemann, M. Kieburg, L. Wei, J. Phys. A: Math. Theor. \textbf{46}, 275205 (2013).
\bibitem{blmp} J.-P. Bouchaud, L. Laloux, M.A. Miceli, M. Potters, European Physical Journal B \textbf{2} 201-207 (2007).
\bibitem{lr} G. Livan, L. Rebecchi, Eur. Phys. J. B \textbf{85}, 213 (2012).
\end{thebibliography}
\end{document}